\documentclass[a4paper,fleqn,usenatbib]{mnras}

\usepackage{hyperref}

\usepackage{newtxtext,newtxmath}

\usepackage[T1]{fontenc}
\usepackage{ae,aecompl}

\usepackage{amsmath}

\usepackage{enumerate}
\usepackage{amsmath}
\usepackage{float}
  \usepackage{longtable}
  \usepackage{lipsum} 
   \usepackage{threeparttable}
  \usepackage{mdframed} 
  \usepackage{multirow}
  \usepackage{color, colortbl}
  \usepackage{xcolor}
\usepackage{graphicx}
\usepackage{natbib}

\usepackage[toc,title,page]{appendix}

\newcommand\raa{RAA}

\title[QNe in the outskirts of galaxies: An explanation of the FRB phenomenon]{Quark-Novae in the outskirts of galaxies: An explanation of the Fast Radio Burst phenomenon}

\author[Ouyed et al.]{
Rachid Ouyed,$^{1}$\thanks{E-mail: rouyed@ucalgary.ca}
Denis Leahy,$^{1}$
and Nico Koning$^{1}$
\\
$^{1}$Department of Physics and Astronomy, University of Calgary, 2500 University Drive NW, Calgary, AB, T2N 1N4, Canada
}

\date{Accepted: 05 November 2020. Received: 09 June 2020; in original form ZZZ}

\pubyear{2021}

\hypersetup{draft}    
\begin{document}
\label{firstpage}
\pagerange{\pageref{firstpage}--\pageref{lastpage}}
\maketitle

\begin{abstract}
We show that old isolated neutron stars in groups and clusters of galaxies  experiencing a Quark-Nova phase (QN: an explosive
transition to a quark star) may be the source of FRBs. Each  of the millions of fragments  of the ultra-relativistic QN ejecta provides a collisionless   plasma for which  the ambient medium (galactic/halo, the intra-group/intra-cluster medium) acts as a relativistic plasma beam.  The Buneman and the  Weibel instabilities, successively induced by the beam in the fragment,  generate particle bunching and observed coherent  emission at  GHz frequency with a corresponding  fluence in the Jy ms range.  The duration, frequency drift and the rate  are  in  agreement with observed properties of  FRBs.    Repeats (on timescales of minutes to months) are due to seeing multiple fragments each beaming at a different direction and coming in at different times. Single (non-repeating) FRBs, occur when only emission from the primary fragment is within the detector's sensitivity.   Key properties of FRB 121102 (its years of activity)  and  of FRB 180916.J0158$+$65 (its $\sim 16$ day period) are recovered.   The spatial and temporal coincidence between SGR 1935$+$2154 and FRB 200428 finds an explanation in our model. We give testable predictions.
\end{abstract}

\begin{keywords}
stars: neutron -- pulsars: general -- Astrophysics - High Energy Astrophysical Phenomena --  acceleration of particles -- plasmas
\end{keywords}



\section{Introduction}
 \label{sec:intro}

FRB science began with the Lorimer burst (\citealt{lorimer_2007}) and followed with  
a decade of discovery of dozens of  intense, millisecond, highly dispersed radio bursts in the GHz range (see http://frbcat.org/; \citealt{petroff_2016}). An FRB may consist of  single or multiple pulses of milliseconds duration.  While most FRBs were  one-off events, a few were repeats  (\citealt{spitler_2016,sholz_2016,chime_2019a,chime_2019b}). 
 FRB dispersion measures (DM) of hundreds of  pc cm$^{-3}$ put them at extra-Galactic to cosmological distances 
which makes them very bright ($> 10^{41}$ erg s$^{-1}$) with their  high brightness temperatures 
requiring a coherent emission mechanism (\citealt{kellermann_1969}; see also \citealt{katz_2014,popov_2018}).

 Observations  and derived properties of FRBs can be found in the literature (\citealt{thornton_2013,spitler_2014,kulkarni_2014,petroff_2016,ravi_2016,gajjar_2018,michilli_2018,lorimer_2018,cordes_2019}).  The large beams of current radio telescopes  makes it difficult to pin-point the host galaxies of most FRBs let alone their association with known astrophysical objects.
This makes it hard to constrain models despite the numerous ideas suggested 
 (\citealt{platts_2019}). The  X-ray activity of the galactic soft $\gamma$-ray repeater (SGR) 1935$+$2154 (\citealt{barthelmy_2020})
coincided spatially with FRB 200428 (\citealt{scholz_2020,chime_2020b,witze_2020,bochenek_2020}). This supports the association of at least some FRBs with SGRs. The repeating nature of FRBs has been used to argue against  catastrophic scenarios, but  we show that is not necessarily the case.

In the QN model,  a massive NS ($\sim 2M_{\odot}$; born from stellar progenitors in the 20-40$M_{\odot}$ mass range) converts spontaneously to a quark star (QS) when quark deconfinement in its core, and the subsequent explosive combustion of neutrons to quarks, is triggered either by: (i) spin-down if  born rapidly
 rotating with a period of a few milliseconds  (\citealt{staff_2006}); (ii) quark nucleation on timescales of hundreds of millions of years if  born slowly rotating.
  During the QN, the outermost layers of the NS are ejected at ultra-relativistic speeds.  From \citet{ouyed_leahy_2009}, the QN ejecta  breaks up into
millions  of dense fragments in a plasma state (hereafter ``chunks").  In case (i), the QN occurs within years of a core-collapse SN (ccSN) explosion of a massive star  
 with the chunks  embedded in the SN. In   case (ii) the QN occurs in the outskirts of galaxies. 
 \citet{ouyed_2020}  showed that the interaction of the chunks with the expanded SN ejecta 
  gives properties (intermittency, light-curve and spectrum) of long-duration Gamma-ray
 bursts (LGRBs). Here, we focus on isolated QNe;  old NSs experiencing the QN phase, outside their birth galaxies. 
 
  Slowly rotating, massive NSs  rely  on quark nucleation in their core to  trigger the QN.
 For nucleation timescales  $\ge10^8$ years  (e.g.  \citealt{bombaci_2004,harko_2004}), a candidate NS with a typical kick velocity of $\sim 300$ km s$^{-1}$  travels a distance 
$> \sim 30$ kpc from its birth place. It would explode  in the intra-group or intra-cluster medium. 
The  chunks  travel through the ambient medium/plasma and expand until they become collisionless. They  experience two 
 inter-penetrating collisionless instabilities:   the Buneman instability (BI) then the Buneman-induced thermal Weibel instability (WI). This triggers particle bunching, coherent synchrotron emission (CSE) and FRBs as shown here.
   The  fragmented nature of the QN ejecta (with every chunk emitting in a specific  direction and at a different time) allows repetition.  Thus FRBs from a QN, a one-off cataclysmic event, are inherently repeaters with single (non-repeating) FRBs, occurring when only emission from the primary chunk is detected.

Hereafter, unprimed quantities are in the chunk's reference frame while the superscripts ``ns" and ``obs." refer to  the NS frame (i.e. the ambient medium)
 and the observer's frame, respectively. The transformation from the local NS frame to the chunk's frame is  $dt^{\rm ns} = \Gamma_{\rm c}dt$  while 
 the transformations from the chunk's frame to the observer's frame are
 $dt^{\rm obs.}= (1+z) dt/D(\Gamma_{\rm c},\theta_{\rm c})$, $\nu^{\rm obs.} = D(\Gamma_{\rm c},\theta_{\rm c})\nu/(1+z)$ with $z$ the source's redshift  and $\theta_{\rm c}$ the  angle between the observer and chunk's velocity vectors; $D(\Gamma_{\rm c},\theta_{\rm c})$ is the chunk's Doppler factor with $\Gamma_{\rm c}$ the Lorentz factor.

 We consider three media: (i) the intra-group medium (IGpM), with number density  $n_{\rm amb.}^{\rm ns}\simeq 10^{-4}$-$10^{-2}$ cm$^{-3}$   (e.g. \citealt{cavaliere_2016}); (ii) the intra-galaxy cluster medium (ICM)  with   $n_{\rm amb.}^{\rm ns}\simeq 10^{-4}$-$10^{-2}$ cm$^{-3}$ 
   (e.g. \citealt{fabian_1994}); (iii) the intergalactic medium (IGM) with $n_{\rm amb.}^{\rm ns}\simeq 10^{-7}$ cm$^{-3}$ (e.g. \citealt{mcquinn_2016}).
        Hereafter ICM refers jointly to the hot diffuse gas  observed in groups 
 and clusters of galaxies. Because the majority of galaxies are in groups (e.g. \citealt{tully_1987}) we take conditions in the IgCM
 with ambient density of $n_{\rm amb.}^{\rm ns}=10^{-3}$ cm$^{-3}$ as our fiducial value.
   The paper focuses on the interaction of the QN chunks with such an ambient medium and 
  is structured as follows: In \S \ref{sec:QN} we give an overview of the  QN  ejecta, and how it becomes collisionless as it travels in the ambient medium.
   We  describe  the  relevant plasma instabilities (BI and WI) and  
    the resulting CSE. The application to FRBs  is done in \S \ref{sec:FRB-ICM-QNe}.   In \S \ref{sec:discussion},  we list our model's 
  predictions and limitations, and conclude in \S \ref{sec:conclusion}.

\section{The QN and its ejecta}
\label{sec:QN}

The energy release during the conversion of
a NS to a QS is  $\sim 3.8\times 10^{53}\ {\rm erg}\times (M_{\rm NS}/2M_{\odot})\times (\Delta E_{\rm dec.}/100\ {\rm MeV})$
for a  NS mass of $M_{\rm NS}=2M_{\odot}$  and a conversion energy release of  $\Delta E_{\rm dec.}=100$ MeV per neutron   (\citet{weber_2005}). This is a fraction of the combined conversion   energy  and gravitational binding energy (\citealt{keranen_2005,niebergal_phd_2011,ouyed_amir_phd_2018,ouyed_2020}). 
A large part is in  kinetic energy of the QN ejecta $E_{\rm QN}\sim 10^{52}$-$10^{53}$ erg when  the converting NS is hot (as in the case of a QN in the wake of a ccSN (\citet{ouyed_2020}). For the case of  an old isolated cold  NS,  the very slow nucleation timescales
 means that most of the conversion  energy is lost to neutrinos yielding a kinetic energy  $E_{\rm QN}\sim 10^{51}$-$10^{52}$ erg.   The QN ejecta consists of the outermost crust layers of the NS with a mass $M_{\rm QN}\sim 10^{-5}M_{\odot}$   and a Lorentz factor $\Gamma_{\rm QN}=E_{\rm QN}/(M_{\rm QN}c^2)\simeq 10^2$-$10^3$.

\subsection{Ejecta properties and statistics}
\label{sec:honeycomb}

With $N_{\rm c}= 10^6$  the number of chunks,  a  typical mass\footnote{Dimensionless quantities
 are defined as $f_{\rm x}= f/10^{x}$ with quantities in cgs units.} is $m_{\rm c}=M_{\rm QN}/N_{\rm c}\simeq  10^{22.3}\ {\rm gm}\times M_{\rm QN, 28.3}/N_{\rm c, 6}$. The chunk's Lorentz factor is taken to be constant  with $\Gamma_{\rm c}=\Gamma_{\rm QN}=10^{2.5}$
  for a QN ejecta's kinetic energy $E_{\rm QN}=\Gamma_{\rm c}M_{\rm QN}c^2\simeq 5.7\times 10^{51}$ erg; i.e.  1\% of the conversion energy is converted to kinetic energy; these values are listed in Table \ref{table:parameters}\footnote{Appendices, tables and
figures in the online supplementary material (SM) for this paper  are denoted by a prefix ``S".}.

In Appendix \ref{appendix:honeycomb} we  summarize  chunk properties:   

\begin{itemize}

  \item The  chunks are equally spaced in solid angle $\Omega$ around the explosion site.
  Defining $N_{\theta_{\rm c}}$ as the number of chunks per angle\footnote{$\theta_{\rm c}$ is also  the angle of the chunk's motion 
with respect to the observer's line-of-sight.} $\theta_{\rm c}$, we write $dN_{\theta_{\rm c}}/d\Omega =const.= N_{\rm c}/4\pi$ with
$d\Omega = 2\pi \sin{\theta_{\rm c}} d\theta_{\rm c}$ so that $dN_{\theta_{\rm c}}/d\theta_{\rm c}= (N_{\rm c}/2) \sin{\theta_{\rm c}}\simeq (N_{\rm c}/2) \theta_{\rm c}$
with $\theta_{\rm c} <<1$. Because $N_{\rm c}\pi \Delta \theta_{\rm c}^2 = 4\pi$, the average angular separation  is 
 \begin{equation}
 \label{eq:dthetas}
 \Delta \theta_{\rm s}=2 \Delta \theta_{\rm c} = \frac{4}{N_{\rm c}^{1/2}}\simeq \frac{4\times 10^{-3}}{N_{\rm c, 6}^{1/2}}\ ,
 \end{equation}
  yielding $\Delta \theta_{\rm s}\sim 1/\Gamma_{\rm c}$  for  $N_{\rm c}=10^6$ and $\Gamma_{\rm c}=10^{2.5}$.
   
   The geometry for chunk angular spacing is a 2-dimensional honeycomb (see Figure \ref{figure:honeycomb}) 
    with 1 primary ($i=1$) chunk at $0 \le \theta_{\rm c} <  \Delta\theta_{\rm c}$ and $6\times (i-1)$ chunks for subsequent,
    and concentric, ``rings"    (with $i\ge 2$;  6 secondary chunks, 12 tertiary chunks etc..). The 
      mean angle of the primary (P), secondary (S) and tertiary (T) chunks are $\bar{\theta}_{\rm P}\simeq  4/(3N_{\rm c, 6}^{1/2})$, 
    $\bar{\theta}_{\rm S}\simeq 2.3 \bar{\theta}_{\rm P}$ and $\bar{\theta}_{\rm T}\sim 6 \bar{\theta}_{\rm P}$ (Eq. (\ref{eq:mean-angles}));

\item    with  $\Gamma_{\rm c}^2>>1$ and $\theta_{\rm c}<<1$, the Doppler factor is $D_{\rm c}(\Gamma_{\rm c},\theta_{\rm c})\simeq 2\Gamma_{\rm c}/f(\theta_{\rm c})$ with
    
 \begin{equation}
 \label{eq:ftheta}
  f(\theta_{\rm c})= 1+(\Gamma_{\rm c}\theta_{\rm c})^2\ ,
  \end{equation}
     and $f(\theta_{\rm c}) << \Gamma_{\rm c}^2$. The average values are
 \begin{align}
 \label{eq:fPfS}
 f(\bar{\theta}_{\rm P})\sim 1 + 0.18\frac{\Gamma_{\rm c, 2.5}^2}{N_{\rm c, 6}}\nonumber\\
  f(\bar{\theta}_{\rm S})\sim 1 + 0.97\frac{\Gamma_{\rm c, 2.5}^2}{N_{\rm c, 6}}\nonumber\\
   f(\bar{\theta}_{\rm T})\sim 1 + 6.67\frac{\Gamma_{\rm c, 2.5}^2}{N_{\rm c, 6}}\ ;
 \end{align}
     
\item   The average change in $f(\theta_{\rm c})$ from one chunk  to another  with respect to the observer is (see Eq. (\ref{eq:Deltac})): 
 \begin{align}
  \label{eq:Deltaf}
 \Delta f^{\rm chunks}\simeq \frac{1.6}{\pi}\times \frac{\Gamma_{\rm c, 2.5}^2}{N_{\rm c, 6}}\ .
 \end{align}

  \end{itemize}

 \begin{figure*}
 \centering
 \includegraphics[scale=0.6]{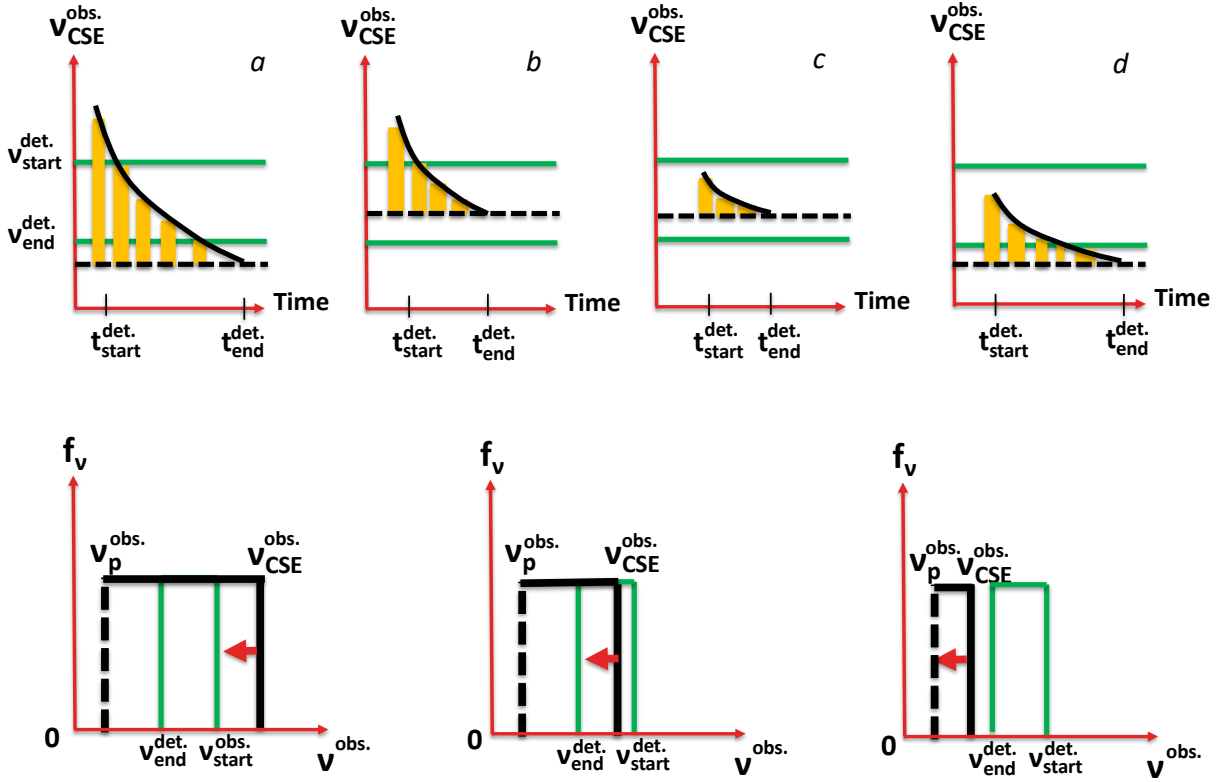}
  \caption{{\bf Top panels}: Schematic representation of frequency drifting in time  (Eq. (\ref{eq:CSE-frequency-time}))
  for a flat emitted spectrum.  The detector has maximum and minimum frequencies $\nu_{\rm max.}^{\rm det.}$ and $\nu_{\rm min.}^{\rm det.}$. 
  The  possible scenarios are: a) $\nu_{\rm CSE, max.}^{\rm obs.}(\theta_{\rm c})>\nu_{\rm max.}^{\rm det.}$
   and $\nu_{\rm p, e}^{\rm obs.}(\theta_{\rm c})<\nu_{\rm min.}^{\rm det.}$;  b) $\nu_{\rm CSE, max.}^{\rm obs.}(\theta_{\rm c})>\nu_{\rm max.}^{\rm det.}$
   and $\nu_{\rm min.}^{\rm det.}<\nu_{\rm p, e}^{\rm obs.}(\theta_{\rm c})<\nu_{\rm max.}^{\rm det.}$; c) $\nu_{\rm max.}^{\rm det.}> \nu_{\rm CSE, max.}^{\rm obs.}(\theta_{\rm c})$
   and $\nu_{\rm p, e}^{\rm obs.}(\theta_{\rm c})>\nu_{\rm min.}^{\rm det.}$; d) $\nu_{\rm max.}^{\rm det.}> \nu_{\rm CSE, max.}^{\rm obs.}(\theta_{\rm c})>\nu_{\rm min.}^{\rm det.}$
   and $\nu_{\rm p, e}^{\rm obs.}(\theta_{\rm c})<\nu_{\rm min.}^{\rm det.}$.   The vertical bands indicate the range of detected frequencies at any given time. 
   The horizontal dashed line is the chunk's plasma frequency  $\nu_{\rm p, e}^{\rm obs.}(\theta_{\rm c})$. 
     {\bf Bottom panels}: The spectrum at three different times for  case ``a". The  arrow shows the  frequency decreasing in time and drifting through the detector's band.}
    \label{figure:drifting-a}
 \end{figure*}

 \subsection{The collisionless QN chunks}
 \label{sec:critical-radii}

The evolution of the QN chunks is given 
in  \citet{ouyed_leahy_2009}. The later evolution is:

(i) The QN chunk becomes optically thin to photons when it expands to a radius $R_{\rm c, opt.}\simeq 2.2\times 10^{10}\ {\rm cm}\times m_{\rm c, 22.3}^{1/2}\kappa_{\rm c, -1}^{1/2}$, using  $R_{\rm c}=1/\kappa_{\rm c}\rho_{\rm c}$; $\kappa_{\rm c}=0.1$ cm$^{2}$ gm$^{-1}$  
is the opacity,  $\rho_{\rm c}=n_{\rm c}m_{\rm H}=(3m_{\rm c}/4\pi R_{\rm c}^3)\times m_{\rm H}$ is its density and $m_{\rm H}$  the hydrogen mass. The baryon number density is $n_{\rm c, opt.}=3m_{\rm c}/4\pi R_{\rm c, opt.}^3m_{\rm H}\simeq 2.8\times 10^{14}\ {\rm cm}^{-3}\times m_{\rm c, 22.3}^{-1/2}\kappa_{\rm c, -1}^{-3/2}$;

 (ii) The chunk is optically thin to hadronic collisions when it expands to a radius $R_{\rm c, HH}\simeq 1.5\times 10^9\ {\rm cm}\times m_{\rm c, 22.3}^{1/2}\sigma_{\rm HH, -27}^{1/2}$,  from $R_{\rm c}= 1/n_{\rm c}\sigma_{\rm HH}$;  $\sigma_{\rm HH, -27}$ is the hadron-hadron cross-section in  milli-barns (\citealt{letaw_1983});

 (iii) A chunk is  subject to electron Coulomb collisions so it thermalizes and expands beyond $R_{\rm c, opt.}$ from
   internal  pressure. The  electron Coulomb collision length   for number density $n_{\rm c}$ and temperature $T_{\rm c}$ is (\citealt{richardson_2019}) 
 $\lambda_{\rm  c, C}\simeq 1.1\times 10^4\ {\rm cm}\times T_{\rm c}^2/n_{\rm c}$ with 
    Coulomb parameter $\ln{\Lambda}= 20$ (\citealt{lang_1999}).  During the early evolution $\lambda_{\rm  c, C}<<R_{\rm c}$.

 After the chunk is optically thin, hadronic collisions continue to heat it.  From Appendix \ref{appendix:heating}, hadronic collisions  with the ambient medium and 
 thermalization from  Coulomb collisions  expand the chunk until it becomes 
  collisionless when $R_{\rm c}=\lambda_{\rm  c, C}$.    Table \ref{table:ICM-equations} lists the  
  number density ($n_{\rm cc}$), radius ($R_{\rm cc}$) and thermal speed  ($\beta_{\rm cc}=v_{\rm cc}/c$) of a typical chunk when
  it becomes collisionless.    At this stage, its interaction with the  ICM 
 triggers the BI and WI (Appendix \ref{appendix:instabilities} and Figure \ref{fig:WI-stages}), yielding particle bunching and CSE (Appendix \ref{appendix:CSE} and Figure \ref{figure:bunch-geometry})  with observed properties (frequency, duration and fluence) similar to  FRBs.

 Chunk electrons bunch up on scales $\lambda_{\rm b}(t)$ yielding CSE at a frequency $\nu_{\rm CSE}(t)=c/\lambda_{\rm b}(t)$.
 It decreases in time, due to bunches merging and increasing in size, as (Eq.  (\ref{eq:CSE-frequency-time})) $\nu_{\rm CSE}(t)=
     \nu_{\rm CSE}(0)\times (1+t/t_{\rm m-WI})^{-\delta_{\rm m-WI}}$ with $t_{\rm m-WI}$ 
     the characteristic merging timescale  (Eq. (\ref{eq:tmWI})). CSE ceases when $\nu_{\rm CSE}(t)$
     drops to the chunk's plasma frequency, $\nu_{\rm p, e} \simeq 9\ {\rm kHz}\times n_{\rm cc}^{1/2}$ (\citealt{lang_1999}), and can no longer escape.      In the observer's frame,   the initial (maximum) CSE frequency is $\nu_{\rm CSE, max.}^{\rm obs.}(\theta_{\rm c})=D(\Gamma_{\rm c},\theta_{\rm c}) \nu_{\rm CSE}(0)/(1+z)$,
  its duration is $\Delta t_{\rm CSE}^{\rm obs.}=(1+z) \Delta t_{\rm CSE}/D(\Gamma_{\rm c},\theta_{\rm c})$,  and the luminosity is $L_{\rm CSE}^{\rm obs.}(\theta_{\rm c})= (D(\Gamma_{\rm c},\theta_{\rm c})^4/(1+z)^2)L_{\rm CSE}$;  $\nu_{\rm CSE}(0), L_{\rm CSE}(0)$ and $\Delta t_{\rm CSE}$ are
     derived in Appendix \ref{appendix:CSE}. I.e.
 
 \begin{align}
 \nu_{\rm CSE, max.}^{\rm obs.}(\theta_{\rm c}) &\simeq      \frac{11.6\ {\rm GHz}}{(1+z)f(\theta_{\rm c})}\times \delta_{\rm CSE, -1}\gamma_{\rm CSE, 1}^2\times 
  \Gamma_{\rm c, 2.5}n_{\rm cc, 1}^{1/2} \label{eq:nu-CSEobs}\\
  L_{\rm CSE}^{\rm obs.}(\theta_{\rm c})&\simeq  \frac{4.7\times 10^{46}\ {\rm erg\ s}^{-1}}{(1+z)^2f(\theta_{\rm c})^4}\times \frac{\zeta_{\rm BI, -1}\beta_{\rm WI, -1}}{\zeta_{\rm m-WI, 2}}     \times\nonumber \\
    &\times        \frac{\Gamma_{\rm c, 2.5}^6 R_{\rm cc, 15}^2 n_{\rm amb., -3}^{\rm ns}\label{eq:LCSEobs}}{\beta_{\rm cc, -2}} \\
   \Delta t_{\rm CSE}^{\rm obs.}(\theta_{\rm c})&\simeq \left(  \left(642.7\delta_{\rm CSE, -1}\gamma_{\rm CSE,1}^2\right)^{\frac{1}{\delta_{\rm m-WI}}} -1\right) \times t_{\rm m-WI}^{\rm obs.}(\theta_{\rm c})\label{eq:nu-CSEobs}\ ,
     \end{align}
         
     with the observed characteristic bunch merging timescale:

\begin{equation}
\label{eq:tmWIobs}
t_{\rm m-WI}^{\rm obs.}(\theta_{\rm c})= \frac{(1+z)}{D(\Gamma_{\rm c},\theta_{\rm c})} t_{\rm m-WI} \simeq 0.24\ {\rm ms}\times (1+z)f(\theta_{\rm c})\times \frac{\zeta_{\rm m-WI,2}}{\Gamma_{\rm c, 2.5}n_{\rm cc, 1}^{1/2}}\ .
\end{equation}
The parameters $\delta_{\rm CSE},\gamma_{\rm CSE},\zeta_{\rm BI},\beta_{\rm WI},\zeta_{\rm m-WI}, \delta_{\rm m-WI}$ are defined and given in Table \ref{table:parameters}.
     
The observed CSE    frequency  decreases in time  as (Eq.  (\ref{eq:CSE-frequency-time})) $\nu_{\rm CSE}^{\rm obs.}(\theta_{\rm c},t^{\rm obs.})=
 \nu_{\rm CSE, max.}^{\rm obs.}(\theta_{\rm c})\times (1+t^{\rm obs.}/t_{\rm m-WI}^{\rm obs.}(\theta_{\rm c}))^{-\delta_{\rm m-WI}}$ reaching  a minimum value of 
     
 \begin{equation}
  \label{eq:nu-plasma-obs}
  \nu_{\rm CSE, min}^{\rm obs.}(\theta_{\rm c})=\frac{D(\Gamma_{\rm c},\theta_{\rm c})}{1+z}\nu_{\rm p, e}\simeq \frac{18\ {\rm MHz}}{(1+z)f(\theta_{\rm c})}\times \Gamma_{\rm c,2.5}n_{\rm cc, 1}^{1/2}\ .
   \end{equation}
            
  Figure \ref{figure:drifting-a} illustrates frequency drifting in time through the detector's band for the case of a flat spectrum.
  The vertical bands  indicate that CSE emerges from the chunk at frequencies $\nu_{\rm CSE, min.}^{\rm obs.}(\theta_{\rm c}) \le  \nu_{\rm CSE}^{\rm obs.}(\theta_{\rm c},t^{\rm obs.})\le \nu_{\rm CSE, max.}^{\rm obs.}(\theta_{\rm c})$. 
       For the steep power-law spectrum case (Figure \ref{figure:drifting-b}), CSE is detectable  around the peak frequency. The frequency bands at a given time are
 narrower in the case of the steep power-law spectrum case, more like observed FRBs.
   
\begin{table}
  \begin{center}
  \caption{Selected FRB detectors}
  \begin{tabular}{|c|c|c|c|}\hline
  Telescope &   Band (MHz) & sensitivity (Jy ms) \\\hline
  Arecibo$^1$ &   $\sim$ 1210-1530  & $\sim 0.1$ \\\hline
  Parkes$^2$ &   $\sim$ 1180-1580 &  $\sim 1$ \\\hline
  ASKAP$^3$ &   $\sim$ 1210-1530 & $\sim$ 10 \\\hline
  CHIME$^4$ &   $\sim$  400-800 &  $\sim 0.1$ \\\hline
  LOFAR$^5$ &   $\sim$ 110-240 &  $> 10^3$\\\hline
   \end{tabular}\\
    \label{table:detectors}
   \end{center}
   $^1$\url{http://www.naic.edu/alfa/gen_info/info_obs.shtml}. \\   
   $^2$\url{https://www.parkes.atnf.csiro.au/cgi-bin/public_wiki/wiki.pl?MB20}.\\
   $^3$\url{https://www.atnf.csiro.au/projects/askap/index.html}. \\
   $^4$\url{https://chime-experiment.ca/instrument}. \\
   $^5$ LOFAR's high-band antenna (\citealt{vanhaarlem_2013}). 
  \end{table}      
     

\begin{table*}
\begin{center}
\caption{{\bf Simulations}: a  repeating FRB (yielding the waterfall plot in Figure \ref{figure:FRBs-waterfall-repeating}) from an ICM-QN at $z=0.2$ with $N_{\rm c} = 10^5,\log{\Gamma_{\rm c}}=2.5, \log{m_{\rm c}}\ {\rm (gm)}=22.55$;
other parameters are the fiducial values listed in Table \ref{table:parameters}. The QN FRB simulator can be run at: \url{http://www.quarknova.ca/FRBSimulator/}.} 
\begin{minipage}{0.5\textwidth}
  \begin{center}
  {\bf Detections} ($\theta_{\rm c}(\# 0)=5.47$E-3)$^{1}$
  \end{center}
\end{minipage}%
  \vskip 0.05in
\begin{tabular}{cccccccccc}\hline
    \#        &            $\Delta \theta_{\rm c}$ (rad)$^{2}$       &  $f(\theta_{\rm c})$      & 
    $t_{\rm OA}^{\rm obs.}$ (days)      & $\Delta t_{\rm OA}^{\rm obs.}$ (days)$^{3}$    & Frequency (MHz)$^{4}$     &  Width (ms) &  Fluence (Jy ms)$^{5}$\\\hline    
     0           &         0.00        &  3.99        &  0.00        &  0.00        &  3.00E3      &  0.93        &  CHIME (64.00)\\\hline
                                                             &              &              &              &              &              &              &  Parkes (5.56)\\\hline 
                  &                                        &              &              &              &              &              &  Arecibo (4.6)\\\hline   
     1           &         2.05E-3     &  6.65        &  9.23        &  9.23        &  1.80E3      &  1.55        &  CHIME (8.27)\\\hline
        &                                       &              &              &              &              &              &  Arecibo (0.6)\\\hline  
     2           &       1.84E-3     &  9.76        &  19.99       &  10.76       &  1.23E3      &  2.27        &  CHIME (1.79)\\\hline
     3           &       7.71E-4     &  11.26       &  25.20       &  5.20        &  1.06E3      &  2.62   &       CHIME (1.01)\\\hline
     \\
\end{tabular}\\
\label{table:FRBs-repeating}
\end{center}
Here, and in all tables in the online supplementary material (SM):\\
$^{1}$ $\theta_{\rm c}(\# 0)$ is the viewing angle in radians of the first detected chunk.\\
$^{2}$ $\Delta \theta_{\rm c}$ is the difference between the current chunk's $\theta_{\rm c}$ and the previous one that arrived.\\
$^{3}$ $\Delta t_{\rm OA}^{\rm obs.}$ is the time-delay (difference in time-of-arrival, $t_{\rm OA}^{\rm obs.}$)
between successive bursts.\\
$^{4}$ Shown  is the maximum CSE frequency $\nu_{\rm CSE}^{\rm obs.}(\theta_{\rm c})$ (Eq. (\ref{eq:nu-CSEobs})). \\
$^{5}$ Only detectors with fluence above sensitivity threshold (see Table \ref{table:detectors}) are shown. 
\end{table*}

\section{FRBs from ICM-QNe}
\label{sec:FRB-ICM-QNe}

 Listed in Table \ref{table:ICM-equations}  are frequency, duration and fluence  of the resulting FRBs\footnote{In Appendix \ref{appendix:CSE-in-Detectors}, we describe the CSE properties (duration, spectrum, fluence) as measured by current detectors (Table \ref{table:detectors}).
   We  derive the  spectrum, flux density, band-integrated flux density and fluence for the case of a power-law spectrum.}. 
  For a detector with  bandwidth $\Delta \nu^{\rm det.}=\nu_{\rm max.}^{\rm det.}-\nu_{\rm min.}^{\rm det.}$ ($\nu_{\rm max.}^{\rm det.}$ and $\nu_{\rm min.}^{\rm det.}$ are the maximum and minimum frequency):

  (i) If  $\nu_{\rm CSE, max.}^{\rm obs.}(\theta_{\rm c})>\nu_{\rm max.}^{\rm det.}$
and $\nu_{\rm CSE, min.}^{\rm obs.}(\theta_{\rm c})<\nu_{\rm min.}^{\rm det.}$, the duration
of the CSE is set by the time it takes emission to drift through the detector's band (Eq. (\ref{eq:dt-detector})): 

\begin{align}
\label{eq:dtdet}
   \Delta t_{\rm CSE}^{\rm det.}(\theta_{\rm c}) &= t_{\rm m-WI}^{\rm obs.}(\theta_{\rm c}) \times\nonumber\\
   &\times  \left(\left( \frac{\nu_{\rm CSE, max.}^{\rm obs.}(\theta_{\rm c})}{\nu_{\rm min.}^{\rm det.}} \right)^{1/\delta_{\rm m-WI}} - \left(\frac{\nu_{\rm CSE, max.}^{\rm obs.}(\theta_{\rm c})}{\nu_{\rm max.}^{\rm det.}} \right)^{1/\delta_{\rm m-WI}}\right)  \ .
   \end{align}

The FRB duration is the minimum between the CSE duration  and the 
drifting time through the detector's band 

\begin{equation}
\label{eq:dtFRB}
\Delta t_{\rm FRB}^{\rm obs.}(\theta_{\rm c})={\rm MIN} \left[ \Delta t_{\rm CSE}^{\rm obs.}(\theta_{\rm c}),\Delta t_{\rm CSE}^{\rm det.}(\theta_{\rm c}) \right] \ ;
\end{equation}

  (ii) The band-integrated fluence for a flat spectrum is 
  $F(\theta_{\rm c},\delta_{\rm m-WI},0)= \mathcal{F}(\theta_{\rm c},0)\times \mathcal{G}(\theta_{\rm c},\delta_{\rm m-WI},0)$
with $\mathcal{F}(\theta_{\rm c},0)$ given by Eq. (\ref{appendix:eq:fluence-flat-3})  as

\begin{align}
\label{appendix:eq:fluence-flat-4}
\mathcal{F}(\theta_{\rm c},0)&\simeq 810\ {\rm Jy\ ms}\   \frac{1}{f(\theta_{\rm c})^2d_{\rm L, 27.5}^2}\times \frac{\zeta_{\rm BI, -1}\beta_{\rm WI, -1}}{\delta_{\rm CSE, -1}\gamma_{\rm CSE,1}^2}\times \nonumber\\
&\times \frac{\Gamma_{\rm c, 2.5}^4 R_{\rm cc, 15}^2 {n_{\rm amb., -3}^{\rm ns}}}{n_{\rm cc, 1}\beta_{\rm cc, -2}}\ ,
\end{align}
with $\mathcal{G}(\theta_{\rm c},\delta_{\rm m-WI},0)$ (Eq. (\ref{appendix:eq:fluence-flat-2})) varying from a few to a few thousands
depending on the
detector's band  (Table \ref{table:G-fluence}); the luminosity distance is $d_{\rm L}\simeq 1$ Gpc
for sources at $z=0.2$.

  Also listed in Table \ref{table:ICM-equations}   is the  timescale between repeats
  $\Delta t_{\rm repeat}^{\rm obs.}$ (emission from two separate chunks)  found  by setting 
   $f(\theta_{\rm c})= (1.6/\pi)\times \Gamma_{\rm c, 2.5}^2/N_{\rm c, 6}$ (Eq. (\ref{eq:Deltaf}))
   in  $t_{\rm cc}^{\rm obs.}$ (Eq. (\ref{appendix:eq:tccobs})):
 
  \begin{align}
 \label{eq:dtrepeat}
 \Delta t_{\rm repeat}^{\rm obs.} \simeq 1.3\ {\rm days}\times (1+z)\times \frac{1}{N_{\rm c, 6}} 
   \times \left( \frac{m_{\rm c, 22.3}\kappa_{\rm c, -1}}{\sigma_{\rm HH, -27}^{3}\Gamma_{\rm c, 2.5} ({n_{\rm amb., -3}^{\rm ns}})^{3}} \right)^{1/5}\ .
 \end{align}

 The delay between two successive CSE bursts  (two different emitting chunks) for an ICM-QN
 depends mainly on $N_{\rm c}$. For fiducial parameter values, typical
 time between repeats is of the order of days in the observer's frame.

Table \ref{table:ICM-QNe-1}  shows examples of FRBs from ICM-QNe
   obtained using  the equations in Table \ref{table:ICM-equations}.
 Because $f(\bar{\theta}_{\rm c})=1+(\Gamma_{\rm c}\bar{\theta}_{\rm c})^2$ (Eq. (\ref{eq:ftheta})) is controlled by $N_{\rm c}$
 ($\theta_{\rm c}\propto 1/N_{\rm c}^{1/2}$) and $\Gamma_{\rm c}$, we  vary these
  parameters and show a range in viewing angles 
 on FRB detections in our model.  Chunks with  $\nu_{\rm CSE, max.}^{\rm obs.}(\theta_{\rm c})> \nu_{\rm max.}^{\rm det.}$ 
   will eventually be detected when the frequency drifts into the detector's band.
  The drift ends when the CSE frequency reaches $\mathbf{{\rm MAX}}(\nu_{\rm min.}^{\rm det.},\nu_{\rm CSE, min.}^{\rm obs.}(\theta_{\rm c}))$.   For fiducial parameters values,  the plasma frequency $\nu_{\rm p, e}^{\rm obs.}(\theta_{\rm c})$ is below the detector's minimum frequency  which implies that CSE drifts through the band (Figure 1).
The fluences per detector  are  given in Table \ref{table:ICM-QNe-1} with the shaded cells showing the values
 within detector's sensitivity (Table \ref{table:detectors}).    Repetition (see Appendix \ref{appendix:repeats-non-repeats}) is set by the angular separation between
emitting chunks which yields a roughly constant time delay between bursts. Boxes A, B and C in Table \ref{table:ICM-QNe-1},  show that typical time delays between bursts within a repeating FRB is $12\ {\rm days}  < \Delta t_{\rm repeat}^{\rm obs.}< 20\ {\rm days}$.

\subsection{Simulations}
\label{sec:simulations}

A parameter survey was performed 
by simulating\footnote{The QN FRB simulator can be run at: \url{http://www.quarknova.ca/FRBSimulator/}} the QN chunks starting from the moment when they become a collisionless plasma within the ICM:
(i) We distribute the chunks on the surface of a unit sphere using the ``Regular Placement"  algorithm
 (\citealt{deserno_2004}).  The chunks are placed  along rings of constant latitude, and  evenly spaced over the  sphere.  The simulation then chooses a random direction vector from which to view the sphere, and calculates the $\theta_{\rm c}$ angle of each chunk based on this vector; (ii) The zero time of arrival $t_{\rm OA}^{\rm obs.}=0$  is set by the chunk which has the
  minimum value of $t_{\rm cc}^{\rm obs.}$.  The time of arrival of subsequent chunks, $t_{\rm OA}^{\rm obs.} (\theta_{\rm c})$, are recorded with respect to the signal  from the first detected  chunk. The time delay between successive chunks we define as $\Delta t_{\rm OA}^{\rm obs.}$     and  $\Delta \theta_{\rm C}$ as the difference between the current chunk's $\theta_{\rm c}$ and the previous one that arrived;  (iii) We take  $E_{\rm QN}=10^{51}$ erg which fixes the chunk's mass for a given $N_{\rm c}$ and $\Gamma_{\rm c}$; $m_{\rm c}=E_{\rm QN}/N_{\rm c}\Gamma_{\rm c}c^2$; (iv)  For non-constant chunk mass simulations, we sample the mass from a Gaussian distribution with a mean  $\bar{m}_{\rm c}=E_{\rm QN}/N_{\rm c}\Gamma_{\rm c}c^2$ and standard deviation $\sigma_{\rm \log (m_{\rm c})}= 1.0$.

 Single FRBs are detector-dependent and  occur when one of the conditions in Eq. (\ref{eq:repeat-condition}) is violated,  which occurs mostly when    $f(\bar{\theta}_{\rm T})>>f(\bar{\theta}_{\rm S})>>f(\bar{\theta}_{\rm P})>>1$ (Appendix \ref{appendix:repeats-non-repeats}).
 This is the case  when  considering fewer QN chunks (typically $N_{\rm c}=10^5$) and higher
 Lorentz factors (typically $\Gamma_{\rm c}=10^3$).   As an example of a typical non-repeating  FRB, we set $z=0.2$ 
with $N_{\rm c} = 10^5,\log \Gamma_{\rm c}=3.0, \log m_{\rm c}\ {\rm (gm)}=22.05$; the other
parameters are the  fiducial values listed in Table \ref{table:parameters}.  The simulation results in a single chunk with a viewing angle 
of $\theta_{\rm c} =5.47\times 10^{-3}$  radians, a frequency of 4.6 GHz, a width of 0.6 ms and was only detectable by CHIME with a fluence of 1.0 Jy ms.

 Repeating FRBs occur   for  lower values of $f(\theta_{\rm c}$) for the secondary and tertiary chunks.    
    An example is  shown in Table \ref{table:FRBs-repeating} with a repeat time of days. 
    Table \ref{table:FRBs-repeating-minutes} shows an example with
    time delay between bursts from minutes to hours to days which 
    requires a wide distribution of the chunk mass $m_{\rm c}$.
    Other simulations of repeating FRBS are shown in the SM.

     The number of chunks $N_{\nu^{\rm obs.}}^{\rm obs.}$ (FRBs per QN) detectable at any frequency is derived
     analytically in    Appendix \ref{appendix:chunks-per-frequency} and   Eq. (\ref{appendix:eq:dNc-nu}).
      It is confirmed by the simulations which show that on average  CHIME detects  5 times more FRBs than
   ASKAP and Parkes.  This is because CSE frequency decreases with an increase in
   $f(\theta_{\rm c})$ (higher viewing angle $\theta_{\rm c}$) making CHIME more sensitive to secondary chunks (a bigger solid angle)   for a given QN   (Tables \ref{table:FRBs-waterfall-scenarios}-\ref{table:FRBs-waterfall-scenarios-3}).

  It is possible to view the QN such that we get FRBs
 from chunks arriving roughly periodically.  An example close to FRB 180916.J0158$+$65 (\citealt{chime_2020a}) is shown in 
 Table \ref{table:FRBs-repeating-16days}  with  a  $\sim$16-day period repeating FRB.
  A 4-day window (a ``smearing" effect)  can  be obtained  by varying
   $m_{\rm c}$ and $\Gamma_{\rm c}$  and/or the ambient number density $n_{\rm amb.}^{\rm ns}$ for a given QN.
 FRB 121102 with its quiescent and active periods on month-long scales   (\citealt{spitler_2014,bassa_2017})   can  be reproduced in our model (Tables \ref{table:FRB121102} and \ref{table:FRB121102-galactic}). FRB 121102's 
  $RM\sim 10^5$ rad m$^{-2}$ (\citealt{michilli_2018}) is induced by the  
     QN chunks; see Eq. (\ref{eq:RMcc}).  These two candidates are studied in Appendix \ref{appendix:case-studies}.

 \subsection{Frequency drifting (waterfall plots)}
 \label{sec:drifting}
 
\begin{figure}
 \centering
 \includegraphics[scale=0.3]{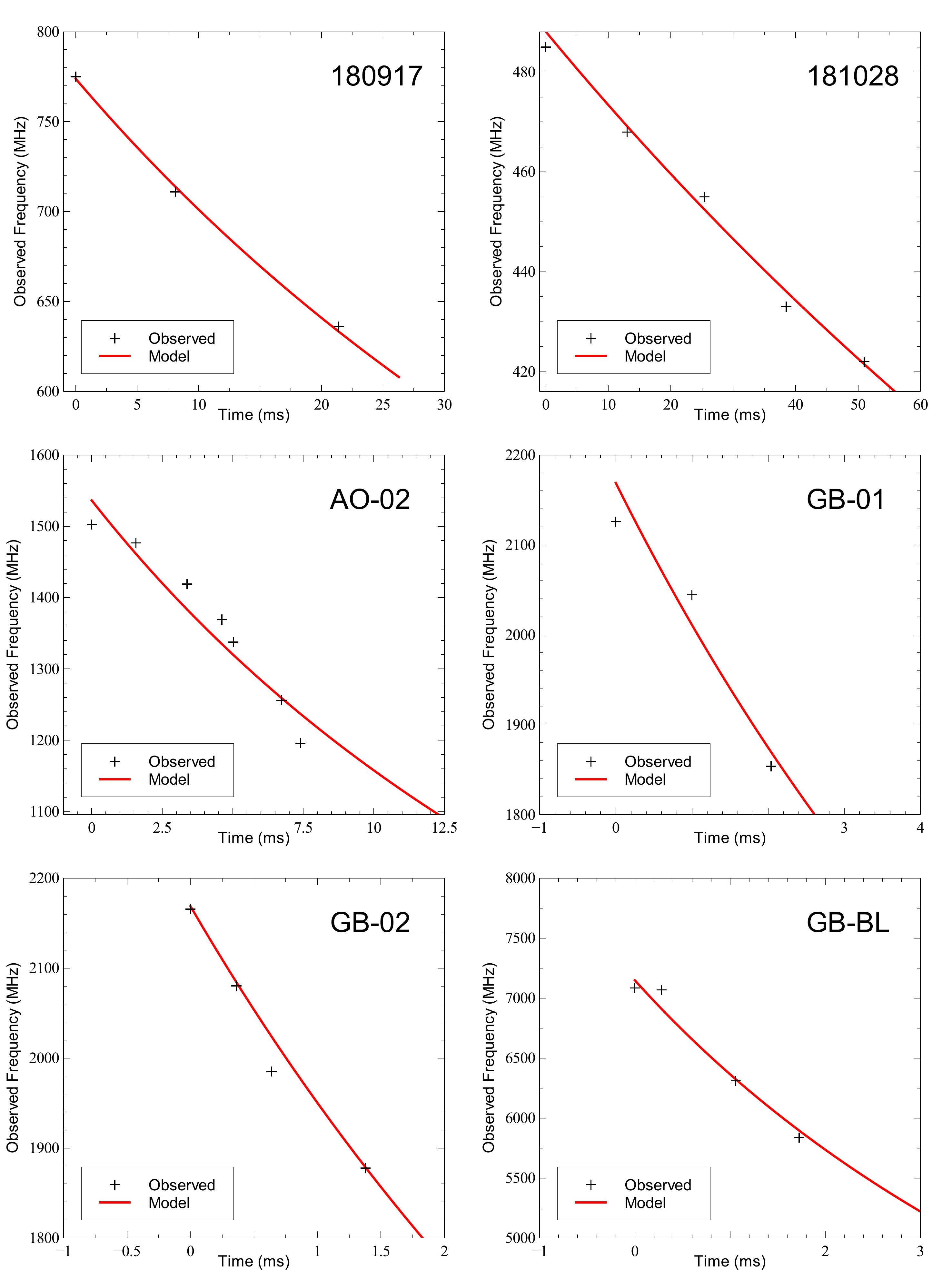}
  \caption{{\bf Sub-pulse frequency drifting fits}: {\bf The two top panels}  show fits (solid lines) to two of CHIME's FRB 180814.J0422$+$73 repeats (the 180917 and 181028 bursts; \citealt{chime_2019a}). The {\bf middle and bottom panels} show fits to four of FRB 121102 bursts (the AO-02, GB-01 GB-02 and GB-BL bursts; \citealt{hessels_2019}).  The fitting parameters are listed in Table \ref{table:drifting-fits} and discussed in \S \ref{sec:drifting}.}
   \label{figure:drifting}
 \end{figure}

 Frequency drifting is  a consequence of the decrease of the CSE frequency in time  during bunch merging.  
  Figure \ref{figure:drifting}, compares our model to two  (180917 and 181028) repeats  of CHIME's FRB 180814.J0422$+$73 (\citealt{chime_2019a}) and four of FRB 121102 bursts (namely, AO-02, GB-01,  GB-02 and GB-BL; \citealt{hessels_2019}). 
  Our  fits to drifting in these FRBs (Table \ref{table:drifting-fits})  yield viewing angles suggestive of secondary 
 and tertiary chunks  ($\bar{\theta}_{\rm S}\simeq 0.008/N_{\rm c, 5}^{1/2}$ and $\bar{\theta}_{\rm T}\simeq 0.02/N_{\rm c, 5}^{1/2}$;  Eq. (\ref{eq:mean-angles})) except for FRB 121102/GB-BL burst which points at a primary chunk ($\bar{\theta}_{\rm P}\simeq 0.004/N_{\rm c, 5}^{1/2}$). We  require  $\zeta_{\rm m-WI}$ 
  of the order of a few thousands which suggests slower bunch merging timescales.
  These two effects combined give longer FRB durations  making these  easier to resolve in time.

\begin{figure*}
 \centering
 \includegraphics[scale=0.25]{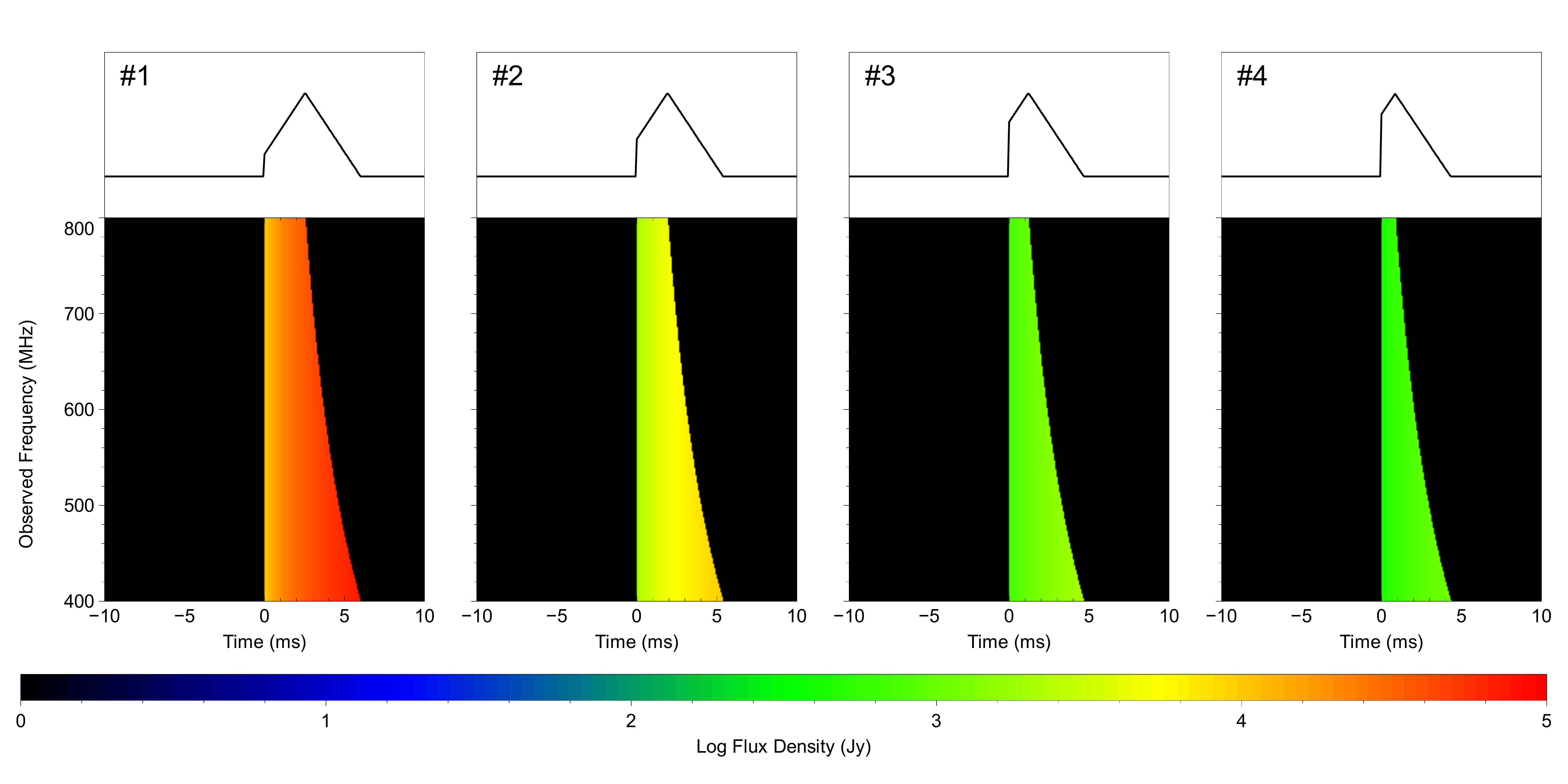}
  \caption{Waterfall plots for the chunks listed in Table \ref{table:FRBs-repeating} and
  detected in CHIME's band.  The palette shows  flux density (Jy) in $\log$ scale.
   The upper sub-panels show  the frequency-summed flux density
  over  the  detector's frequency band  $\nu_{\rm min.}^{\rm det.} \le \nu^{\rm det.} \le \nu_{\rm max.}^{\rm det.}$.}
  \label{figure:FRBs-waterfall-repeating}
 \end{figure*}

 Figure \ref{figure:FRBs-waterfall-repeating} is the  frequency-time plot (``waterfall" plot; Appendix \ref{sec:waterfall-plots}) for the simulation shown in   Table \ref{table:FRBs-repeating}. The band(frequency)-summed flux density is in  the upper sub-panels and matches  the  analytically derived one (Appendix \ref{appendix:flux-fluence} and  Figure \ref{figure:band-integrated-flux}). 
Figures \ref{figure:FRBs-waterfall-scenarios} and \ref{figure:FRBs-waterfall-scenarios-2} show
waterfall plots for the repeating FRBs listed in Tables \ref{table:FRBs-waterfall-scenarios} and  \ref{table:FRBs-waterfall-scenarios-2}.
 Figure \ref{figure:FRBs-waterfall-scenarios-3} shows an example where 
  the maximum CSE frequency falls within the detector's band (here CHIME); see Table \ref{table:FRBs-waterfall-scenarios-3}
  for the corresponding simulations.  Our model can reproduce the  cases in the upper panels in Figure  \ref{figure:drifting-a} and  Figure  \ref{figure:drifting-b}.
  
\begin{table}
\begin{center}
 \caption{{\bf Drifting in repeating FRBs}: Fits to frequency drifting in time for selected CHIME FRB 180814.J0422$+$73
 and FRB 121102 bursts (see related Figure \ref{figure:drifting}). Other parameters are kept to their fiducial values   listed in Table \ref{table:parameters}.}
 \begin{tabular}{|c|c|c|c|c|c|}\hline
  FRB  & $z$ & $\theta_{\rm c}$ &  $\zeta_{\rm m-WI}$  &  Chunk\\\hline
  CHIME (18/09/17)$^a$  & 0.1 & 0.012  & 2700  &  Tertiary\\\hline
   CHIME (18/10/28)$^a$  & 0.1 & 0.016 & 5700  &  Tertiary\\\hline
   121102 (AO-02)$^b$  & 0.2 & 0.008 & 1700  &  Secondary\\\hline
   121102 (GB-01)$^b$  & 0.2 & 0.007 & 1000  &  Secondary\\\hline 
   121102 (GB-02)$^b$  & 0.2 & 0.007 & 700  &  Secondary\\\hline  
   121102 (GB-BL)$^b$  & 0.2 & 0.002 & 2100  &  Primary\\\hline
 \end{tabular}\\
  \label{table:drifting-fits}
 \end{center}
 $^a$  See \citet{chime_2019a}.\\
 $^b$  See \citet{hessels_2019}.
 \end{table}

 \section{Discussion}
 \label{sec:discussion}

\subsection{Rate}
\label{sec:FRB-rate}

Assuming that the progenitors of ICM-QNe are old  massive NSs, we  estimate the
ICM-QN occurrence rate. Slowly rotating, massive NSs are the most likely to experience
quark deconfinement via nucleation and undergo a QN phase. We count only 
NSs  with birth periods greater than $\sim 100$ ms and  stellar progenitors with masses 20-40 M$_{\sun}$.  We use  a lognormal initial magnetic field distribution 
(mean of 12.5 and standard deviation of 0.5) and a normal distribution for initial period 
(mean of 300 ms and standard deviation of 150 ms); 
from \citet{faucher_2006}. We assume that ICM-QNe 
occur after  a nucleation timescale of $\sim 10^8$ years.

 Integration of the initial mass function 
(\citealt{salpeter_1955}) and the initial period distribution (assuming period  and 
magnetic field are independent) gives $\sim$ 10\% of all neutron stars as QN candidates
in the ICM. For a galactic core-collapse SN rate of $\sim 1/50$ years, over $10^{10}$ years
about $\sim 2\times 10^8$ NSs would have formed. This yields 
\begin{equation}
 r_{\rm FRB}\sim  2\times 10^{-3}\ {\rm yr}^{-1}\ ,
 \end{equation}
or  a few FRBs per thousand year per galaxy,
  consistent with the observed rate (\citealt{champion_2016,petroff_2019}).

\subsection{Predictions (also Appendix \ref{appendix:predictions})}
\label{sec:predictions}

\begin{itemize}

\item {\bf ``Periodicity" (Repeats vs ``non-repeats")}:  All FRBs are repeats 
because every chunk emits an FRB beamed in a specific direction. 
 Non-repeaters are a consequence of observing limitations when emission from 
the secondary and tertiary chunks 
is too faint or when the corresponding frequency is below the bandwidth (Eq. (\ref{eq:repeat-condition})). 
All FRBs, if viewed at the right angle, will appear periodic in time with  period (Eq.(\ref{eq:dtrepeat})) set by the roughly constant angular separation between chunks (Eq. (\ref{eq:Deltaf})). This ``periodicity" may be washed out
 with a  variation in chunks parametersf ($m_{\rm c}$ and $\Gamma_{\rm c}$)
 and/or  in the ambient density ($n_{\rm amb.}^{\rm ns}$);

 \item {\bf The  halo/ICM low dispersion measure (DM)}:  Recent studies (\citealt{caleb_2019,ravi_2019}) concluded that FRBs sources
must  repeat  in order to account for the high FRB volumetric rate.
This constraint is relaxed in our model given the   low DM, and thus larger volume, of the ambient medium (galactic halo, ICM, IGM) and our estimated rate of FRBs  ($\sim$ 10\% of the core-collapse SN rate).  Within uncertainties on the observed rate, our model is in the allowed region (Figure 2 in  \citealt{caleb_2019}) with no need for sources to repeat;

 \item {\bf The solid angle effect}:  CHIME (sub-GHz)  is
  more sensitive to higher angle chunks and should detect more FRBs per QN  than ASKAP and Parkes (GHz) detectors (Eq. (\ref{eq:CHIME-vs-ASKAP})). CHIME FRBs should be dimmer and will 
  be associated with duration (burst width) on average longer (but with variations) than ASKAP and Parkes FRBs;
  
   \item {\bf Super FRBs}: We expect  detection of super FRBs (with fluence in the
thousands to tens of thousands of Jy ms) from  ICM-QNe due to  chunks very close to the
observer's line-of-sight. Monster FRBs from IGM-QNe with a fluence in the millions of Jy ms,
may be detected by LOFAR's low-band antenna. 
FRBs from IGM-QNe (Appendix \ref{appendix:IGM-QNe} and Table \ref{table:IGM-equations}) are extremely rare.

\end{itemize}

\subsection{Model's limitations (also Appendix \ref{appendix:limitations})}
\label{sec:limitations}

\begin{itemize}

\item {\bf Parameter fitting}: Table \ref{table:parameters} lists 13 parameters.
In developing the model we varied all of the parameters in order to obtain reasonable values for them.
It is a challenge to constrain all of them and 
 we restrict our investigations
to varying a few basic parameters;   $m_{\rm c}$, $\Gamma_{\rm c}$ and $N_{\rm c}$.
We also varied $n_{\rm amb.}^{\rm ns}$  which allowed
  us to explore the  evolution of the QN chunks in different environments 
  (the Galactic halo, the ICM and the IGM). Other parameters (related to the BI-WI instabilities) will need to be surveyed in future studies;

\item {\bf Bunching mechanism}: The exact mechanism for bunching is unclear.  In Appendix \ref{appendix:bunch-geometry} we speculate that bunching occurs in the
periphery and along the Weibel filaments.   Our reasoning is that bunching (and CSE) would  not occur  
 inside filaments where the currents reside and the magnetic field is weaker.
    Regardless, bunches emit the BI heating efficiently and promptly as CSE
    during the bunch merging phase; 
    
\item     {\bf QN compact remnant as magnetars (and the FRB association)}:  The QS is born with a  surface magnetic field of 
  $\sim 10^{14}$-$10^{15}$ G  (\citealt{iwazaki_2005}). 
During the QS spin-down, vortices (and their magnetic field) are expelled (\citealt{ouyed_2004,niebergal_2010b})
leading to X-ray activity similar to that of Soft $\gamma$-ray Repeaters (SGRs) and Anomalous X-ray Pulsars (AXPs)  (\citealt{ouyed_2007a,ouyed_2007b}).
Thus   FRB activity for magnetars  (e.g.  \citealt{metzger_2017}) 
   would be applicable to a QN compact remnant. The coincidence and association between FRB 200428 and  SGR 1935$+$2154
 (\citealt{barthelmy_2020,scholz_2020,chime_2020b,witze_2020,bochenek_2020}) may 
    find an explanation  in our model. 
 
\end{itemize}

\section{Conclusion}
\label{sec:conclusion}

We present a model for FRBs involving old, slowly rotating and isolated NSs converting explosively to QSs (experiencing a QN event) in the ICM of galaxy groups and clusters. The NSs are embedded in the ICM when the QN occurs. The millions of QN chunks  expand, due to heating by hadronic collisions with ambient protons, and become collisionless as they propagate  away from the QN.  The interaction of the collisonless  chunks (the background plasma) with the ambient medium   (the plasma beam), successively triggers the BI and WI  yielding electron bunching and CSE with properties  of repeating and non-repeating FRBs such as  the GHz frequency, the milli-second duration and a fluence in the Jy ms range.

There are  three classes of FRBs:  those  from ICM-QNe (i.e. galaxy group and cluster FRBs; \S \ref{sec:FRB-ICM-QNe}),  from galactic/halo-QNe (Appendix \ref{appendix:FRB121102}),  and a third  class, but the least 
likely one, corresponding to FRBs from IGM-QNes (Appendix \ref{appendix:IGM-QNe}) with
 frequencies at the lower limit of LOFAR's low-band antenna. 
 The distribution of NS natal kick velocities controls the ratio of galactic versus extra-galactic QNe  (and their FRBs). We estimate an FRB rate  of about 10\% of that of ccSNe.    
  Because of the low DM of the ambient medium, their volumetric
rate can be explained without the need for the FRB sources to repeat  (\S \ref{sec:predictions}).

 Our model can be used to probe collisionless plasma instabilities (see Appendix \ref{appendix:implications-plasmas}) and the
QCD phase diagram (see Appendix \ref{appendix:implications-QCD}).  It has implications to Astrophysics (see Appendix \ref{appendix:implications-astro}) such as the existence of QSs forming mainly from  old NSs exploding as QNe in the outskirts of galaxies.  If the model is a correct representation of FRBs then it would  strengthen the idea  that QNe within a few years of a core-collapse SN of massive stars may  be the origin of LGRBs (\citealt{ouyed_2020}).  Thus the same engine, the exploding NS, is responsible
for GRBs and FRBs. For the FRBs case, the QN occurs hundreds of million of years after the SN.

We demonstrated that FRBs can be caused by a cataclysmic
event, the QN.    The millions of  emitting chunks per QN is key  with repeats  a consequence of seeing multiple chunks coming in at different times 
and beamed in different directions.  Single FRBs occur when only the primary chunk is within detector's sensitivity.
Our model  relies on the feasibility of  an explosive transition of a  NS to a QS. While such a transition is indicated by analytical studies (\citealt{keranen_2005,vogt_2004,ouyed_leahy_2009,ouyed_2020}) and by one-dimensional numerical simulations (\citealt{niebergal_2010a,ouyed_amir_2018a,ouyed_amir_2018b}), detailed multi-dimensional simulations are required to confirm or refute it (\citealt{niebergal_phd_2011,ouyed_amir_phd_2018}).

\section*{Acknowledgements}

This research is supported by operating grants from the Natural Science and Engineering Research Council of Canada.

\section*{Data availability}

The data underlying this article are available in the article and in its online supplementary material.



\appendix
\renewcommand{\thesection}{S\Alph{section}}
\setcounter{section}{0}
 
\onecolumn

\newpage
\vspace*{\fill}
\begingroup
\hspace{2.0in}{\Huge Supplementary online material.}
\endgroup
\vspace*{\fill}
\newpage


\section{Ejecta properties and statistics}
\label{appendix:honeycomb}

As described in \citet{ouyed_leahy_2009}, the QN ejecta breaks up into
millions  of chunks. Here we adopt $N_{\rm c}= 10^6$ as our fiducial value 
for the number of fragments which yields a  typical chunk mass\footnote{Dimensionless quantities
 are defined as $f_{\rm x}= f/10^{x}$ with quantities in cgs units.} of $m_{\rm c}=M_{\rm QN}/N_{\rm c}\simeq  10^{22.3}\ {\rm gm}\times M_{\rm QN, 28.3}/N_{\rm c, 6}$. The chunk's Lorentz factor is taken to be constant  with $\Gamma_{\rm c}=\Gamma_{\rm QN}=10^{2.5}$
  corresponding to a fiducial QN ejecta's kinetic energy $E_{\rm QN}=\Gamma_{\rm c}M_{\rm QN}c^2\simeq 5.7\times 10^{51}$ erg; i.e. roughly 1\% of the conversion energy is converted to the kinetic energy of the QN ejecta; these fiducial values are listed in Table \ref{table:parameters}.

\begin{figure*}
\vskip 1.5in
 \centering
  \includegraphics[scale=0.4]{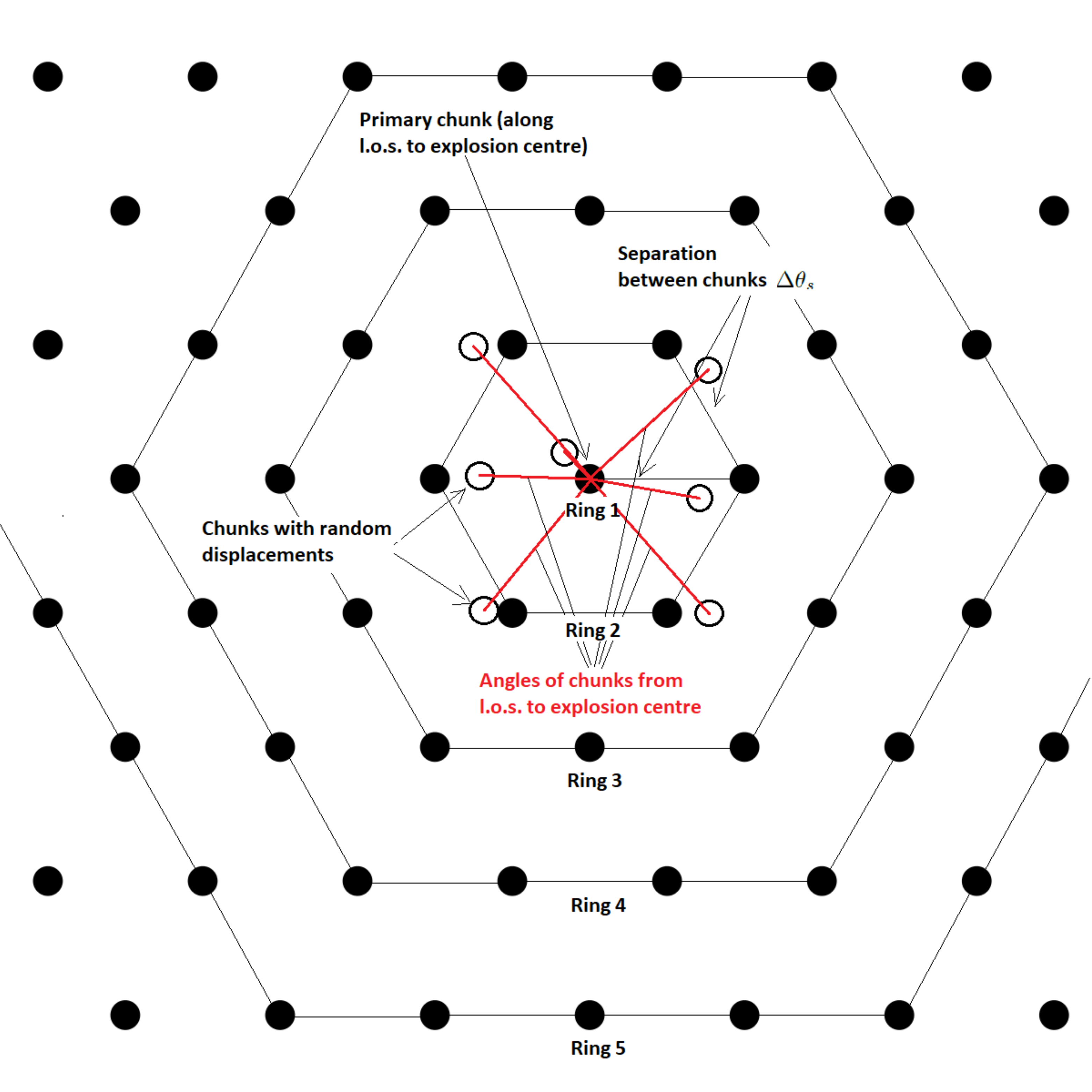}
  \caption{{\bf Honeycomb geometry}: Illustration of observer's face on view of distribution of chunks.
{\bf Uniformly spaced case}: The solid black circles represent evenly spaced chunks travelling radially outward from the QN site with the primary chunk (labelled ``ring" 1) travelling at angle $0^{\circ}$ with respect to the line-of-sight (l.o.s) to the QN.
The 6 chunks in ``ring" 2 all travel at angle $\Delta \theta_s$ (see Eq. (\ref{eq:dthetas})) from the l.o.s.. The 12 chunks in ``ring" 3 all travel at angle  $2\Delta \theta_s$ from the l.o.s, and so on for successively outward rings. The time delays of received FRB emission by the observer are  determined by the angular time delays which depends on angle from the l.o.s.. Thus the FRB emission from the chunks in ``ring" 2 all arrive at the same time. {\bf Randomly spaced case}: The open circles represent randomly spaced chunks (only illustrated for ``ring" 1 and ``ring" 2), which are offset at small random angles and directions from the uniformly spaced case. In this case the arrival times of the different chunks in a given ring (e.g. ``ring" 2) are different, again depending on the l.o.s. angle of each chunk.}
 \label{figure:honeycomb}
 \end{figure*}

  Below we  summarize some  general properties of the QN ejecta (see details in \S 2.1 and Appendix B.1 in \citealt{ouyed_2020}).
  Hereafter, unprimed quantities are in the chunk's reference frame
while the superscripts ``ns" and ``obs." refer to quantities  in the NS frame (i.e. the ambient medium)
 and the observer's frame, respectively. The chunk's Lorentz factor $\Gamma_{\rm c}$  does not vary in time,
 during the FRB phase, in our model. The transformation
 from the local NS frame to the chunk's frame is given by  $dt^{\rm ns} = \Gamma_{\rm c}dt$ while 
 the transformations from the chunk's frame to the observer's frame (where the emitted light is being observed) are
 $dt^{\rm obs.}= (1+z) dt/D(\Gamma_{\rm c},\theta_{\rm c})$, $\nu^{\rm obs.} = D(\Gamma_{\rm c},\theta_{\rm c})\nu/(1+z)$ with $z$ the source's redshift  and $\theta_{\rm c}$ the  viewing angle (the angle between the observer and chunk's velocity vector); $D(\Gamma_{\rm c},\theta_{\rm c})$ is the chunk's Doppler factor.
 We note the following:

\begin{itemize}

  \item The  chunks are equally spaced in solid angle $\Omega$ around the explosion site.
  Defining $N_{\theta}$ as the number of chunks per angle $\theta$, we write $dN_{\theta}/d\Omega =const.= N_{\rm c}/4\pi$ with
$d\Omega = 2\pi \sin{\theta} d\theta$ so that $dN_{\theta}/d\theta= (N_{\rm c}/2) \sin{\theta}\simeq (N_{\rm c}/2) \theta$
with $\theta_{\rm c}<<1$. Because $N_{\rm c}\pi \Delta \theta_{\rm c}^2 = 4\pi$ when the chunks first form, 
the average angular separation between them  is 
 \begin{equation}
 \label{eq:dthetas}
 \Delta \theta_{\rm s}=2 \Delta \theta_{\rm c} = \frac{4}{N_{\rm c}^{1/2}}\simeq \frac{4\times 10^{-3}}{N_{\rm c, 6}^{1/2}}\ ,
 \end{equation}
  yielding $\Delta \theta_{\rm s}\sim 1/\Gamma_{\rm c}$ 
   for our fiducial value of $N_{\rm c}=10^6$ and $\Gamma_{\rm c}=10^{2.5}$.
   
   A simplistic geometry to visualize the spatial distribution is a 2-dimensional honeycomb (see Figure \ref{figure:honeycomb}) 
    with 1 primary (the $i=1$) chunk at $0 \le \theta_{\rm c} <  \Delta\theta_{\rm c}$ and $6\times (i-1)$ chunks for subsequent,
    and concentric, ``rings"\footnote{A group of chunks with roughly the same $\theta_{\rm c}$ but different azimuths
    as illustrated in Figure \ref{figure:honeycomb}.}
    (with $i\ge 2$; i.e.  6 secondary chunks, 12 tertiary chunks etc..). In this geometry the 
     range  and mean angle of the primary (P), secondary (S) and tertiary (T) chunks are
   
    \begin{align}
    \label{eq:mean-angles}
   0 &\le \theta_{\rm P} <  \Delta\theta_{\rm c}\ , \quad \bar{\theta}_{\rm P}= \frac{\int_{0}^{\Delta\theta_{\rm c}}\theta^2 d\theta}{\int_{0}^{\Delta\theta_{\rm c}}\theta d\theta}= \frac{4}{3N_{\rm c}^{1/2}}\nonumber\\
 \Delta\theta_{\rm c}  &\le \theta_{\rm S} <  3 \Delta\theta_{\rm c}\ , \quad \bar{\theta}_{\rm S}= \frac{\int_{\Delta\theta_{\rm c}}^{3\Delta\theta_{\rm c}}\theta^2 d\theta}{\int_{\Delta\theta_{\rm c}}^{3\Delta\theta_{\rm c}}\theta d\theta}=\frac{28}{9N_{\rm c}^{1/2}}\nonumber\\
   3 \Delta\theta_{\rm c}  &\le \theta_{\rm T} < 5 \Delta\theta_{\rm c}\ ,  \quad \bar{\theta}_{\rm T}=  \frac{\int_{3\Delta\theta_{\rm c}}^{5\Delta\theta_{\rm c}}\theta^2 d\theta}{\int_{3\Delta\theta_{\rm c}}^{5\Delta\theta_{\rm c}}\theta d\theta}=\frac{49}{6N_{\rm c}^{1/2}}\ .
   \end{align}
   We see that $\bar{\theta}_{\rm S}\simeq 2.3 \bar{\theta}_{\rm P}$ and $\bar{\theta}_{\rm T}\sim 6 \bar{\theta}_{\rm P}$;

\item    Because  $\Gamma_{\rm c}^2>>1$ and $\theta_{\rm c}<<1$ applies here, 
  we write the Doppler factor as $D_{\rm c}(\Gamma_{\rm c},\theta_{\rm c})\simeq 2\Gamma_{\rm c}/f(\theta_{\rm c})$ with\footnote{Because we take the observer to be located at large distance, compared
to the chunks distance from the explosion centre, $\theta_{\rm c}$ is also  the angle of the chunk's motion 
with respect to the observer's line-of-sight.}
    
    \begin{equation}
    \label{eq:ftheta}
     f(\theta_{\rm c})= 1+(\Gamma_{\rm c}\theta_{\rm c})^2\ ,
     \end{equation}
     and $f(\theta_{\rm c}) << \Gamma_{\rm c}^2$. This yields:

  \begin{align}
 \label{eq:tPtS}
  1 \le f(\theta_{\rm P}) < 1 + 0.4\frac{\Gamma_{\rm c, 2.5}^2}{N_{\rm c, 6}}\nonumber\\
  1 + 0.4\frac{\Gamma_{\rm c, 2.5}^2}{N_{\rm c, 6}} \le f(\theta_{\rm S}) <  1 + 3.6\frac{\Gamma_{\rm c, 2.5}^2}{N_{\rm c, 6}}\nonumber\\
   1 + 3.6\frac{\Gamma_{\rm c, 2.5}^2}{N_{\rm c, 6}} \le  f(\theta_{\rm T}) <  1 + 10\frac{\Gamma_{\rm c, 2.5}^2}{N_{\rm c, 6}}\ ,
 \end{align}
 and average values
 \begin{align}
 \label{eq:fPfS}
 f(\bar{\theta}_{\rm P})\sim 1 + 0.18\frac{\Gamma_{\rm c, 2.5}^2}{N_{\rm c, 6}}\nonumber\\
  f(\bar{\theta}_{\rm S})\sim 1 + 0.97\frac{\Gamma_{\rm c, 2.5}^2}{N_{\rm c, 6}}\nonumber\\
   f(\bar{\theta}_{\rm T})\sim 1 + 6.67\frac{\Gamma_{\rm c, 2.5}^2}{N_{\rm c, 6}}\ ;
 \end{align}
     
\item      The change in $f(\theta_{\rm c})$ between two successive ``rings"  is 
$\Delta f_{i+1, i}^{\rm rings}= f(\theta_{\rm ring,  i+1})-f(\theta_{\rm ring, i})= \Gamma_{\rm c}^2\times (\theta_{\rm ring, i+1}^2-\theta_{\rm ring, i}^2)=
 \Gamma_{\rm c}^2\times (\theta_{\rm ring, i+1}-\theta_{\rm ring, i})(\theta_{\rm ring, i+1}+\theta_{\rm ring, i})$.
 Using $\theta_{\rm ring,i+1}=\theta_{\rm ring,i}+\Delta \theta_{\rm s}$ one finds 
 
 \begin{equation}
 \label{eq:dfrings}
 \Delta f_{i+1, i}^{\rm rings} = 1.6\times \frac{\Gamma_{\rm c, 2.5}^2}{N_{\rm c, 6}}\times (1+\frac{2\theta_{\rm ring, i}}{\Delta \theta_{\rm s}})\ ,
 \end{equation}
 with $\theta_{\rm ring, i}/\Delta \theta_{\rm s}=i$ for equally spaced``rings".

 The average change in the radial angle from one random chunk to another (i.e. the actual 
 separation projected onto the radial direction; see Figure \ref{figure:honeycomb}) with respect to the observer is
 $\Delta f_{i+1, i}^{\rm chunks}\simeq \Delta f_{i+1, i}^{\rm rings}/N_i$.
  The number of chunks per ring, of perimeter $2\pi\times (i\Delta \theta_{\rm s})$, is $N_{\rm i}=2\pi\times (i\Delta \theta_{\rm s}/\Delta \theta_{\rm s})=2\pi i$.
 Thus,
 \begin{align}
  \label{eq:Deltac}
 \Delta f_{i+1, i}^{\rm chunks}\simeq 1.6\times \frac{\Gamma_{\rm c, 2.5}^2}{N_{\rm c, 6}}\times \frac{(1+2i)}{2\pi i} \sim \frac{1.6}{\pi}\times \frac{\Gamma_{\rm c, 2.5}^2}{N_{\rm c, 6}}\ ,
 \end{align}
 where the last expression applies for higher $i > \sim  2$.
 Because the chunks are not precisely equally spaced, the variation in $f(\theta_{\rm c})$ 
between chunks is somewhat variable.

  \end{itemize}

   \section{Chunk expansion and the onset of the collisionless plasma regime}
\label{appendix:heating}

 We define $R_{\rm c}$ as the chunk's radius. The photon transparency
radius for the chunk is $R_{\rm c, opt.}\simeq 2.2\times 10^{10}\ {\rm cm}\times m_{\rm c, 22.3}^{1/2}\kappa_{\rm c, -1}^{1/2}$ with 
   $m_{\rm c, 22.3}$ its mass (in units of $10^{22.3}$ gm) and $\kappa_{\rm c, -1}$   its opacity (in units
   of 0.1 cm$^2$ gm$^{-1}$). For $R_{\rm c}>R_{\rm c, opt.}$,  
  the thermal and dynamical evolution of the chunk is governed by heating ($Q_{\rm HH} $) from hadronic collisions  with the ambient
  medium   and thermalization  due to electron Coulomb collisions followed by adiabatic cooling (PdV expansion). 
 The  heat transfer equations describing  the time evolution of the chunk's radius $R_{\rm c}$ and
 its temperature $T_{\rm c}$ are 

\begin{align}
\frac{d R_{\rm c}}{d t} &= c_{\rm c, s}\nonumber\\
C_{\rm v}\frac{d T_{\rm c}}{d t} &=  Q_{\rm HH} -  p_{\rm c} \frac{d V_{\rm c}}{d t} \ ,
\end{align}
where $c_{\rm c, s}= \sqrt{\gamma_{\rm ad.}k_{\rm B}T_{\rm c}/\mu_{\rm e}m_{\rm H}}$
is the chunk's sound speed, $p_{\rm c}=n_{\rm c}k_{\rm B}T_{\rm c}$ its 
pressure, $V_{\rm c}=(4\pi/3)R_{\rm c}^3$ its volume and   
 $C_{\rm V}=(m_{\rm c}/m_{\rm H})\times (3k_{\rm B}/2)$ its heat capacity. The adiabatic index we take to be $\gamma_{\rm ad.}=5/3$ with 
 a mass per electron  $\mu_{\rm e}=2$; $k_{\rm B}$ is the Boltzmann constant.

Equations above can be combined into 

\begin{align}
\label{appendix:eq:heating}
\frac{d c_{\rm c, s}^2}{d t} &=   \frac{2\gamma_{\rm ad.}Q_{\rm HH}}{3\mu_{\rm e}m_{\rm c}} -   2\frac{c_{\rm c, s}^3}{R_{\rm c}} \ .
\end{align}

 The optical depth to hadronic collisions is  $\tau_{\rm HH}=n_{\rm c}\sigma_{\rm HH} R_{\rm c} = 3m_{\rm c}\sigma_{\rm HH}/m_{\rm H}4A_{\rm c}<<1$
where $A_{\rm c}=\pi R_{\rm c}^2$ is the chunk's area and $n_{\rm c}=3m_{\rm H}/4\pi R_{\rm c}^3m_{\rm H}$ its baryon number density. Thus, heating due to hadronic collisions can be written as $Q_{\rm HH}= \tau_{\rm HH}\times (\Gamma_{\rm c}m_{\rm H}c^2)\times (A_{\rm c}\Gamma_{\rm c} n_{\rm amb.}^{\rm ns} c)=m_{\rm c}\sigma_{\rm HH}\Gamma_{\rm c}^2n_{\rm amb.}^{\rm ns}c^3$. The term $A_{\rm c}\Gamma_{\rm c} n_{\rm amb.}^{\rm ns} c$
is the number of ambient protons swept-up by the chunk per unit time; here $c$ is the light speed.  This yields

\begin{equation}
\label{appendix:eq:QHH}
Q_{\rm HH}\simeq 5.4\times 10^{28}\ {\rm erg\ s}^{-1}\times m_{\rm c, 22.3}\sigma_{\rm HH, -27}\Gamma_{\rm c, 2.5}^2n^{\rm ns}_{\rm amb., -3}\ ,
\end{equation}
where $\sigma_{\rm HH, -27}$ is the proton
hadronic collision cross-section in units of milli-barns (e.g. \citealt{letaw_1983,tanabashi_2018}).
Eq. (\ref{appendix:eq:heating}) becomes 

\begin{align}
\label{appendix:eq:heating-2}
\frac{d c_{\rm c, s}^2}{d t} &=      q_{\rm HH} -   2\frac{c_{\rm c, s}^3}{R_{\rm c}} \ ,
\end{align}
where $q_{\rm HH}=2\gamma_{\rm ad.}Q_{\rm HH}/3\mu_{\rm e}m_{\rm c}\simeq 1.5\times 10^{6}\times \sigma_{\rm HH, -27}\Gamma_{\rm c, 2.5}^2 n_{\rm amb.,-3}^{\rm ns}$ is the specific heating term due to hadronic collisions.

The solution of the system above is $R_{\rm c}(t)= R_{\rm c, 0} (t/t_0)^{3/2}$ and $c_{\rm c, s}(t)= c_{\rm c, s, 0} (t/t_0)^{1/2}$
with $c_{\rm c, s, 0} = 3R_{\rm c, 0}/2 t_{\rm 0}$ and $t_{\rm 0}= (27R_{\rm c, 0}^2/2 q_{\rm HH})^{1/3}$.
For $R_{\rm c, 0}=R_{\rm c, opt.}$, we get
   
   \begin{equation}
   \label{appendix:eq:heating-t0}
   t_0 \simeq 1.6\times 10^5 \ {\rm s}\times \left(\frac{m_{\rm c, 22.3} \kappa_{\rm c, -1}}{\sigma_{\rm HH, -27}\Gamma_{\rm c, 2.5}^2 n_{\rm amb.,-3}^{\rm ns}}\right)^{1/3}\ .
   \end{equation}
   The initial chunk temperature, found from $k_{\rm B}T_0=\mu_{\rm e}m_{\rm H}c_{\rm c, s, 0}^2$, is

   \begin{equation}
   \label{appendix:eq:T0}
   T_0\simeq  10^3\ {\rm K}\times (m_{\rm c, 22.3}\kappa_{\rm c, -1})^{1/3} \times \left(\sigma_{\rm HH, -27}\Gamma_{\rm c, 2.5}^2 n_{\rm amb.,-3}^{\rm ns} \right)^{2/3}\ .
   \end{equation}

The chunk becomes collisionless when the electron Coulomb collision length inside the chunk,
 $\lambda_{\rm  Coul., e}\simeq 1.1\times 10^4\ {\rm cm}\times T_{\rm c, e}^2/n_{\rm c, e}$ (\citet{richardson_2019} with a Coulomb parameter $\ln{\Lambda}= 20$; e.g. \citealt{lang_1999}) is of the order of the chunk's radius $R_{\rm c}$. 
   Setting $\lambda_{\rm  Coul., e}(t_{\rm cc})= R_{\rm c}(t_{\rm cc})$ with $n_{\rm c}(t)= 3 m_{\rm c}/4\pi R_{\rm c}(t)^3m_{\rm H}$, $R_{\rm c}(t)= R_{\rm c, 0} (t/t_0)^{3/2}$
  and $T(t)=T_0 (t/t_0)$ yields
   
   \begin{equation}
   \frac{t_{\rm cc}}{t_0}\simeq 878.8\times T_{\rm 0, 3}^{-2/5}\ ,
   \end{equation}
   where the chunk's initial temperature  (when it becomes optically thin) is in units of $10^3$ K. The subscript ``cc" stands
   for collisionless chunk.
   
   The chunk's temperature, radius and number density  when it enters the collisionless regime are
   
   \begin{align}
   T_{\rm cc}&\simeq  8.8\times 10^5\ {\rm K}\times T_{\rm 0, 3}^{3/5} \quad  \simeq 8.8\times 10^5 \ {\rm K}\times (m_{\rm c, 22.3}\kappa_{\rm c, -1})^{1/5}
    (\sigma_{\rm HH, -27}\Gamma_{\rm c, 2.5}^2 {n_{\rm amb.,-3}^{\rm ns}})^{2/5} \label{appendix:eq:cc1}\\
    R_{\rm cc}&\simeq 5.9\times 10^{14}\ {\rm cm}\times \frac{m_{\rm c, 22.3}^{1/2}\kappa_{\rm c, -1}^{1/2}}{T_{\rm 0, 3}^{3/5}} \quad  \simeq 5.9\times 10^{14}\ {\rm cm}\times\frac{(m_{\rm c, 22.3}\kappa_{\rm c, -1})^{3/10}}{ (\sigma_{\rm HH, -27}\Gamma_{\rm c, 2.5}^2{n_{\rm amb.,-3}^{\rm ns}})^{2/5}}\label{appendix:eq:cc2}\\
   n_{\rm cc}&\simeq 14.6\ {\rm cm}^{-3}\times \frac{T_{\rm 0, 3}^{9/5}}{m_{\rm c, 22.3}^{1/2}\kappa_{\rm c, -1}^{3/2}}
    \quad  \simeq 14.6\ {\rm cm}^{-3}\times \frac{m_{\rm c, 22.3}^{1/10}}{\kappa_{\rm c, -1}^{9/10}}\times (\sigma_{\rm HH, -27}\Gamma_{\rm c, 2.5}^{2}{n_{\rm amb.,-3}^{\rm ns}})^{6/5}\label{appendix:eq:cc3}\ .
     \end{align}
   This yields an estimate of the chunk electron thermal speed  $\beta_{\rm cc}=v_{\rm cc}/c= \sqrt{\gamma_{\rm ad.}k_{\rm B}T_{\rm cc}/m_{\rm e}c^2}$ as
   \begin{equation}
   \label{appendix:eq:betaecc}
    \beta_{\rm cc}\simeq 1.6\times 10^{-2} \times T_{\rm 0, 3}^{3/10} \quad   \simeq 1.6\times 10^{-2} \times (m_{\rm c, 22.3}\kappa_{\rm c, -1})^{1/10} (\sigma_{\rm HH, -27}\Gamma_{\rm c, 2.5}^{2}{n_{\rm amb.,-3}^{\rm ns}})^{1/5} \ .
   \end{equation}

   The time, since the QN event, it takes the chunk to become collisionless in the chunk frame is:
   \begin{equation}
    \label{appendix:eq:tcc}
   t_{\rm cc}\simeq 1.7\times 10^3\ {\rm days}\times T_{\rm 0, 3}^{-2/5}\times \left(\frac{m_{\rm c, 22.3}\kappa_{\rm c, -1}}{\sigma_{\rm HH, -27}\Gamma_{\rm c, 2.5}^2 n_{\rm amb.,-3}^{\rm ns}}\right)^{1/3} \quad  \simeq 1.7\times 10^3\ {\rm days}\times  \frac{(m_{\rm c, 22.3}\kappa_{\rm c, -1})^{1/5}}{(\sigma_{\rm HH, -27}\Gamma_{\rm c, 2.5}^{2}{n_{\rm amb.,-3}^{\rm ns}})^{3/5}}\ ,
   \end{equation}
   which is $t_{\rm cc}^{\rm obs.}=(1+z)t_{\rm cc}/D(\Gamma_{\rm c},\theta_{\rm c})$ in the observer's frame
   with the Doppler factor $D(\Gamma_{\rm c},\theta_{\rm c})\simeq 2\Gamma_{\rm c}/f(\theta_{\rm c})$ and $f(\theta_{\rm c})$ given by Eq. (\ref{eq:ftheta}). I.e., 
   
  \begin{equation}
  \label{appendix:eq:tccobs}
   t_{\rm cc}^{\rm obs.}\simeq 2.6\ {\rm days}\times \frac{(1+z)f(\theta_{\rm c})}{\Gamma_{\rm c, 2.5}}\times T_{\rm 0, 3}^{-2/5}\times \left(\frac{m_{\rm c, 22.3}\kappa_{\rm c, -1}}{\sigma_{\rm HH, -27} \Gamma_{\rm c, 2.5}^2 n_{\rm amb.,-3}^{\rm ns}}\right)^{1/3} \quad  \simeq 2.6\ {\rm days}\times \frac{(1+z)f(\theta_{\rm c})}{\Gamma_{\rm c, 2.5}}\times  \frac{(m_{\rm c, 22.3}\kappa_{\rm c, -1})^{1/5}}{(\sigma_{\rm HH, -27}\Gamma_{\rm c, 2.5}^{2}{n_{\rm amb.,-3}^{\rm ns}})^{3/5}} \ .
   \end{equation}
    In the NS (ambient) frame $t_{\rm cc}^{\rm ns} = \Gamma_{\rm c} t_{cc} \sim$  1472.8 yr.

   \subsection{The ionization stage}
   \label{appendix:ionized-chunk}
   
   The chunk becomes ionized at time $t=t_{\rm ic}$ when hadronic collisions heats it up to $T_{\rm ic}=13.6$ eV prior to becoming collisionless; ``ic" stands for ionized chunk.  For $t_{\rm ic}\le t < t_{\rm cc}$, 
 we can associate  a thermal Bremsstrahlung (TB) luminosity to the chunk
   $L_{\rm TB}(t)= 1.43\times 10^{-27} n_{\rm c, e}(t)n_{\rm c, i}(t)T_{\rm c}(t)^{1/2}V_{\rm c}(t) Z^2 g$ (e.g. \citealt{lang_1999}). 
   In our case we have $Z=1$, $n_{\rm c, e}=n_{\rm c, i}= n_{\rm c, 0} (t/t_0)^{-9/2}$ (with $n_{\rm c, 0}=3m_{\rm c}/4\pi R_{\rm c, opt.}^3m_{\rm H}$), 
   $T_{\rm c}(t)=T_0 (t/t_0)$ and, $V_{\rm c}
   =(4\pi/3)R_{\rm c}(t)^3$ the chunk's volume; 
   $g\simeq 1.2$ is the frequency averaged Gaunt factor. We get 
   
   \begin{equation}
   L_{\rm TB}(t)\simeq 1.7\times 10^{35}\ {\rm erg\ s}^{-1}\times \frac{m_{\rm c, 22.3}^{1/2}T_{\rm 0, 3}^{1/2}}{\kappa_{\rm c, -1}^{3/2}}
   \times \left(\frac{t}{t_0}\right)^{-4}\ .
   \end{equation}

   Setting $T_{\rm ic}=T_0 (t_{\rm ic}/t_0)$ gives us 
   
   \begin{equation}
   \frac{t_{\rm ic}}{t_0}\simeq 157.8\times T_{\rm 0, 3}^{-1}\ ,
   \end{equation}
   and a maximum (i.e. initial) thermal Bremsstrahlung luminosity (at $t=t_{\rm ic}$)
   
   \begin{equation}
   L_{\rm TB, max.}\simeq 2.7\times 10^{26}\ {\rm erg\ s}^{-1}\times \frac{m_{\rm c, 22.3}^{1/2}T_{\rm 0, 3}^{9/2}}{\kappa_{\rm c, -1}^{3/2}}\ .
   \end{equation}
   
   The above is negligible compared to heating from hadronic collision ($Q_{\rm HH}$; see Eq.(\ref{appendix:eq:QHH})).
    When    the chunk enters the collisionless phase at $t_{\rm cc}$, 
   with $t_{\rm cc}/t_{\rm ic}\simeq (878.8/157.8)\times T_{\rm 0, 3}^{3/5}$, the thermal Bremsstrahlung  is even smaller with 
     $L_{\rm TB}(t_{\rm cc})\simeq 10^{-3} \times L_{\rm TB}(t_{\rm ic})$.
     Although negligible compared to hadronic heating, thermal Bremsstrahlung (when $t_{\rm ic}\le t \le t_{\rm cc}$) is boosted to a maximum observed luminosity 
   
       \begin{equation}
       \label{appendix:eq:Lic}
   L_{\rm TB, max.}^{\rm obs.}\simeq 4.4\times 10^{37}\ {\rm erg\ s}^{-1}\times \frac{1}{(1+z)^2f(\theta_{\rm c})^4}\times \sigma_{\rm HH, -27}^{3}\Gamma_{\rm c, 2.5}^{10}m_{\rm c, 22.3}^2{n_{\rm amb., -3}^{\rm ns}}^{3} \ .
   \end{equation}
  The TB phase lasts  for $t_{\rm cc}^{\rm obs.}$ which is of the order of days for fiducial parameter values.

  %
  %
  
  \section{Interaction with the ambient plasma: the relevant instabilities}
\label{appendix:instabilities}

 \subsection{The background plasma and the beam}
 \label{appendix:plasmas}

   We use results from Particle-In-Cell (PIC) and laboratory studies of instabilities in inter-penetrating plasmas,  
  to identify the relevant plasmas:

  \begin{itemize}
  
  \item {\bf The  background ${\it (e^{-},p^{+})}$ plasma}:  is the collisionless  ionized chunk material
  dissociated into hadronic constituents during its early evolution  when  interacting with the ambient medium 
 in the close vicinity of the QN site. When the chunk becomes collisionless, its radius,
   baryon number density and temperature are (see Eqs. (\ref{appendix:eq:cc1}), (\ref{appendix:eq:cc2}) and ((\ref{appendix:eq:cc3}) in Appendix \ref{appendix:heating}): $R_{\rm cc}\sim 10^{15}$ cm, $n_{\rm cc}\sim 10$ cm$^{-3}$
 and $T_{\rm cc}\sim 0.1$ keV, respectively. Here, the subscript ``cc" stands for ``collisionless chunk" defining the start of the collisionless phase. This occurs at time $t_{\rm cc}^{\rm obs.}\simeq 2.6\ {\rm days}\times (1+z)f(\theta_{\rm c})\times  (m_{\rm c, 22.3}\kappa_{\rm c, -1})^{1/5}/(\sigma_{\rm HH, -27}^3\Gamma_{\rm c, 2.5}^{11}({n_{\rm amb., -3}^{\rm ns}})^{3})^{1/5}$ after the QN (Eq.(\ref{appendix:eq:tcc}));

  \item  {\bf The ${\it (e^{-},p^{+})}$ plasma beam}:  is the ionized ambient medium (e.g. ICM) incident
  on the QN chunks as they travel. Its  baryon number density is $n_{\rm amb.}^{\rm ns}$ in the NS frame.

\end{itemize}

The parameters that define the regimes of collisionless instabilities are:
 
 \begin{itemize}

  \item {\bf Ultra-relativistic motion}:  $\Gamma_{\rm c}>> 1$;
  
  \item {\bf Density ratio}:  The beam (ambient medium) to background plasma (collisionless chunk) baryon number density ratio
  in the chunk frame is
\begin{equation}
\label{eq:alpha}
\alpha_{\rm cc}= \frac{\Gamma_{\rm c}n_{\rm amb.}^{\rm ns}}{n_{\rm cc}}= 10^{-1.5} \frac{\Gamma_{\rm c, 2.5}{n_{\rm amb.,-3}^{\rm ns}}}{n_{\rm cc, 1}}\ .
\end{equation}
From $n_{\rm cc}\propto {\Gamma_{\rm c}^{12/5}(n_{\rm amb.}^{\rm ns})}^{6/5}$ (Eq. (\ref{appendix:eq:cc2})), we have
 $\alpha_{\rm cc}\propto {\Gamma_{\rm c}^{-7/5}(n_{\rm amb.}^{\rm ns})}^{-1/5}$ which is weakly dependent on $n_{\rm amb.}^{\rm ns}$ but  rather more on the chunk's Lorentz factor $\Gamma_{\rm c}$;

  \item {\bf Magnetization ($B_{\rm cc}^2/8\pi  n_{\rm cc} m_{\rm H}c^2$)}:    The evolution of the chunk's magnetic field  is estimated
  by   flux conservation $B_{\rm NS} (4\pi R_{\rm NS}^2/N_{\rm c})= B_{\rm cc} \pi R_{\rm cc}^2$ where $B_{\rm NS}$ and
$R_{\rm NS}$ are the NS's magnetic field and radius, respectively.  This gives

\begin{equation}
\label{eq:Bcc}
B_{\rm cc}\simeq 1.3\times 10^{-11}\ {\rm G}\times \frac{B_{\rm NS, 12.5}}{N_{\rm c, 6}}\times  \left(\frac{R_{\rm NS, 6}}{R_{\rm cc, 15}}\right)^2\ .
\end{equation}
With $n_{\rm cc}\sim 10$ cm$^{-3}$ (see Eq. (\ref{appendix:eq:cc2})),  one has $B_{\rm cc}^2/8\pi << n_{\rm cc}m_{\rm H}c^2$ when the chunk becomes collisionless, effectively becoming a non-magnetized plasma when experiencing the inter-penetrating
 instabilities discussed below.

\end{itemize}

  \subsection{The instabilities}
\label{appendix:BI-WI}
 
 The above parameter ranges imply that at the onset of the collisionless stage,   
  the Buneman instability (BI) dominates the dynamics (e.g. Table 1 and Figure 5 in \citealt{bret_2009}). 
    The BI induces an anisotropy in the chunk's electron 
    temperature distribution, triggering the thermal Weibel instability (WI; \citealt{weibel_1959}),  which has the
    effect of isotropizing the temperature.   The thermal WI requires
    only a temperature anisotropy to exist and is beam-independent. The Weibel filamentation instability (FI)  
    on the other hand requires a beam to exist (\citealt{fried_1959}). However, 
 the FI dominates   only when $\alpha_{\rm cc}\sim 1$ (see Figure 5 in \citealt{bret_2009}), which is not the case 
    here because we have $\alpha_{\rm cc} << 1$ as expressed in Eq. (\ref{eq:alpha}).
 
The beam (i.e. the ICM plasma) triggers the longitudinal BI (with wave vector aligned with the beam). 
 This creates the needed anisotropy since the BI yields efficient heating of electrons
 in the longitudinal  direction (parallel to the beam). The scenario is  a parallel plasma temperature
   which exceeds the perpendicular plasma temperature, allowing the thermal
   WI to act (even in the weak anisotropy case).  During the development of the WI, the beam  continues to feed the BI
  by continuous excitation of electrostatic waves.

These two instabilities are discussed in more detail below. We define
$\beta_{\parallel}=v_{\parallel}/c$ and $\beta_{\perp}=v_{\perp}/c$, where $c$ is the light speed, as the  chunk
 electron's  speed  in the direction parallel and perpendicular to the beam,  respectively.
  When a QN chunk becomes collisionless, it has 
   $\beta_{\parallel}=\beta_{\perp}= \beta_{\rm cc}\sim 10^{-2}$ (Eq. (\ref{appendix:eq:betaecc})):

\begin{itemize}
\item {\bf The Buneman Instability (BI)} is an electro-static instability (i.e.  excitations of electrostatic waves). 
 It is an electron-ion two-stream instability caused by the resonance between the plasma oscillation of the  chunk electrons and plasma oscillation of the  ambient medium protons (\citealt{buneman_1958,buneman_1959}).  In our case,  it arises 
when the relative drift velocity between the beam (i.e. ICM) protons and the plasma (i.e. chunk) electrons exceeds the chunk's electron thermal velocity. Its wave vector is parallel to the beam propagation direction
  and generates stripe-like patterns (density stripes perpendicular to the
  beam; e.g. \citealt{bret_2010}). The BI gives rise to  rapid electron heating (e.g. \citealt{davidson_1970,davidson_1974,hirose_1978})
by transferring a percentage  of the beam's kinetic energy into thermal (electron) energy of the background plasma (here the QN chunk)
by turbulent (electric field) heating. The end result is an increase in $\beta_{\parallel}$ with $\beta_{\perp}$ unchanged.
The wavelength of the dominant mode is

\begin{equation}
\label{eq:lambdaBI}
\lambda_{\rm BI} = \alpha_{\rm cc} \times \frac{c}{\nu_{\rm p, e}}\ ,
\end{equation}
 where  $\nu_{\rm p, e}= (4\pi e^2n_{\rm cc, e}/m_{\rm e})^{1/2}\simeq (9\ {\rm kHz})\times n_{\rm cc}^{1/2}$ is
  the non-relativistic electron plasma frequency of the chunk and $n_{\rm cc, e}=n_{\rm cc}$  the chunk's
  electron density; $m_{\rm e}$ and $e$ are the electron's mass and charge, respectively. The e-folding growth timescale is

\begin{equation}
\label{eq:tBI}
t_{\rm BI}=  \frac{2^{4/3}}{3^{1/2}} \left(\frac{m_{\rm p}}{m_{\rm e}}\right)^{1/3}\times \frac{1}{\nu_{\rm p, e}}\simeq \frac{17.8}{\nu_{\rm p, e}}\simeq
\frac{0.63\ {\rm ms}}{n_{\rm cc, 1}^{1/2}}\ ,
\end{equation}
where $m_{\rm p}$ is the proton mass. The above is much shorter than the chunk's crossing time $R_{\rm cc}/c\sim 3.3\times 10^4\ {\rm s}\times R_{\rm cc, 15}$ allowing plenty of time for the BI to grow and saturate locally throughout the collisionless chunk.

BI heating occurs by transferring beam electron energy to heating chunk's electrons. 
BI saturation occurs much before the beam kinetic energy is depleted because of trapping of electrons by turbulence;
i.e. BI saturates at a particular electric field (e.g. \citealt{hirose_1978}). The heat gain by the chunk's electrons is
   $Q_{\rm BI}=(\zeta_{\rm BI}\Gamma_{\rm c}m_{\rm e}c^2)\times (A_{\rm cc}\Gamma_{\rm c}n_{\rm amb.}^{\rm ns}c)$, where $A_{\rm cc}=\pi R_{\rm cc}^2$, or

  \begin{equation}
  \label{eq:QBI}
   Q_{\rm BI}\simeq 7.6\times 10^{35}\ {\rm erg\ s}^{-1}\times \zeta_{\rm BI, -1} \Gamma_{\rm c, 2.5}^2 R_{\rm cc, 15}^2 n_{\rm amb., -3}^{\rm ns}\ ,
  \end{equation}
  expressed in terms of the BI saturation,  parameter $\zeta_{\rm BI}$ (here free). It is 
   the fraction of the electrons kinetic energy (in the beam)
 converted to an electrostatic field and subsequently to heating the chunk's electrons (to increasing $\beta_{\parallel}$).
  At saturation, the electron energy gain is $\sim$ 10\% of the beam electron kinetic energy;
  the protons energy gain is much less than that of chunk electrons
  (e.g. \citet{dieckmann_2012,moreno_2018}).

\item {\bf The thermal Weibel Instability (WI)}  is an electro-magnetic instability which occurs
in plasmas with an anisotropic electron temperature distribution
(\citealt{weibel_1959}; see \citealt{fried_1959} for Weibel FI). 
Its wave vector is  perpendicular to the  high temperature axis 
which corresponds to the beam propagation direction  induced by the BI heating.  
The WI  can efficiently generate magnetic fields. The corresponding currents are  in the direction parallel to  the beam
with the resulting magnetic field  perpendicular to it. As the BI accelerates electrons (increasing $\beta_{\parallel}$), 
 the WI heats up the chunk's electrons via  particle scattering by the generated magnetic field,  
   accelerating them  in the transverse direction (increasing $\beta_{\perp}$), 
   as to reduce the BI-induced thermal anisotropy.

 The WI was studied theoretically in the non-relativistic and relativistic regimes (\citealt{weibel_1959,fried_1959,yoon_1987}; see also \citealt{medvedev_1999,gruzinov_2001}) and numerically using  
 PIC simulations (e.g. \citealt{kato_2007,spitkovsky_2008,nishikawa_2009}). 
Its key phases  which are of relevance to our model for FRBs are:

(i) {\bf Electron-WI (e-WI)}:  Because of the small  inertia, chunk electrons dominate the dynamics setting the characteristic
   correlation length  of magnetic field  and its growth rate. 
     We use Eqs. (A4a) and (A4b) in \citet{medvedev_1999}, which describe the mode with the largest growth rate,
  valid for  $\beta_{\perp}=v_{\perp}/c<<1$  and for $\beta_{\parallel}=v_{\parallel}/c$ which is arbitrary in our model.  
   In the notation of \citet{medvedev_1999}, our scenario corresponds to $\hat{\gamma}=1$
   and $G(\beta_{\perp})=(2\beta_{\perp})^{-1}\ln{(1+\beta_{\perp})/(1-\beta_{\perp})}\simeq (2\beta_{\perp})^{-1}\ln{(1+2\beta_{\perp})}\simeq 1$.
  The wavenumber $k_{\rm max.}$ and growth rate $\Gamma_{\rm max.}$ of the dominant mode are, respectively,

   \begin{equation}
   \label{appendix:eq:ML}
   k_{\rm max}^2c^2 \simeq \nu_{\rm p, e}^2 \frac{\beta_{\parallel}}{\sqrt{2}\gamma_{\parallel}\beta_{\perp}} \left(1- \frac{\beta_{\parallel}\beta_{\perp}}{\sqrt{2}} - \frac{\sqrt{2}\beta_{\perp}}{\beta_{\parallel}}\right)
    \quad\quad\quad \Gamma_{\rm max.}^2\simeq  \nu_{\rm p, e}^2 \frac{\beta_{\parallel}^2}{\gamma_{\parallel}} \left(1-2\sqrt{2}\frac{\beta_{\perp}}{\beta_{\parallel}} \right)\ ,
   \end{equation}
   where $\nu_{\rm p, e}$ is the electron plasma frequency. The above shows that the WI saturates when $\beta_{\parallel}=2\sqrt{2}\beta_{\perp}$.

  The dominant wavelength in the non-relativistic regime (with $\gamma_{\parallel}=1$) is: 
 
 \begin{align}
\label{eq:lambdaWI}
 \lambda_{\rm e-WI}  &\simeq \left(\frac{\beta_{\perp}}{\beta_{\parallel}}\right)^{1/2}\times \frac{c}{\nu_{\rm p,e}} = \beta_{\rm WI}^{1/2}\times \frac{c}{\nu_{\rm p,e}}\nonumber\\
  &\simeq 3.5\times 10^5\ {\rm cm}\times \frac{\beta_{\rm WI, -1}^{1/2}}{n_{\rm cc, 1}^{1/2}}\ ,
 \end{align}
 where we defined $\beta_{\rm WI}=\beta_{\perp}/\beta_{\parallel}<1$. 
 From Eq. (\ref{appendix:eq:ML}), we must have
  $\beta_{\parallel}> 2\sqrt{2}\beta_{\perp}$ for the WI to be triggered.  Hereafter we set 
   $\beta_{\parallel}=10\beta_{\perp}$ (i.e. $\beta_{\rm WI}=\beta_{\perp}/\beta_{\parallel}=0.1$ as the fiducial value) with  $\beta_{\perp}=\beta_{\rm cc}\sim 10^{-2}$  given by the  initial conditions   when the chunk first becomes collisionless (i.e. Eq.(\ref{appendix:eq:betaecc})).

 The WI current filament structures have a transverse width of the order of $\lambda_{\rm e-WI}$ 
 and are elongated in the beam's direction (Appendix \ref{appendix:bunch-geometry} and Figure \ref{figure:bunch-geometry}). This dominant mode 
   grows  on an e-folding timescale of 
  
  \begin{align}
   \label{eq:teWI}
  t_{\rm e-WI} &\simeq \frac{1}{\beta_{\parallel}}\times  \frac{1}{\nu_{\rm p, e}}=  \frac{\beta_{\rm WI}}{\beta_{\perp}}\times  \frac{1}{\nu_{\rm p, e}}\nonumber\\
    &\simeq  0.35\ {\rm ms}\times \frac{\beta_{\rm WI, -1}}{\beta_{\rm cc, -2}}\times  \frac{1}{n_{\rm cc, 1}^{1/2}}\ \ .
  \end{align}

 In the linear regime we estimate the saturation time of the e-WI, $ t_{\rm e-WI, s}$, by setting $B_{\rm e-WI, s}= B_{\rm cc}e^{(t_{\rm e-WI, s}/t_{\rm e-WI})}$ with $B_{\rm e-WI, s}^2/8\pi\sim n_{\rm cc}m_{\rm e}c^2$
    the magnetic field strength at saturation. This is equivalent to writing 
     $\nu_{\rm B}\sim \nu_{\rm p, e}$ at e-WI saturation  with here $\nu_{\rm B}=eB_{\rm e-WI, s}/m_{\rm e}c$, 
   the  electron cyclotron frequency at e-WI saturation. We get
    
    \begin{equation}
    \label{eq:teWIs}
    t_{\rm e-WI, s}\simeq \left(21.2 +\ln{\frac{n_{\rm cc, 1}^{1/2}}{B_{\rm cc, -11}}}\right)\times t_{\rm e-WI} \ ,
    \end{equation}
    with $B_{\rm cc}$ given in Eq. (\ref{eq:Bcc});

   (ii) {\bf Proton-WI (p-WI)}:  After the e-WI stage, and still in the linear regime, follows the   p-WI stage 
   which grows more slowly than the e-WI,  on timescales $t_{\rm p-WI}=\sqrt{m_{\rm p}/m_{\rm e}} t_{\rm e-WI}$.  
    The  magnetic field  is further amplified to a saturation value  $B_{\rm p-WI, s}=\sqrt{m_{\rm p}/m_{\rm e}} B_{\rm e-WI, s}$ (\citealt{bret_2016}
  and references therein). The saturation time $t_{\rm p-WI, s}$ of the p-WI phase is found by setting $B_{\rm p-WI, s}= B_{\rm e-WI, s}\times e^{(t_{\rm p-WI, s}/t_{\rm p-WI,})}$.
      This gives $t_{\rm p-WI, s}= (m_{\rm p}/m_{\rm e})^{1/2}\ln{(\sqrt{m_{\rm p}/m_{\rm e}})}\times t_{\rm e-WI}$, or, 
    
    \begin{equation}
     \label{eq:tpWIs}
    t_{\rm p-WI, s}\sim 161\times t_{\rm e-WI}\ .
    \end{equation}
    
\subsection{Non-linear regime: filament merging (m-WI)}
\label{appendix:merging}
  
 In the non-linear regime, following the saturation of the p-WI stage,  the  filaments start merging and grow
  in size  increasing $\lambda_{\rm e-WI}$. The merging is a result of 
 the attractive force between parallel currents   (\citealt{lee_2001,frederiksen_2004,kato_2005,medvedev_2005,milosavljevic_2006}).
 Recent theoretical  (e.g. \citet{achterberg_2007}) and numerical (e.g. \citet{takamoto_2018}) studies 
 suggest a slow and a complex merging process. 
  PIC simulations of filament merging in 3-dimensions (\citet{takamoto_2018,takamoto_2019}) find that during filament merging:
 {\bf (a)} Electrons are stochastically accelerated by the magnetic turbulence generated by the WI up to a Lorentz factor of 
$\gamma_{\rm e}\sim 10$; {\bf (b)} this heating sustains the WI saturated magnetic field 
 for at least hundreds of ion plasma oscillations.  Relying on these studies, we set a typical 
 merging timescale as

 \begin{align}
 \label{eq:tmWI}
 t_{\rm m-WI}&\simeq \frac{10^2 \zeta_{\rm m-WI, 2}}{\nu_{\rm p, p}}\simeq  \frac{4.3\times 10^3\ \zeta_{\rm m-WI,2}}{\nu_{\rm p, e}}\nonumber\\ 
 &\simeq 0.15\ {\rm s}\times \frac{\zeta_{\rm m-WI,2}}{n_{\rm cc, 1}^{1/2}}\ ,
 \end{align}
 where $\nu_{\rm p, p}=\sqrt{m_{\rm p}/m_{\rm e}}\nu_{\rm p, e}$ is the proton plasma
 frequency and $\zeta_{\rm m-WI}=10^2$ a parameter which allows us to adjust   
  the merging timescale. The time evolution of the filament size we consider to be a power law 

\begin{equation}
\label{eq:filament-merging}
\lambda_{\rm F}(t)= \lambda_{\rm e-WI} \times \left(1+ \frac{t}{t_{\rm m-WI}}\right)^{\delta_{\rm m-WI}}\ ,
\end{equation}
with $\lambda_{\rm F}(0)=\lambda_{\rm e-WI}$ the filament's transverse 
size during the linear regime and $\delta_{\rm m-WI} >0$ (simulations suggest $\delta_{\rm m-WI} \sim 0.76$; e.g. \citet{takamoto_2019}). Hereafter we adopt $\delta_{\rm m-WI}=1$ as our fiducial value.

{\bf Proton trapping and shock formation}: The Weibel shock occurs when the  protons are trapped by the
growing filaments; i.e. when the filament size becomes of the order of the beam's  proton's 
Larmor radius. The shock quickly converts the   chunk's kinetic energy
    to internal energy by sweeping ambient protons leading to full chunk slowdown 
   and shutting off the BI-WI process.

\end{itemize}

We close this appendix by discussing a few points:

\begin{itemize}

\item Table \ref{table:parameters} lists the parameters related to the BI and WI
instabilities and the fiducial values we adopted in this work. For the BI we have  $\zeta_{\rm BI}$ 
which is the percentage of the beam's electron energy (in the
chunk fame) converted by the BI to heating the chunk electrons. The WI-related parameters are: 
(i) $\beta_{\rm WI}=\beta_{\perp}/\beta_{\parallel}$, the ratio of transverse to longitudinal 
 thermal speed of chunk electrons (Eq. (\ref{eq:lambdaWI})) at the onset of the WI; (ii)
      $\zeta_{\rm m-WI}$  the filament merging characteristic timescale (Eq. (\ref{eq:tmWI})) and; (iii) 
       $\delta_{\rm m-WI}$  the power index of the filament merging rate as given in Eq. (\ref{eq:filament-merging});

  \item   We adopt     $\beta_{\parallel}=10\beta_{\perp}$ (i.e. $\beta_{\rm WI}=\beta_{\perp}/\beta_{\parallel}=0.1$) during the linear stages of the WI instability which keeps   $\lambda_{\rm e-WI}$ constant. While $\beta_{\parallel}$ grows due to the BI, the WI
   increases $\beta_{\perp}$ accordingly, as to the keep $\beta_{\rm WI}$ constant. However, because 
   $\lambda_{\rm BI}/\lambda_{\rm e-WI}=\alpha_{\rm cc}/\beta_{\rm WI}^{1/2}<<1$, the BI deposits energy (i.e. heats up and accelerates electrons) in layers that are much narrower than those of the WI;

 \item With $t_{\rm BI}\sim 17.8/\nu_{\rm p, e}$ and $t_{\rm e-WI}=(\beta_{\rm WI}/\beta_{\perp})/\nu_{\rm p, e}\sim 10/\nu_{\rm p, e}$
 being of the same order, the BI heat deposited within $\lambda_{\rm BI}$ is quickly mixed into  much larger scales 
 given by $\lambda_{\rm e-WI}$;

\item  The Oblique mode instability (when both longitudinal and transversal waves components
 are present at the same time) dominates  when $\alpha_{\rm cc} > (m_{\rm e}/m_{\rm p}) \Gamma_{\rm c}$
 (e.g. \citealt{bret_2009}). In our case this translates to $n_{\rm amb.}^{\rm ns} > (m_{\rm e}/m_{\rm p})n_{\rm cc}\simeq 0.5\ {\rm cm}^{-3}\times n_{\rm cc, 1}$. Since the ICM's density is $n_{\rm amb.}^{\rm ns} <<1$, the BI will always dominate;
 
 \item The BI heat is partly converted to amplifying the magnetic field (i.e. to magnetic energy
 density $B_{\rm p-WI,s}^2$), partly  to turbulence with energy density $\delta B_{\rm p-WI,s}^2$ and,
to currents. During filament merging, electrons are accelerated by dissipation
of turbulent energy and currents while the WI saturated magnetic field is
preserved (\citealt{takamoto_2018}). The BI energy   harnessed   during the linear regime is $E_{\rm BI}\sim Q_{\rm BI}t_{\rm p-WI,s}$.  With $Q_{\rm BI}$  given by Eq. (\ref{eq:QBI})  and  $t_{\rm p-WI,s}$  given by Eqs. (\ref{eq:tpWIs}) and (\ref{eq:teWI}), respectively, we get
 
  \begin{align}
  \label{eq:EBI}
  E_{\rm BI}\simeq 4.4\times 10^{34}\ {\rm ergs}\times \zeta_{\rm BI, -1} \times \frac{\beta_{\rm WI, -1}}{\beta_{\rm cc, -2}} \times 
  \frac{\Gamma_{\rm c, 2.5}^2R_{\rm cc, 15}^2 {n_{\rm amb., -3}^{\rm ns}}}{n_{\rm cc, 1}^{1/2}}\ ;
  \end{align}

 \item The top panel in Figure \ref{fig:WI-stages} is a schematic representation of the evolution of $\beta_{\parallel}$ during the
 linear and non-linear WI stages ($\beta_{\perp}= 0.1\beta_{\parallel}$ follows the evolution of $\beta_{\parallel}$). 
 The increase in $\beta_{\parallel}$ is due to the BI  and proceeds until the end of the p-WI stage, when the magnetic field saturates.
  At this point the BI excitations are converted entirely to heating  electrons with the consequence
 that $\beta_{\parallel}$ increases rapidly  following p-WI saturation. The BI shuts off when $\gamma_{\parallel}\sim 2$ because
  it acts only when the relative drift between the beam electrons
 and the chunk protons (here $c$) exceeds the chunk's electrons thermal speed. Despite the BI shutting-off, the electrons continue to be
 accelerated  by magnetic turbulence and by current dissipation during filament merging 
 yielding $\gamma_{\parallel}\sim \gamma_{\perp}\sim 10$ (\citealt{takamoto_2018}). As discussed below,
 the increase in electron Lorentz factor during the merging phase, provides conditions
 favorable for coherent synchrotron emission (CSE) to occur  in the  WI-amplified magnetic field layers of the chunk.

 \end{itemize}

   \section{Coherent synchrotron emission (CSE)}
\label{appendix:CSE}
\setcounter{equation}{0}

A relativistic electron beam moving in a circular orbit  can radiate coherently if the characteristic wavelength of the 
  incoherent synchrotron emission (ISE), $\lambda_{\rm ISE}$,  exceeds the length of the electron bunch $\lambda_{\rm b}$.  
 The near field of the radiation from each electron overlaps the entire bunch structure, resulting in a
coherent interaction yielding a CSE frequency $\nu_{\rm CSE}=c/\lambda_{\rm b}$.
 With $N_{\rm e, b}$  the number of electrons in a bunch, the 
  intensity  of  CSE scales as $N_{\rm e, b}^2$ instead of $N_{\rm e, b}$ as in 
   the incoherent case (\citealt{schiff_1946,schwinger_1949,motz_1951,nodvick_1954,ginzburg_1965}).

The total power per bunch is estimated as  $N_{\rm e, b}^2 (\nu F_{\nu})_{\nu_{\rm CSE}}$ where
$F_{\nu}= (\sqrt{3} \nu_{\rm B} e^2/c)\times F(\nu/\nu_{\rm ISE})$ is the  incoherent synchrotron frequency distribution (in erg s$^{-1}$ Hz$^{-1}$)
 at the  characteristic frequency $\nu_{\rm ISE}=(3/2)\gamma_{\rm e}^2\nu_{\rm B}$ with $\nu_{\rm B}= eB/m_{\rm e}c$
 the cyclotron frequency and $\gamma_{\rm e}$ the electrons' Lorentz factor. 
At $\nu_{\rm CSE}\sim c/\lambda_{\rm b} << \nu_{\rm ISE}$, we have $F(\nu/\nu_{\rm ISE})\sim 2.15 (\nu/\nu_{\rm ISE})^{1/3}$ which gives a total power per bunch of

\begin{equation}
\label{eq:CSE-ouyed}
L_{\rm b} \simeq  3.3\times 10^{-29} \times N_{\rm e, b}^2 \nu_{\rm CSE}^2 \frac{1}{\gamma_{\rm e}^{2/3}} \left( \frac{\nu_{\rm B}}{\nu_{\rm CSE}}\right)^{2/3}\ .
\end{equation}
This agrees within a factor of a few with expressions given in the literature (e.g. \citealt{murphy_1997} and references therein). 
  The spectrum of CSE is the same as the incoherent one except for the $N_{\rm e, b}$ boosting
  and a decrease in the maximum (peak) frequency.

  \subsection{CSE properties in our model}

 During the linear phase of the WI (up to p-WI saturation), CSE is unlikely to occur because  BI heating
cannot yield relativistic electrons ($\gamma_{\rm CSE}<2$; see top panel in Figure \ref{fig:WI-stages}). Furthermore, bunching cannot be induced by the BI during filament
 merging because the instability does not grow if the background (i.e. chunk) electrons 
 are so hot ($\gamma_{\rm CSE}>2$) that their thermal velocity spread exceeds the drift velocity relative to the
  beam (i.e. streaming ambient) ions.  Instead, bunching is related to (i.e. entangled with) the WI filaments and 
  CSE is likely to be triggered during filament merging when electrons are accelerated
 by magnetic turbulence and current dissipation to $\gamma_{\rm CSE}>>1$.

  \subsubsection{Frequency and duration}
  \label{sec:nu-CSE}
  
  With $\nu_{\rm B}\sim  \sqrt{m_{\rm p}/m_{\rm e}}\nu_{\rm p, e}$ after p-WI saturation and during the filament
  merging phase, we calculate the  chunk's  magnetic field strength to be 
  
  \begin{equation}
  \label{eq:Bcc,s}
  B_{\rm p-WI, s}\sim 0.12\ {\rm G}\times n_{\rm cc, 1}^{1/2}\ , 
  \end{equation}
  and the  characteristic ISE frequency to be $\nu_{\rm ISE}= 3/2\times \gamma_{\rm CSE}^2\sqrt{m_{\rm p}/m_{\rm e}}\nu_{\rm p, e}$. 

   The CSE frequency, $\nu_{\rm CSE}(t)=c/\lambda_{\rm b}(t)$, evolves in time due to the scaling of the bunch size
   $\lambda_{\rm b}(t)$  with that of the WI filament $\lambda_{\rm F}(t)$ which is expressed in Eq. (\ref{eq:filament-merging}). 
   We find the  CSE frequency  to decrease in time during the filament merging phase 
   at a rate given by
  
   \begin{align}
  \label{eq:CSE-frequency-time}
  \nu_{\rm CSE}(t) = \frac{c}{\lambda_{\rm b}(t)} = \nu_{\rm CSE}(0)\times \left(1+\frac{t}{t_{\rm m-WI}}\right)^{-\delta_{\rm m-WI}}\ ,
 \end{align}
 with $\delta_{\rm m-WI}>0$ and $ \nu_{\rm CSE}(0)=c/\lambda_{\rm e-WI}$; $\lambda_{\rm e-WI}$  given by Eq. (\ref{eq:lambdaWI})
 is the  filament's transverse size during the linear phase.
 
 Because $\nu_{\rm CSE}<<\nu_{\rm ISE}$ we set the initial (also the maximum) CSE frequency as 
  \begin{equation}
  \label{eq:nu-CSE}
  \nu_{\rm CSE}(0)=\delta_{\rm CSE}\nu_{\rm ISE}\ ,
  \end{equation}
  with $\delta_{\rm CSE}<<1$. The CSE frequency decreases in time until
  it reaches the chunk's plasma frequency $\nu_{\rm p, e}$ shutting-off emission. The range in CSE frequency
  from a collisionless QN chunk is thus
  
  \begin{equation}
\nu_{\rm p, e} \le \nu_{\rm CSE} \le \nu_{\rm CSE}(0)=\delta_{\rm CSE}  \nu_{\rm ISE}\ .
  \end{equation}

The duration of CSE is found from $\nu_{\rm p, e}= \nu_{\rm CSE}(0)\times (1+\Delta t_{\rm CSE}/t_{\rm m-WI})^{-\delta_{\rm m-WI}}$
giving us:

\begin{equation}
\label{eq:dtCSE}
\Delta t_{\rm CSE} =\left(  \left(642.7\delta_{\rm CSE, -1}\gamma_{\rm CSE,1}^2\right)^{\frac{1}{\delta_{\rm m-WI}}} -1\right) \times t_{\rm m-WI}\ ,
\end{equation}

with $\delta_{\rm m-WI}=1.0$, $\delta_{\rm CSE}=0.1$ and $\gamma_{\rm CSE}=10$ the fiducial values listed in Table \ref{table:parameters}.

\subsubsection{Luminosity}
\label{sec:luminosity}

  Most of the BI-induced heat is harnessed during the linear regime and up until the start of filament merging.
  Once the electrons thermal energy becomes relativistic (with $\gamma_{\rm CSE}>2$),  the BI shuts-off. Effectively, the electrostatic energy
  deposited by the BI inside the chunk  during the linear regime is $E_{\rm BI}\simeq Q_{\rm BI} t_{\rm p-WI,s}$ (see Eq. (\ref{eq:EBI})) where
  $t_{\rm p-WI,s}$ is the p-WI saturation timescale. This energy is converted by the WI to:
  (i) magnetic field amplification with $B_{\rm p-WI,s}\sim \sqrt{m_{\rm p}/m_{\rm e}}B_{\rm e-WI,s}$ at saturation;
  (ii) magnetic turbulence; (iii) currents. Filament merging converts
  about 2/3 of the BI energy (by turbulence acceleration and current dissipation) to accelerating 
  electrons (e.g. \citet{takamoto_2018}). The energy gained by the chunk electrons during filament
  merging is  re-emitted as CSE luminosity expressed as $L_{\rm CSE}\sim (2/3)E_{\rm BI}/t_{\rm m-WI}$:

   \begin{equation}
   \label{eq:LCSE}
   L_{\rm CSE}\simeq 1.9\times 10^{35}\ {\rm ergs}\times \frac{\zeta_{\rm BI, -1}\beta_{\rm WI, -1}}{\zeta_{\rm m-WI, 2}} \times \frac{\Gamma_{\rm c, 2.5}^2 R_{\rm cc, 15}^2 n_{\rm amb., -3}^{\rm ns}}{\beta_{\rm cc, -2}}\ .
   \end{equation}

  \subsection{Bunch geometry and CSE luminosity}
  \label{appendix:bunch-geometry}  
   
 As illustrated in Figure \ref{figure:bunch-geometry} here, the Weibel filament extend across the
collisionless chunk with length $2R_{\rm cc}$. The  initial filament's diameter is $\lambda_{\rm F}(0)=\lambda_{\rm e-WI}$  as expressed in Eq. (\ref{eq:lambdaWI}). Bunching would  manifest itself  in a narrow region around the Weibel filaments where the magnetic field amplification 
is expected to occur and not inside filaments where the currents reside and the magnetic field is weaker. In other words,  a typical bunch, where CSE occurs, would resemble a cylindrical shell around the Weibel filament  with initial thickness $\lambda_{\rm b}(0)$, initial area $A_{\rm b}(0)=2\pi \lambda_{\rm e-WI} \lambda_{\rm b}(0)$ and, extending
  across the chunk.  We have 
  
   \begin{equation}
  \label{eq:deltab}
  \lambda_{\rm b}(0)\simeq \delta_{\rm b}\times \lambda_{\rm e-WI}\ ,
  \end{equation} 
  and because  the maximum CSE frequency is expressed as 
  $\nu_{\rm CSE}(0)=c/\lambda_{\rm e-WI}= \delta_{\rm CSE}\nu_{\rm ISE}$ (see Eq.(\ref{eq:nu-CSE})), this implies
  
  \begin{equation}
  \delta_{\rm b}= \frac{4.9\times 10^{-3}}{\beta_{\rm WI, -1}^{1/2}\delta_{\rm CSE, -1}\gamma_{\rm CSE, 1}^2} << 1.0\ .
  \end{equation}

During filament merging the filament's diameter (and thus
 the associated bunch thickness $\lambda_{\rm b}(t)=\delta_{\rm b}\lambda_{\rm F}(t)$) increases in time as $\lambda_{\rm F}(t)=\lambda_{\rm e-WI}\times (1+t/t_{\rm m-WI})^{-\delta_{\rm m-WI}}$
     (see Eq. (\ref{eq:filament-merging})) with  $t_{\rm m-WI}$, given by Eq. (\ref{eq:tmWI}), the characteristic filament merging timescale.  
There is one bunch per filament which implies that the total number of bunches per chunk is $N_{\rm b, T} =\pi R_{\rm cc}^2/\pi \lambda_{\rm F}(t)^2$
 and decreases in time at a rate given by 
 
\begin{equation}
\label{eq:NbT}
N_{\rm b, T}(t) \simeq 9\times 10^{18} \times \frac{R_{\rm cc, 15}^2 n_{\rm cc, 1}}{\beta_{\rm WI, -1}}\times \left(1+\frac{t}{t_{\rm m-WI}}\right)^{-2\delta_{\rm m-WI}} \ .
\end{equation}

 The corresponding number of electrons per bunch is $N_{\rm e, b}(t)= V_{\rm b}(t) n_{\rm cc}$
 with $V_{\rm b}(t)= (2\pi\lambda_{\rm F}(t)\lambda_{\rm b}(t)) \times 2R_{\rm cc} = \delta_{\rm b}\times (2\pi\lambda_{\rm F}(t)^2) \times 2R_{\rm cc}$ the volume. Thus

\begin{equation}
\label{eq:particles-per-bunch}
N_{\rm e, b}(t)\simeq 1.4\times 10^{26}\times R_{\rm cc, 15}\times (\delta_{\rm b, -2}\beta_{\rm WI, -1}) \times  \left(1+\frac{t}{t_{\rm m-WI}}\right)^{+2\delta_{\rm m-WI}}  \ .
\end{equation}

The luminosity per bunch $L_{\rm b}(t)$ is given by inserting 
Eq. (\ref{eq:particles-per-bunch}) into Eq. (\ref{eq:CSE-ouyed}) with  $\nu_{\rm B}\simeq \sqrt{m_{\rm p}/m_{\rm e}}\nu_{\rm p, e}$
 at proton-WI (p-WI) saturation. We get

\begin{align}
 L_{\rm b}(t)\simeq 1.6\times 10^{36}\ {\rm erg\ s}^{-1}\times R_{\rm cc, 15}^2n_{\rm cc, 1}\times \gamma_{\rm CSE, 1}^2\delta_{\rm CSE,-1}^{4/3}\times (\delta_{\rm b, -2}\beta_{\rm WI, -1})^2\times \left(1+\frac{t}{t_{\rm m-WI}}\right)^{+\frac{8}{3}\delta_{\rm m-WI}} \ .
 \end{align}
 The corresponding cooling timescale of a bunch  $t_{\rm b}=N_{\rm e, b}\gamma_{\rm CSE}m_{\rm e}c^2/L_{\rm b}(t)$
 can be shown to be extremely fast  compared to the duration of CSE $\Delta t_{\rm CSE}$ (see Eq.  (\ref{eq:dtCSE})). With
    $t_{\rm b}<< {\Delta t_{\rm CSE}}$ it points to the fact that a given bunch has a very low duty cycle and
 emits only once (i.e. a single pulse) during the duration of the CSE, $\Delta t_{\rm CSE}$. It also has the consequence that 
 the fraction of bunches emitting at any give time during the  CSE phase  is $t_{\rm b}(t)/\Delta t_{\rm CSE}$.  The total CSE luminosity  is thus $(N_{\rm b, T}(t)\times t_{\rm b}(t)/\Delta t_{\rm CSE})\times L_{\rm b}(t)= \frac{N_{\rm b, T}(t) N_{\rm e, b}(t) \gamma_{\rm CSE}m_{\rm e}c^2}{\Delta t_{\rm CSE}}$, or

        \begin{align}
        \label{eq:LCSE-2}
        L_{\rm CSE}\simeq  10^{37}\ {\rm erg\ s}^{-1}\times \frac{R_{\rm cc, 15}^3 n_{\rm cc, 1}\gamma_{\rm CSE, 1}\delta_{\rm b, -2}}{\Delta t_{\rm CSE, 3}} \ ,
        \end{align}  
        which is a constant because $N_{\rm b, T}(t)\propto \left(1+\frac{t}{t_{\rm m-WI}}\right)^{-2\delta_{\rm m-WI}}$ and $N_{\rm e, b}(t)\propto \left(1+\frac{t}{t_{\rm m-WI}}\right)^{+2\delta_{\rm m-WI}}$. The CSE duration in the chunk frame is given in units of $10^3$ s for fiducial parameter values (see Eq.  (\ref{eq:dtCSE})). Comparing the equation above to  Eq. (\ref{eq:LCSE})  which gives $L_{\rm CSE}\simeq  10^{33}$-$10^{34}\ {\rm erg\ s}^{-1}$ suggests that  the length of a bunch does not extend across the entire chunk and that it may instead be  a small fraction of
  the chunk's radius; i.e. $\sim (10^{-3}$-$10^{-2})R_{\rm cc}$. However, this has no consequence to
  our findings here since the bunches are very effective at releasing the heat harnessed during the BI phase regardless
  of their shape and size.
             
\subsection{Summary}  
     
\begin{figure*}
 \centering
  \includegraphics[scale=0.4]{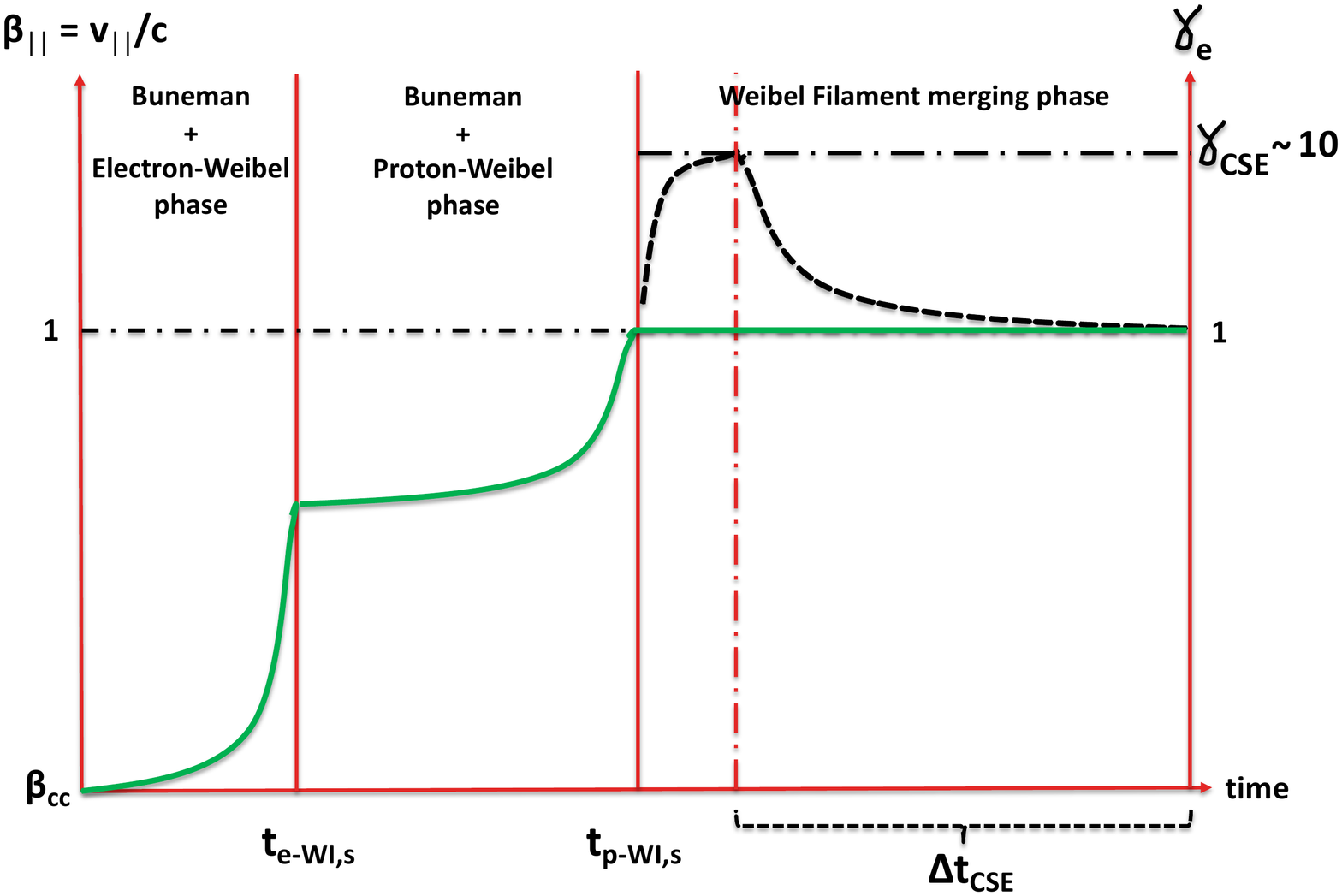}\\
\hskip -0.8cm  \includegraphics[scale=0.405]{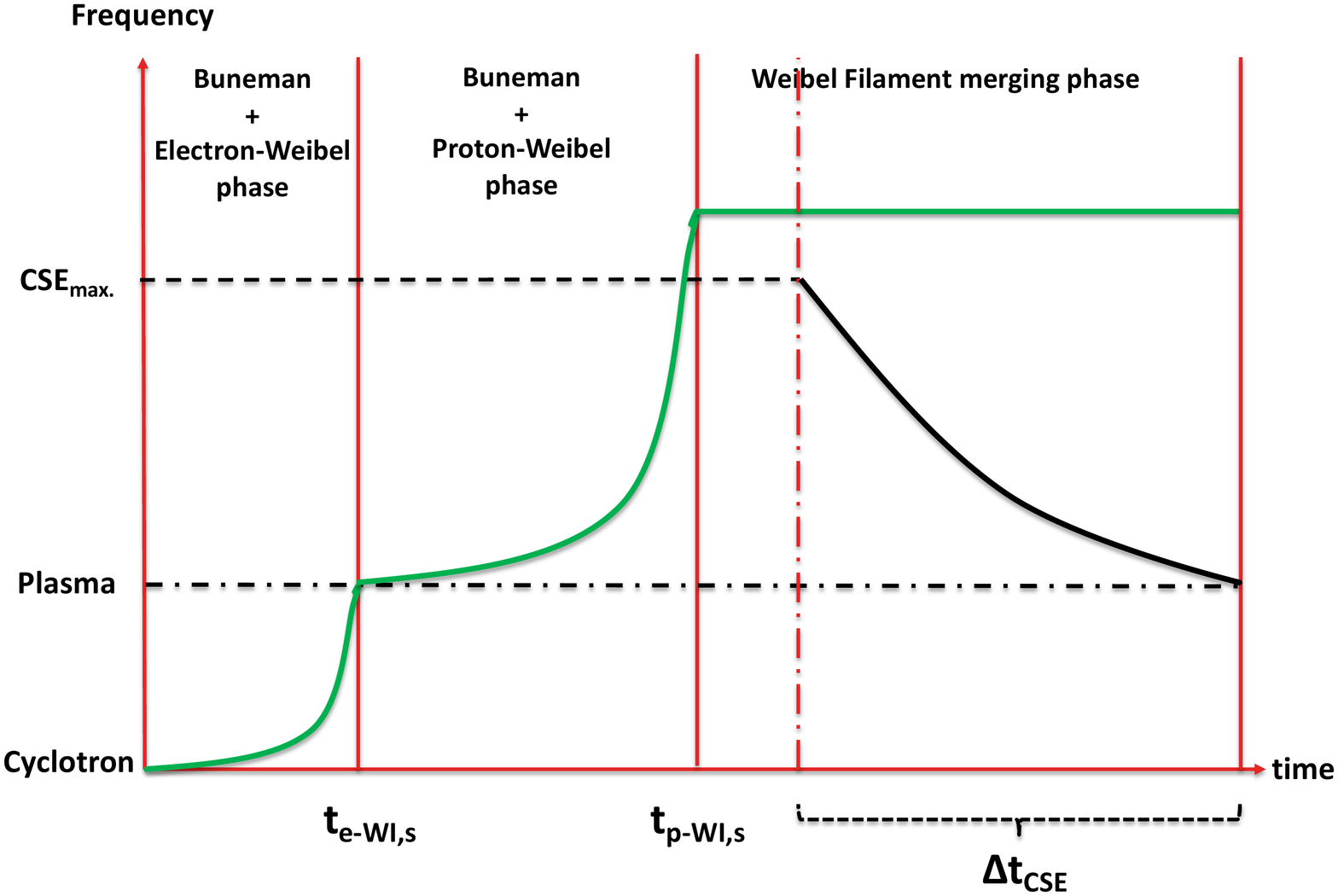}
 \caption{{\bf Top panel}: A schematic representation of the evolution of  $\beta_{\parallel}=v_{\parallel}/c$
during the BI-WI phases; $\beta_{\perp}=0.1\beta_{\parallel}$ (not shown here). 
In the linear phase for $t<t_{\rm p-WI,s}$ (i.e. up to the saturation of the proton-Weibel phase; see Appendix \ref{appendix:instabilities}), the BI heating of chunk electrons (i.e. the increase in $\beta_{\parallel}$) is converted by 
WI  into magnetic field amplification, into magnetic turbulence and into  currents.
In this regime, $\beta_{\parallel}$ increases from $\beta_{\parallel}\sim \beta_{\rm cc}$
(where $v_{\rm cc}$ is the electron thermal speed when the chunk become collisionless; see Eq. (\ref{appendix:eq:betaecc}))
to $\beta_{\parallel}\sim 1$. During filament merging,  magnetic turbulence and current
dissipation accelerates electrons to relativistic speed, $\gamma_{\rm CSE}>>1$, shutting-off
 the BI. The BI requires the drift velocity (here the light speed $c$) between the beam protons and the chunk's
 electrons to exceed the thermal speed of the chunk's electrons (see  
 Appendix \ref{appendix:instabilities}). The decrease in $\gamma_{\rm CSE}$ is due to Coherent Synchrotron Emission (CSE) cooling. 
{\bf Lower panel}: A schematic representation of the evolution of the different frequencies during the BI-WI process in our model. The electron plasma frequency ($\nu_{\rm p, e}=\sqrt{4\pi n_{\rm cc}e^2/m_{\rm e}}$, dot-dashed horizontal line) remains constant. 
  The electron cyclotron frequency ($\nu_{\rm B}=eB_{\rm c}/m_{\rm e}c$, thick green line) saturates
 first during the e-WI phase when $\nu_{\rm B}\sim \nu_{\rm p, e}$ (i.e. $B_{\rm c}=B_{\rm e-WI,s}$) and later  
   at the end of the p-WI phase with $\nu_{\rm B}\sim \sqrt{m_{\rm p}/m_{\rm e}}\nu_{\rm p, e}$ (i.e. $B_{\rm c}=B_{\rm p-WI,s}$).    
 CSE at frequency $\nu_{\rm CSE}$ is triggered throughout the filament merging phase when  $\nu_{\rm CSE}<< \gamma_{\rm CSE}^2 \sqrt{m_{\rm p}/m_{\rm e}}\nu_{\rm p, e}$ is satisfied (see Appendix \ref{appendix:CSE}).
 The CSE frequency $\nu_{\rm CSE}$ decreases over time (the thick black line) due to the increase in bunch size during filament merging.  
 CSE ceases when its frequency drops to the chunk's plasma frequency ($\nu_{\rm p, e}$).
 The end of filament merging occurs when the filaments grow to a size of the
 order of the beam's protons Larmor radius. The trapping of the protons is followed by the formation
 of the Weibel shock (not shown here), quickly decelerating the chunk and putting an end to the BI-WI process.}
 \label{fig:WI-stages}
 \end{figure*}

\begin{figure*}
 \centering
 \includegraphics[scale=0.8]{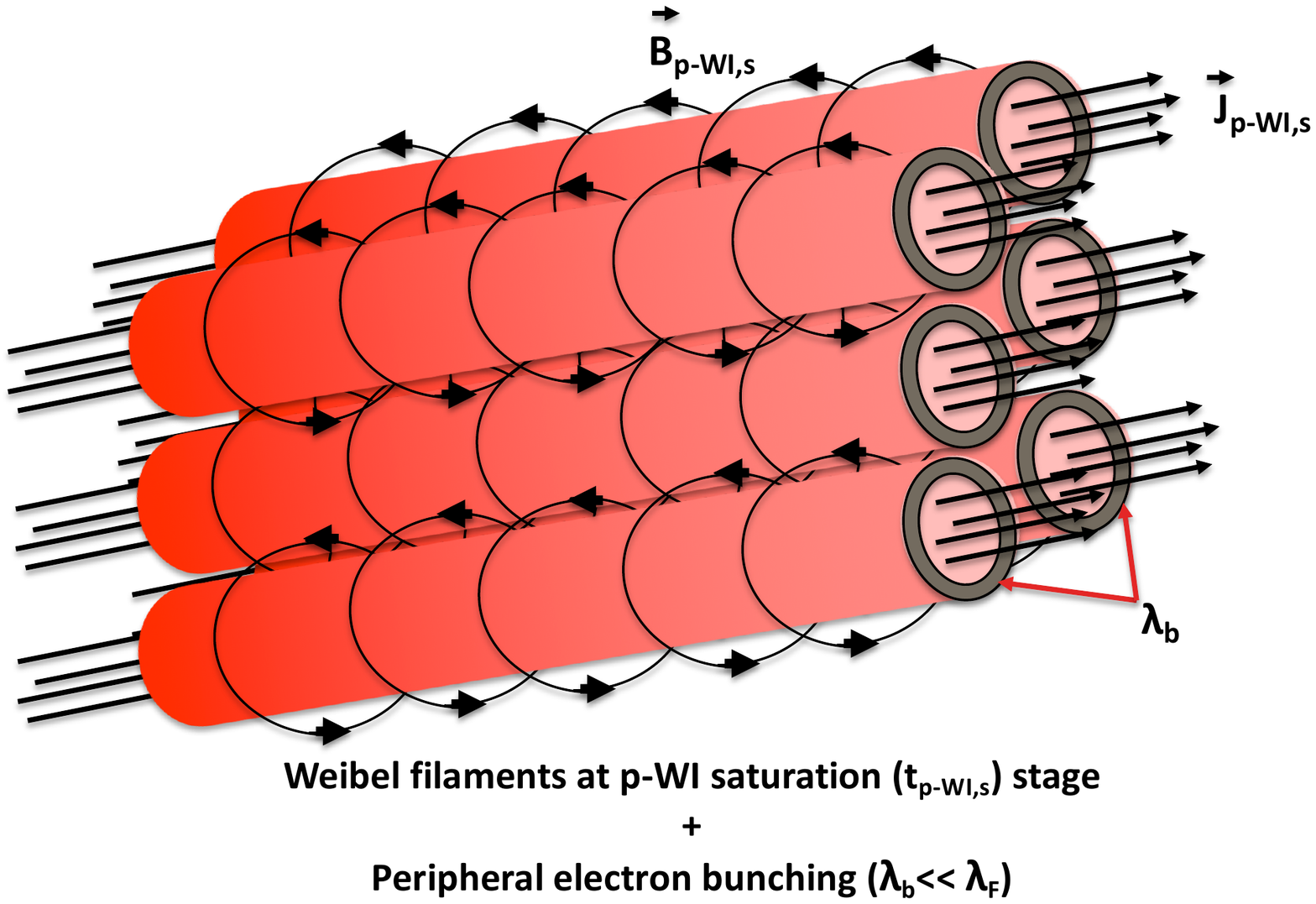}
  \caption{{\bf A schematic representation of Weibel filament and electron bunch geometry}: Bunches are shown as cylindrical shells with 
  thickness $\lambda_{\rm b}$ (shown by the dark-grey shadding) around the Weibel cylindrical filaments of diameter $\lambda_{\rm F}$ ($\lambda_{\rm b}<<\lambda_{\rm F}$).
  The bunches, tied to the Weibel filaments, extend across the QN chunk (the background plasma) in the direction parallel to the beam's direction (here the ICM). Also illustrated are the Weibel saturated magnetic field, $\vec{B}_{\rm p-WI, s}$ (see Appendix  \ref{appendix:instabilities} and Eq. (\ref{eq:Bcc,s}))
  reached at the end of the proton-Weibel (p-WI) phase, and the corresponding filament currents, $\vec{J}_{\rm p-WI, s}$.}
   \label{figure:bunch-geometry}
 \end{figure*}

 Illustrated in the lower panel in Figure \ref{fig:WI-stages} are the key phases of the
 BI-WI episode. The  depicted key frequencies are:
 
 (i) The electron plasma frequency
  ($\nu_{\rm p, e}=\sqrt{4\pi n_{\rm cc}e^2/m_{\rm e}}$) which remains constant during the entire BI-WI process.
  This also sets the minimum observed CSE frequency  as $\nu_{\rm p, e}^{\rm obs.}=D(\Gamma_{\rm c},\theta_{\rm c})\nu_{\rm p, e}/(1+z)$;
  
  (ii) The electron cyclotron frequency ($\nu_{\rm B}=eB_{\rm c}/m_{\rm e}c$; with $B_{\rm c}=B_{\rm cc}$
  at the start of the BI-WI process). It increases  in time as $B_{\rm c}$ increases first during the e-WI phase reaching saturation at $B_{\rm c}= B_{\rm e-WI, s}$ when the cyclotron  frequency is $\nu_{\rm B}\sim\nu_{\rm p, e}$.   During the p-WI phase, the magnetic field  grows further to a saturation value of  $B_{\rm p-WI,s}=\sqrt{m_{\rm p}/m_{\rm e}} B_{\rm e-WI, s}$ when $\nu_{\rm B}\sim\sqrt{m_{\rm p}/m_{\rm e}}\nu_{\rm p, e}$ at time  $t_{\rm p-WI, s}$; 
     
     (iii)  The BI shuts-off in the early stages of filament merging phase once the chunk's electrons are so hot that their thermal velocity
spread exceeds their drift velocity relative to the beam's ions (when $\gamma_{\rm CSE}>2$); during filament merging, electron
acceleration is due to dissipation of magnetic turbulence and currents; 
  
 (iv) Once CSE is triggered, electrons in bunches cool rapidly with the
 cooling timescale of a bunch $t_{\rm b}(t)<<\Delta t_{\rm CSE}$ (see Appendix \ref{appendix:bunch-geometry}).
 Each bunch emits once during filament merging  with  bunches  emitting uniformly spaced in time during this phase; 
   
  (v) Beyond the CSE phase,  the filaments continue to grow in size until they are of the order of 
 the beam's proton Larmor radius. Once the protons are trapped, the Weibel shock develops 
 slowing down the  chunk drastically (in a matter of seconds in the observer's
 frame; see Eq. (\ref{eq:tsw-obs}) in Appendix \ref{sec:additional-predictions}) and putting an end to the BI-WI process.

  \section{FRBs in current detectors}
  \label{appendix:CSE-in-Detectors}
  
  \subsection{Number of FRBs per frequency ($N_{\nu^{\rm obs.}}^{\rm obs.}$)}
\label{appendix:chunks-per-frequency} 
  
     Here we estimate the number of chunks (i.e. FRBs per QN) detectable at any frequency $\nu^{\rm obs.}$ and 
     at any given time $t^{\rm obs.}$.
      Appendix \S \ref{appendix:honeycomb}  describes the spatial distribution of the QN chunks with $N_{\theta}$
     the number of chunks per angle $\theta$. We have 
      $dN_{\theta}/d\nu^{\rm obs.} = (d N_{\theta}/d\theta_{\rm c})\times (d \theta_{\rm c} /d\nu^{\rm obs.})$ where 
     $dN_{\theta}/d\Omega= N_{\rm c}/4\pi$ and $d \Omega/d\theta_{\rm c}=2\pi\theta_{\rm c}$ (for $\theta_{\rm c}<<1$) so that 
     $d N_{\theta}/d\theta_{\rm c} = d N_{\nu^{\rm obs.}}^{\rm obs.}/d\Omega\times  d \Omega/d\theta_{\rm c} =  (N_{\rm c}/2)\times \theta_{\rm c}$.

    Furthermore, because at any given time $\nu^{\rm obs.}(\theta_{\rm c}) = D(\Gamma_{\rm c},\theta_{\rm c}) \nu^{\rm obs.}(0)$
    where $\nu^{\rm obs.}(0)$ is the frequency at $\theta_{\rm c}=0$,  then
    for a given QN (i.e. for a fixed $\Gamma_{\rm c}$) we can write
        
        \begin{equation}
        \frac{d\nu^{\rm obs.}}{d \theta_{\rm c}}=  \frac{d\nu^{\rm obs.}}{d D(\Gamma_{\rm c},\theta_{\rm c})}\times 
        \frac{d D(\Gamma_{\rm c},\theta_{\rm c})}{d f(\theta_{\rm c})}\times  \frac{d f(\theta_{\rm c})}{d \theta_{\rm c}}
        = \nu^{\rm obs.}(0)\times \left( -\frac{2\Gamma_{\rm c}}{f(\theta_{\rm c})^2}\right) \times \left( 2\Gamma_{\rm c}^2\theta_{\rm c}\right)
        = - \frac{{\nu^{\rm obs.}}^2}{\nu^{\rm obs.}(0)}\times (\Gamma_{\rm c}\theta_{\rm c})\ ,
        \end{equation}
        where $D(\Gamma_{\rm c},\theta_{\rm c}) \simeq 2\Gamma_{\rm c}/f(\theta_{\rm c})$ and $f(\theta_{\rm c})=1+(\Gamma_{\rm c}\theta_{\rm c})^2$.
          We arrive at
     
     \begin{equation}
     \label{appendix:eq:dNc-nu}
     \frac{dN_{\nu^{\rm obs.}}^{\rm obs.}}{d\nu^{\rm obs.}}=\frac{dN_{\theta}/d\theta_{\rm c}}{d\nu^{\rm obs.}/d \theta_{\rm c}}= - \frac{N_{\rm c}}{2\Gamma_{\rm c}}\times \frac{\nu^{\rm obs.}(0)}{{\nu^{\rm obs.}}^2}\ .
     \end{equation}

  \subsection{FRB duration}
\label{appendix:duration}
     
\begin{figure*}
 \vskip 1.5in
 \centering
 \includegraphics[scale=0.65]{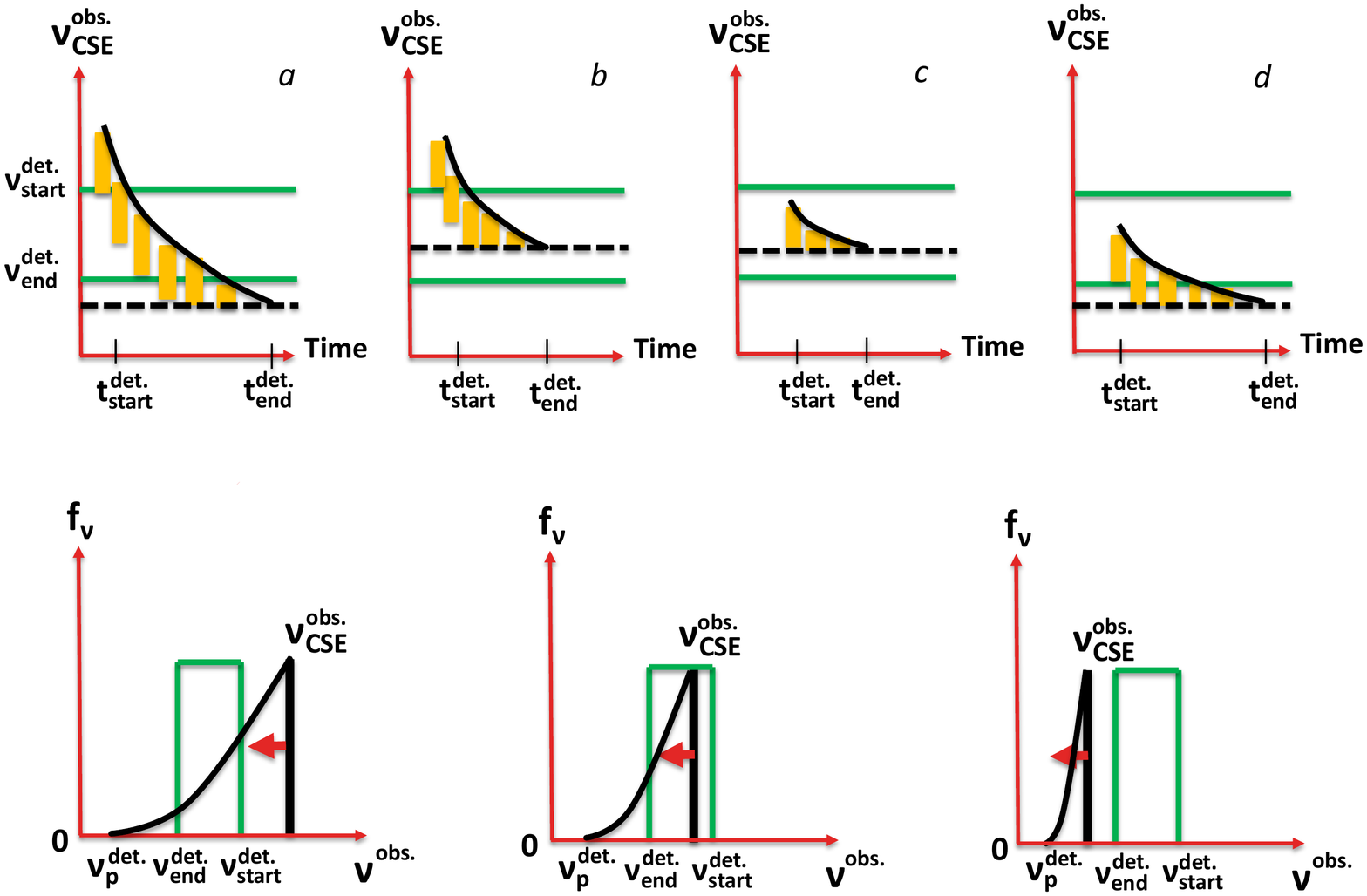}
  \caption{Same as in Figure \ref{figure:drifting-a}  in the main paper but here for the  case of a 
  power-law spectrum with positive index $\alpha_{\rm CSE}$. The difference is that for a steep spectrum only the emission
   near the peak frequency (the narrower vertical bands) is detected  at a given time.}
    \label{figure:drifting-b}
 \end{figure*}

     We define $\nu_{\rm max.}^{\rm det.}$ and $\nu_{\rm min.}^{\rm det.}$ as  the maximum  and minimum frequencies of 
   the detector's band with $t_{\rm start}^{\rm det.}$ and $t_{\rm end}^{\rm det.}$  the times corresponding to the start
  (at $\nu_{\rm max.}^{\rm det.}$) and end of detection (at $\nu_{\rm min.}^{\rm det.}$).
      When  the chunk's plasma frequency, $\nu_{\rm p, e}\simeq 9\ {\rm kHz}\times n_{\rm cc}^{1/2}$ (e.g. \citealt{lang_1999}), is such that $\nu_{\rm p, e}^{\rm obs.}(\theta_{\rm c})<\nu_{\rm min.}^{\rm det.}$, 
     the CSE frequency will drift through the entire detector's band with  $\nu_{\rm CSE}^{\rm obs.}(\theta_{\rm c},t^{\rm obs.})=\nu_{\rm CSE, max.}^{\rm obs.}(\theta_{\rm c})(1+ t^{\rm obs.}/t_{\rm m-WI}^{\rm obs.})^{-\delta_{\rm m-WI}}$ (see \S \ref{sec:drifting});  this is illustrated in Figure \ref{figure:drifting-a} in the main paper and 
     Figure \ref{figure:drifting-b} here.
      In this case,  the detector's CSE (i.e. FRB) duration
      $\Delta t_{\rm CSE, detector}^{\rm obs.}=(t_{\rm end}^{\rm det.}-t_{\rm start}^{\rm det.})$ can be found by combining $\nu_{\rm min.}^{\rm det.}=\nu_{\rm CSE, max.}^{\rm obs.}(\theta_{\rm c})(1+ t_{\rm end}^{\rm det.}/t_{\rm m-WI}^{\rm obs.})^{-\delta_{\rm m-WI}}$ and $\nu_{\rm max.}^{\rm det.}=\nu_{\rm CSE, max.}^{\rm obs.}(\theta_{\rm c})(1+ t_{\rm start}^{\rm det.}/t_{\rm m-WI}^{\rm obs.})^{-\delta_{\rm m-WI}}$ giving us
    
   \begin{align}
   \label{eq:dt-detector}
   \Delta t_{\rm CSE}^{\rm det.} &= t_{\rm m-WI}^{\rm obs.}\times  \left( \left( \frac{\nu_{\rm CSE, max.}^{\rm obs.}(\theta_{\rm c})}{\nu_{\rm min.}^{\rm det.}} \right)^{1/\delta_{\rm m-WI}} - \left(\frac{\nu_{\rm CSE, max.}^{\rm obs.}(\theta_{\rm c})}{\nu_{\rm max.}^{\rm det.}} \right)^{1/\delta_{\rm m-WI}}\right)\nonumber\\
    &\simeq 0.24\ {\rm ms}\times (1+z)f(\theta_{\rm c})\times \frac{\zeta_{\rm m-WI,2}}{\Gamma_{\rm c, 2.5}n_{\rm cc, 1}^{1/2}}
    \times \left(\left( \frac{\nu_{\rm CSE, max.}^{\rm obs.}(\theta_{\rm c})}{\nu_{\rm min.}^{\rm det.}} \right)^{1/\delta_{\rm m-WI}} - \left(\frac{\nu_{\rm CSE, max.}^{\rm obs.}(\theta_{\rm c})}{\nu_{\rm max.}^{\rm det.}} \right)^{1/\delta_{\rm m-WI}}\right)\ ,
   \end{align}
   with $\nu_{\rm CSE, max.}^{\rm obs.}(\theta_{\rm c})$ given by Eq. (\ref{eq:nu-CSEobs}) and 
     $t_{\rm m-WI}^{\rm obs.}$ given by  Eq. (\ref{eq:tmWIobs}). There are three other possible scenarios, depicted in Figures \ref{figure:drifting-a} and  \ref{figure:drifting-b},  which could  make the duration shorter than the one given in Eq. (\ref{eq:dt-detector}).

\subsection{Band-integrated flux density and corresponding fluence}
\label{appendix:flux-fluence}

With regards to the spectrum,  each  bunch  emits at all frequencies within $0\le \nu \le \nu_{\rm CSE}$ even though radiation below the plasma frequency is re-absorbed by the chunk material.  
 Because $I_{\nu}^{\rm obs.}(t)/{\nu^{\rm obs.}}^3= I_{\nu}(t)/\nu^3$ is an invariant,
the flux density is found from  (e.g. \citealt{ryden_2016}) $f_{\nu^{\rm obs.}}(\theta_{\rm c},t)=I_{\nu}^{\rm obs.}(t)\times A_{\rm cc}/4\pi d_{\rm L}^2 = D(\Gamma_{\rm c},\theta_{\rm c})^3L_{\nu}(t)/(1+z))4\pi d_{\rm L}^2$ with $L_{\nu}(t)=I_{\nu}(t)  A_{\rm cc}$ the spectral luminosity
and $A_{\rm cc}$ the chunk's area which is also invariant; $z$ is the redshift  and $d_{\rm L}$  the luminosity distance. In the emitter's frame (i.e. the QN chunk), we assume a spectrum with positive index $\alpha_{\rm CSE}$
\begin{equation}
L_{\nu}(t)=(\nu/\nu_{\rm CSE})^{\alpha_{\rm CSE}} L_{\nu_{\rm CSE}}(t) \ ,
\end{equation}
so that $L_{\rm CSE}(t)=\int_0^{\nu_{\rm CSE}(t)} L_{\nu}(t)d\nu = \nu_{\rm CSE}(t)L_{\nu_{\rm CSE}}(t)/(\alpha_{\rm CSE}+1)$ with $\alpha_{\rm CSE}>-1$;
here $L_{\nu_{\rm CSE}}(t)$ is the spectral luminosity at maximum frequency $\nu_{\rm CSE}(t)$.

The flux density, in the observer's frame, can then be recast into
    
    \begin{align}
    \label{appendix:eq:pureflux}
    f_{\nu^{\rm obs.}} (\theta_{\rm c},t) 
    &=  \frac{D(\Gamma_{\rm c},\theta_{\rm c})^3L_{\rm CSE}(t)}{(1+z)4\pi d_{\rm L}^2\nu_{\rm CSE}(t)} \times
    (\alpha_{\rm CSE}+1)\left( \frac{\nu}{\nu_{\rm CSE}(t)}\right)^{\alpha_{\rm CSE}}\nonumber\\
    &= \frac{D(\Gamma_{\rm c},\theta_{\rm c})^4L_{\rm CSE}(t)}{(1+z)^24\pi d_{\rm L}^2\nu_{\rm CSE}^{\rm obs.}(t)} \times
    (\alpha_{\rm CSE}+1)\left( \frac{\nu}{\nu_{\rm CSE}(t)}\right)^{\alpha_{\rm CSE}}\ .
    \end{align}
    As expected, 
    $\int_0^\infty f_{\nu^{\rm obs.}}(t)d\nu^{\rm obs.} = (D(\Gamma_{\rm c},\theta_{\rm c})^4/(1+z)4\pi d_{\rm L}^2) \int_0^\infty L_{\nu}(t)d\nu = D(\Gamma_{\rm c},\theta_{\rm c})^4L_{\rm CSE}(t)/(1+z)^24\pi d_{\rm L}^2$ with  $\nu^{\rm obs.}=D(\Gamma_{\rm c},\theta_{\rm c})\nu/(1+z)$.

  To compare to FRB data, we define  $f_{\rm \nu, band}(\theta_{\rm c},t^{\rm obs.})=   \frac{1}{\Delta \nu^{\rm det.}} \int_{\nu_{\rm min.}^{\rm obs.}}^{\nu_{\rm max.}^{\rm det.}} f_{{\nu^{\rm obs.}}}(\theta_{\rm c},t^{\rm obs.}) d\nu^{\rm obs.}$ as the band-averaged flux density
  with $\Delta \nu^{\rm det.}= \nu_{\rm max.}^{\rm det.}-\nu_{\rm min.}^{\rm det.}$; i.e. a frequency summed flux
  over  the  detector's frequency band  $\nu_{\rm min.}^{\rm det.} \le \nu^{\rm det.} \le \nu_{\rm max.}^{\rm det.}$.
  I.e.
  
  \begin{equation}
   \label{appendix:eq:fluxdensity-0}
   f_{\rm \nu, band}(\theta_{\rm c},t^{\rm obs.}) = \frac{D(\Gamma_{\rm c},\theta_{\rm c})^4L_{\rm CSE}(t)}{(1+z)^24\pi d_{\rm L}^2\Delta \nu^{\rm det.}}\times  (\alpha_{\rm CSE}+1) \int_{\nu_{\rm lower}^{\rm obs.}}^{\nu_{\rm upper}^{\rm obs.}}
   \left(\frac{\nu}{\nu_{\rm CSE}(t)}\right)^{\alpha_{\rm CSE}} d\left( \frac{\nu^{\rm obs.}}{\nu_{\rm CSE}^{\rm obs.}(t)} \right)\
  \end{equation}
  where  $\nu_{\rm lower}^{\rm obs.}={\rm max}\left(\nu_{\rm min.}^{\rm det.},\nu_{\rm p, e}^{\rm obs.}(\theta_{\rm c})\right)$ and 
  $\nu_{\rm upper}^{\rm obs.}={\rm min}\left(\nu_{\rm max.}^{\rm det.},\nu_{\rm CSE}^{\rm obs.}(\theta_{\rm c},t^{\rm obs.}))\right)$.

 With $\nu/\nu_{\rm CSE}(t)=\nu^{\rm obs.}/\nu_{\rm CSE}^{\rm obs.}(\theta_{\rm c},t^{\rm obs.})$, Eq. (\ref{appendix:eq:fluxdensity-0}) becomes
  
  \begin{equation}
  \label{appendix:eq:fluxdensity}
  f_{\rm \nu, band}(\theta_{\rm c},t^{\rm obs.}) = \frac{D(\Gamma_{\rm c},\theta_{\rm c})^4L_{\rm CSE}(t)}{(1+z)^24\pi d_{\rm L}^2\Delta \nu^{\rm det.}}\times 
  \begin{cases}
   \frac{{\nu_{\rm max.}^{\rm det.}}^{\alpha_{\rm CSE}+1}-{\nu_{\rm min.}^{\rm det.}}^{\alpha_{\rm CSE}+1}}{{\nu_{\rm CSE}^{\rm obs.}(\theta_{\rm c},t)}^{\alpha_{\rm CSE}+1}} & \text{if }   \nu_{\rm CSE}^{\rm obs.}(\theta_{\rm c},t^{\rm obs.}) > \nu_{\rm max.}^{\rm det.}\\
  \frac{{\nu_{\rm CSE}^{\rm obs.}}(\theta_{\rm c},t^{\rm obs.})^{\alpha_{\rm CSE}+1}-{\nu_{\rm min.}^{\rm det.}}^{\alpha_{\rm CSE}+1}}{{\nu_{\rm CSE}^{\rm obs.}(\theta_{\rm c},t)}^{\alpha_{\rm CSE}+1}},& \text{if }  \nu_{\rm lower}^{\rm obs.}< \nu_{\rm CSE}^{\rm obs.}(\theta_{\rm c},t^{\rm obs.}) \le \nu_{\rm max.}^{\rm det.}\\
    0,              & \text{if}\  \nu_{\rm CSE}^{\rm obs.}(\theta_{\rm c},,t^{\rm obs.}) \le \nu_{\rm lower}^{\rm obs.}\ .
    \end{cases}
  \end{equation}
 
 The above means that once $\nu_{\rm CSE}^{\rm obs.}(\theta_{\rm c},t)$ drops below the detector's maximum
frequency $\nu_{\rm max.}^{\rm det.}$, the band-averaged flux density starts to drop with time until the CSE frequency
 exits the detector's band at $\nu_{\rm min.}^{\rm det.}$ or when the plasma frequency is reached;  this is illustrated in Figure \ref{figure:drifting-a} in the main paper and in 
  Figure  \ref{figure:drifting-b} here.

     The  fluence based on the band-averaged flux density  is $F(\theta_{\rm c},\delta_{\rm m-WI},\alpha_{\rm CSE})=\int_{t_{\rm start}^{\rm det.}}^{t_{\rm end}^{\rm det.}} f_{\rm \nu, band}(\theta_{\rm c},t^{\rm obs.})dt^{\rm obs.}$ and 
    with the substitutions $dt^{\rm obs.}=(1+z)dt/D(\Gamma_{\rm c},\theta_{\rm c})$ and
  ${\nu_{\rm CSE}^{\rm obs.}(\theta_{\rm c},t)} = D(\Gamma_{\rm c},\theta_{\rm c}) \nu_{\rm CSE}(t)/(1+z)$, it can then be expressed as
    
  \begin{align}
  \label{appendix:eq:fluence-1}
 F(\theta_{\rm c},\delta_{\rm m-WI},\alpha_{\rm CSE})= \mathcal{F}(\theta_{\rm c},\alpha_{\rm CSE})\times \mathcal{G}(\theta_{\rm c},\delta_{\rm m-WI},\alpha_{\rm CSE})\ ,
 \end{align}
 with
 
 \begin{equation}
\label{appendix:eq:fluence-2}
\mathcal{F}(\theta_{\rm c},\alpha_{\rm CSE})= \frac{D(\Gamma_{\rm c},\theta_{\rm c})^{3}}{(1+z)4\pi d_{\rm L}^2}\times   \frac{L_{\rm CSE}(t)t_{\rm m-WI}}{\Delta \nu^{\rm det.}}\times  \frac{{\nu_{\rm max.}^{\rm det.}}^{\alpha_{\rm CSE}+1}-{\nu_{\rm min.}^{\rm det.}}^{\alpha_{\rm CSE}+1}}{{\nu_{\rm CSE, max.}^{\rm obs.}(\theta_{\rm c})}^{\alpha_{\rm CSE}+1}}\ ,
\end{equation}
with $L_{\rm CSE}(t)=L_{\rm CSE}(0)$  a constant in our model (see Eqs. (\ref{eq:LCSE})) and  $\nu_{\rm CSE, max.}^{\rm obs.}(\theta_{\rm c})$ the maximum
CSE frequency given by (\ref{eq:nu-CSEobs}); $L_{\rm CSE}(0)t_{\rm m-WI}=E_{\rm BI}$ expresses the energy harnessed from BI heating
during the BI-WI phase prior to filament merging (see Eq. (\ref{eq:EBI})). Also,

\begin{equation}
\label{appendix:eq:fluence-3}
 \mathcal{G}(\theta_{\rm c},\delta_{\rm m-WI},\alpha_{\rm CSE}) =
\begin{cases}
    \int_{x_{\rm lower}}^{x_{\rm upper}} x^{\delta_{\rm m-WI}(\alpha_{\rm CSE}+1)} dx,& \text{if }\nu_{\rm CSE}^{\rm obs.}(\theta_{\rm c},x) > \nu_{\rm max.}^{\rm det.}\    [{\rm if}\ x < x_{\rm start}]\\
    \int_{x_{\rm lower}}^{x_{\rm upper}} x^{\delta_{\rm m-WI}(\alpha_{\rm CSE}+1)}  \times \frac{\left(\frac{x}{x_{\rm end}}\right)^{-\delta_{\rm m-WI}(\alpha_{\rm CSE}+1)} -1}{\left(\frac{x_{\rm start}}{x_{\rm end}}\right)^{-\delta_{\rm m-WI}(\alpha_{\rm CSE}+1)} -1}dx,& \text{if }  \nu_{\rm lower}^{\rm obs.}< \nu_{\rm CSE}^{\rm obs.}(\theta_{\rm c},x) \le \nu_{\rm max.}^{\rm det.}\  [{\rm if}\ x_{\rm start}\le x < x_{\rm lower}]\\
    0,              & \text{if}\  \nu_{\rm CSE}^{\rm obs.}(\theta_{\rm c},x) \le \nu_{\rm lower}^{\rm obs.}\  [{\rm if}\ x \ge x_{\rm lower}]\ .
    \end{cases}
\end{equation}  

where we defined $x=1+t/t_{\rm m-WI}$ so that   $\nu_{\rm CSE}(t)=\nu_{\rm CSE}(0)\times x^{-\delta_{\rm m-WI}}$.
The term $((x/x_{\rm end})^{-\delta_{\rm m-WI}(\alpha_{\rm CSE}+1)} -1)/((x_{\rm start}/x_{\rm end})^{-\delta_{\rm m-WI}(\alpha_{\rm CSE}+1)} -1)$  is due to $\nu_{\rm CSE}(t^{\rm obs.})$ drifting through the detector's band.
The relevant $x$-values are
     
     \begin{align}
     \label{appendix:eq:fluence-4}
     x_{\rm end} &= \left(\frac{\nu_{\rm CSE, max.}^{\rm obs.}(\theta_{\rm c})}{\nu_{\rm min.}^{\rm det.}}\right)^{1/\delta_{\rm m-WI}}  \nonumber \\
x_{\rm start} &= \left( \frac{\nu_{\rm CSE, max.}^{\rm obs.}(\theta_{\rm c})}{\nu_{\rm max.}^{\rm det.}}\right)^{1/\delta_{\rm m-WI}}\nonumber \\ 
x_{\rm p, e} &= \left( \frac{\nu_{\rm CSE, max.}^{\rm obs.}(\theta_{\rm c})}{\nu_{\rm p, e}^{\rm obs.}(\theta_{\rm c})}\right)^{1/\delta_{\rm m-WI}} \ .
\end{align}
The limits of integration in $ \mathcal{G}(\theta_{\rm c},\delta_{\rm m-WI},\alpha_{\rm CSE})$ are 
\begin{align}
x_{\rm lower}&={\rm max}\left(x_{\rm start},1.0\right)\nonumber\\
x_{\rm upper}&={\rm min}\left(x_{\rm end},x_{\rm p, e}\right)\ .
     \end{align}
     
     \subsection{Flat spectrum}
     \label{appendix:flat-spectrum}

For the case of a flat spectrum (i.e. $\alpha_{\rm CSE}=0$)
with  $F(\theta_{\rm c},0)= \mathcal{F}(\theta_{\rm c},0)\times \mathcal{G}(\theta_{\rm c},\delta_{\rm m-WI},0)$,  Eqs. (\ref{appendix:eq:fluence-1}) and
(\ref{appendix:eq:fluence-2}) above become

 \begin{equation}
\label{appendix:eq:fluence-flat-1}
\mathcal{F}(\theta_{\rm c},0)= \frac{D(\Gamma_{\rm c},\theta_{\rm c})^{3}}{(1+z)4\pi d_{\rm L}^2}\times   \frac{L_{\rm CSE}(0)t_{\rm m-WI}}{{\nu_{\rm CSE}^{\rm obs.}(0)}}\ .
\end{equation}

\begin{equation}
\label{appendix:eq:fluence-flat-2}
 \mathcal{G}(\theta_{\rm c},\delta_{\rm m-WI},0) =
\begin{cases}
    \int_{x_{\rm lower}}^{x_{\rm upper}} x^{\delta_{\rm m-WI}} dx,& \text{if }\nu_{\rm CSE}^{\rm obs.}(\theta_{\rm c},x) > \nu_{\rm max.}^{\rm det.}\\
    \int_{x_{\rm lower}}^{x_{\rm upper}} x^{\delta_{\rm m-WI}}  \times \frac{\left(\frac{x}{x_{\rm end}}\right)^{-\delta_{\rm m-WI}} -1}{\left(\frac{x_{\rm start}}{x_{\rm end}}\right)^{-\delta_{\rm m-WI}} -1}dx,& \text{if }  \nu_{\rm lower}^{\rm obs.}< \nu_{\rm CSE}^{\rm obs.}(\theta_{\rm c},x) \le \nu_{\rm max.}^{\rm det.}\\
    0,              & \text{if}\  \nu_{\rm CSE}^{\rm obs.}(\theta_{\rm c},x) \le \nu_{\rm lower}^{\rm obs.}\ .
    \end{cases}
\end{equation}

       CSE is so efficient that it radiates most of the BI energy ($E_{\rm BI}\sim L_{\rm CSE}(0)t_{\rm m-WI}$; see Eq. (\ref{eq:EBI})) 
       during filament merging. Eq. (\ref{appendix:eq:fluence-flat-1}) becomes
      
\begin{align}
\label{appendix:eq:fluence-flat-3}
\mathcal{F}(\theta_{\rm c},0)\simeq 810\ {\rm Jy\ ms}\   \frac{1}{f(\theta_{\rm c})^2d_{\rm L, 27.5}^2}\times  \frac{\zeta_{\rm BI, -1}\beta_{\rm WI, -1}}{\delta_{\rm CSE, -1}\gamma_{\rm CSE,1}^2} \times \frac{\Gamma_{\rm c, 2.5}^4 R_{\rm cc, 15}^2 {n_{\rm amb., -3}^{\rm ns}}}{n_{\rm cc, 1}\beta_{\rm cc, -2}}\ ,
\end{align}
 after making use of  $\nu_{\rm CSE, max.}^{\rm obs.}(\theta_{\rm c})=D(\Gamma_{\rm c},\theta_{\rm c})\nu_{\rm CSE}(0)/(1+z)$ 
 and $\nu_{\rm CSE}(0)=\delta_{\rm CSE}\times (3/2)\gamma_{\rm CSE}^2\sqrt{m_{\rm p}/m_{\rm e}}\nu_{\rm p, e}$ 
 (see Appendix \ref{appendix:CSE}); the
luminosity distance $d_{\rm L}$ is in units of Giga-parsecs. 
 
 Our calculations of $\mathcal{G}(\theta_{\rm c},\delta_{\rm m-WI},0)$  is detector's dependent 
via $x_{\rm end}$ and $x_{\rm start}$ (see Eq. (\ref{appendix:eq:fluence-flat-2})) and varies from a value of 
a few for ASKAP, Parkes and Arecibo detectors to about a few hundreds for  CHIME's
and even higher for the LOFAR's detectors (see Table \ref{table:G-fluence}).

 \subsection{``Waterfall" plots}
 \label{sec:waterfall-plots}

\begin{figure*}
 \vskip 1.5in
 \centering
 \includegraphics[scale=0.7]{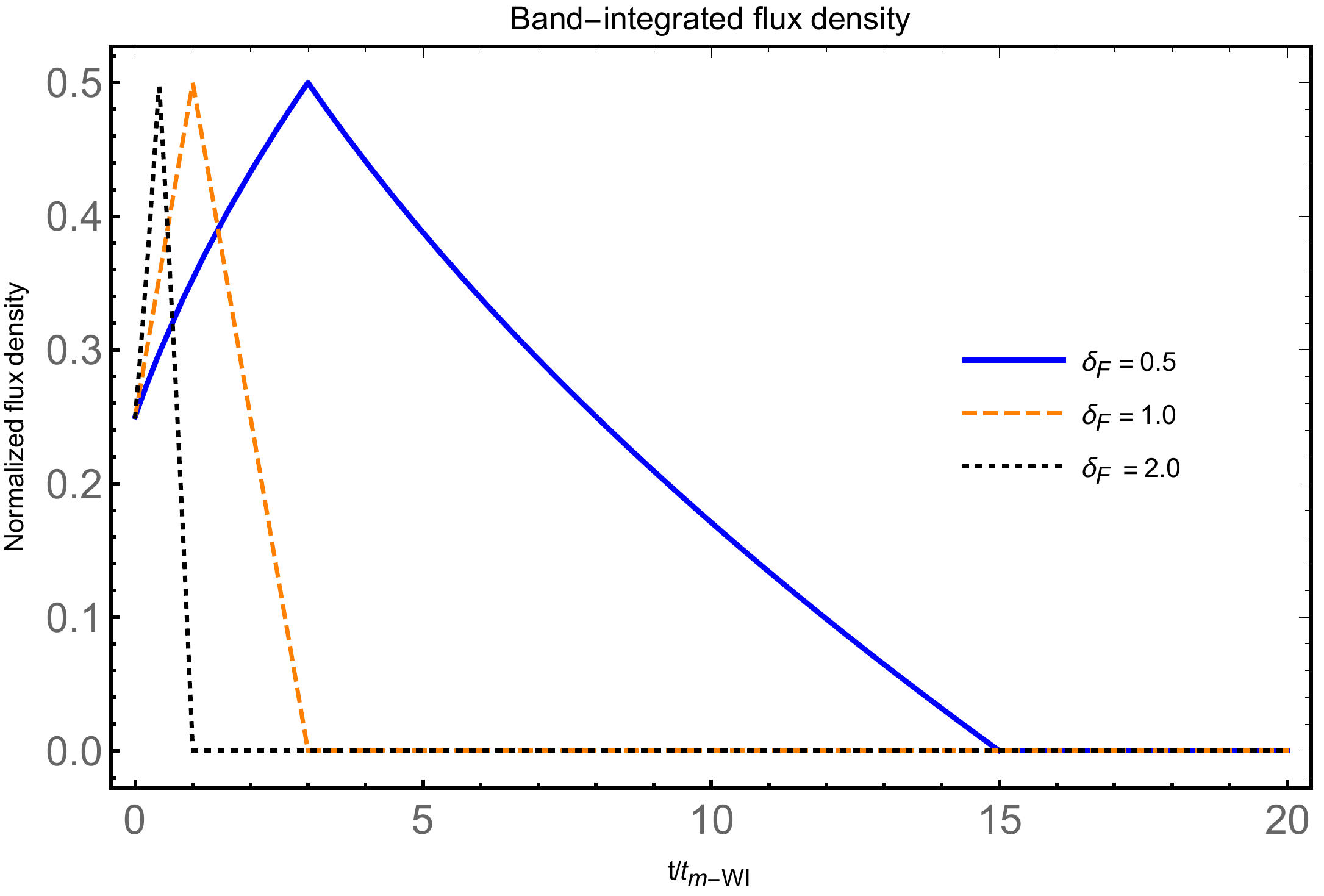}
  \caption{The analytical and normalized band-integrated flux density (given by Eq. (\ref{appendix:eq:fluxdensity}) in Appendix \ref{appendix:flux-fluence}) versus normalized time, $t/t_{\rm m-WI}=t^{\rm obs.}/t_{\rm m-WI}^{\rm obs.}(\theta_{\rm c})$; $t_{\rm m-WI}^{\rm obs.}(\theta_{\rm c})$ is the characteristic filament merging timescale (Eq. (\ref{eq:tmWIobs})).   Shown here is the case ``a" in the top panel of Figure \ref{figure:drifting-a}   applied to CHIME's detector  with $\nu_{\rm max.}^{\rm det.}=800$ MHz and $\nu_{\rm min.}^{\rm det.}=400$ MHz and 
   with $\nu_{\rm CSE, max.}^{\rm obs.}(0)=2\nu_{\rm max.}^{\rm det.}$ and $\nu_{\rm CSE, min.}^{\rm obs.}(0)=\nu_{\rm min.}^{\rm det.}/2$. 
  Three different filament merging rates are shown ($\delta_{\rm m-WI}=0.5,1.0,2.0$)  with
  the filament size evolving in ime as $\lambda_{\rm F}(t)=\lambda_{\rm e-WI}\times (1+t/t_{\rm m-WI})^{\delta_{\rm m-WI}}$ (Eq. (\ref{eq:lambdaWI})).}
   \label{figure:band-integrated-flux}
 \end{figure*}

 The analytical and normalized band-integrated flux density is given by Eq. (\ref{appendix:eq:fluxdensity}).
  Figure \ref{figure:band-integrated-flux}  shows examples of the band-integrated flux in our model for the CHIME detector
when $\nu_{\rm CSE, max.}^{\rm obs.}(0)=2\nu_{\rm max.}^{\rm det.}$ and $\nu_{\rm p, e.}^{\rm obs.}(0)=\nu_{\rm min.}^{\rm det.}/2$.
  The three different curves show different filament merging rates defined
  by the parameter $\delta_{\rm m-WI}$  (see Eq. (\ref{eq:lambdaWI})).

\begin{figure*}
 \vskip 2.0in
 \centering
 \includegraphics[scale=0.25]{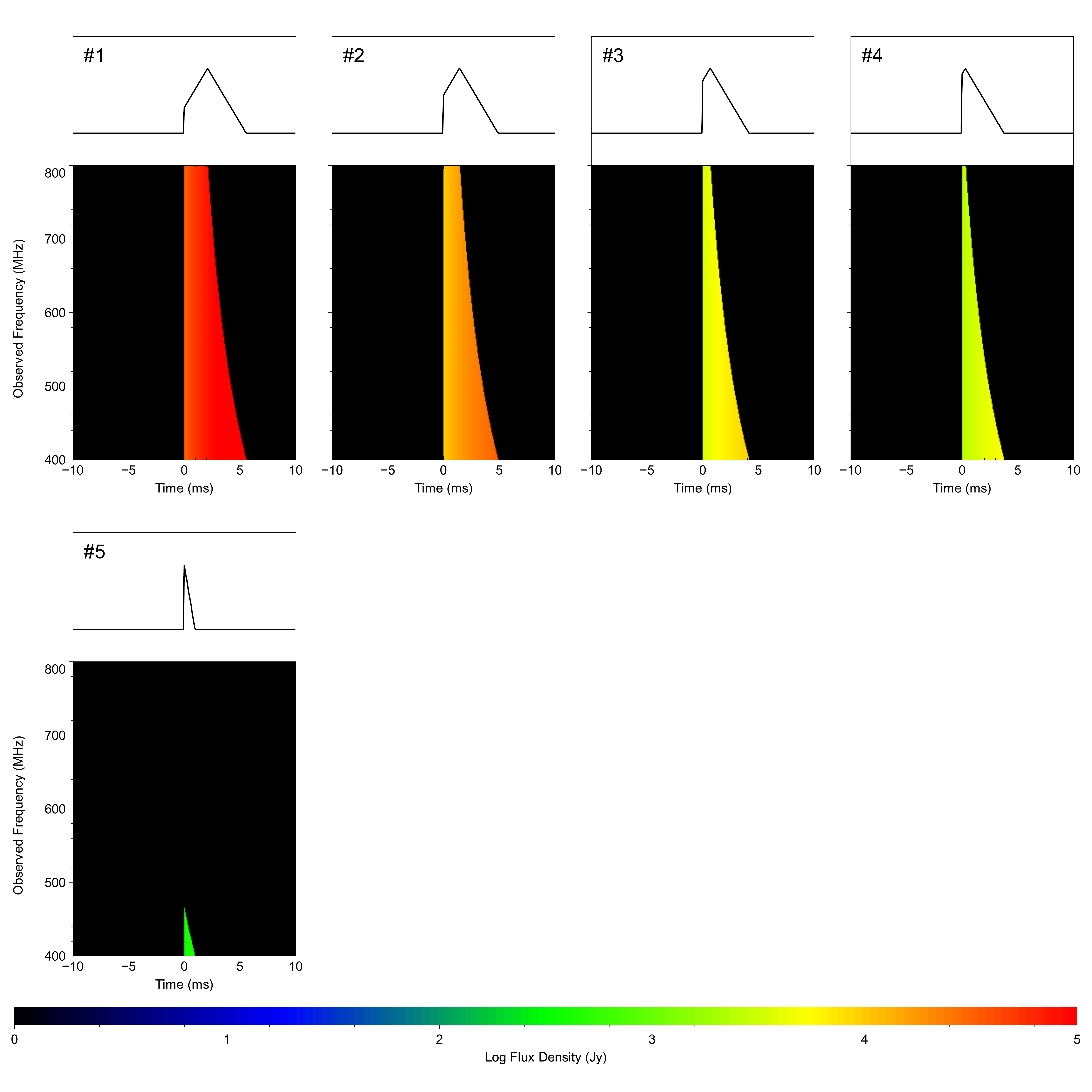}
  \caption{Same as in Figure \ref{figure:FRBs-waterfall-repeating} for the chunks listed 
   in Table \ref{table:FRBs-waterfall-scenarios}.}
  \label{figure:FRBs-waterfall-scenarios}
 \end{figure*}

\begin{figure*}
 \vskip 2.0in
 \centering
 \includegraphics[scale=0.25]{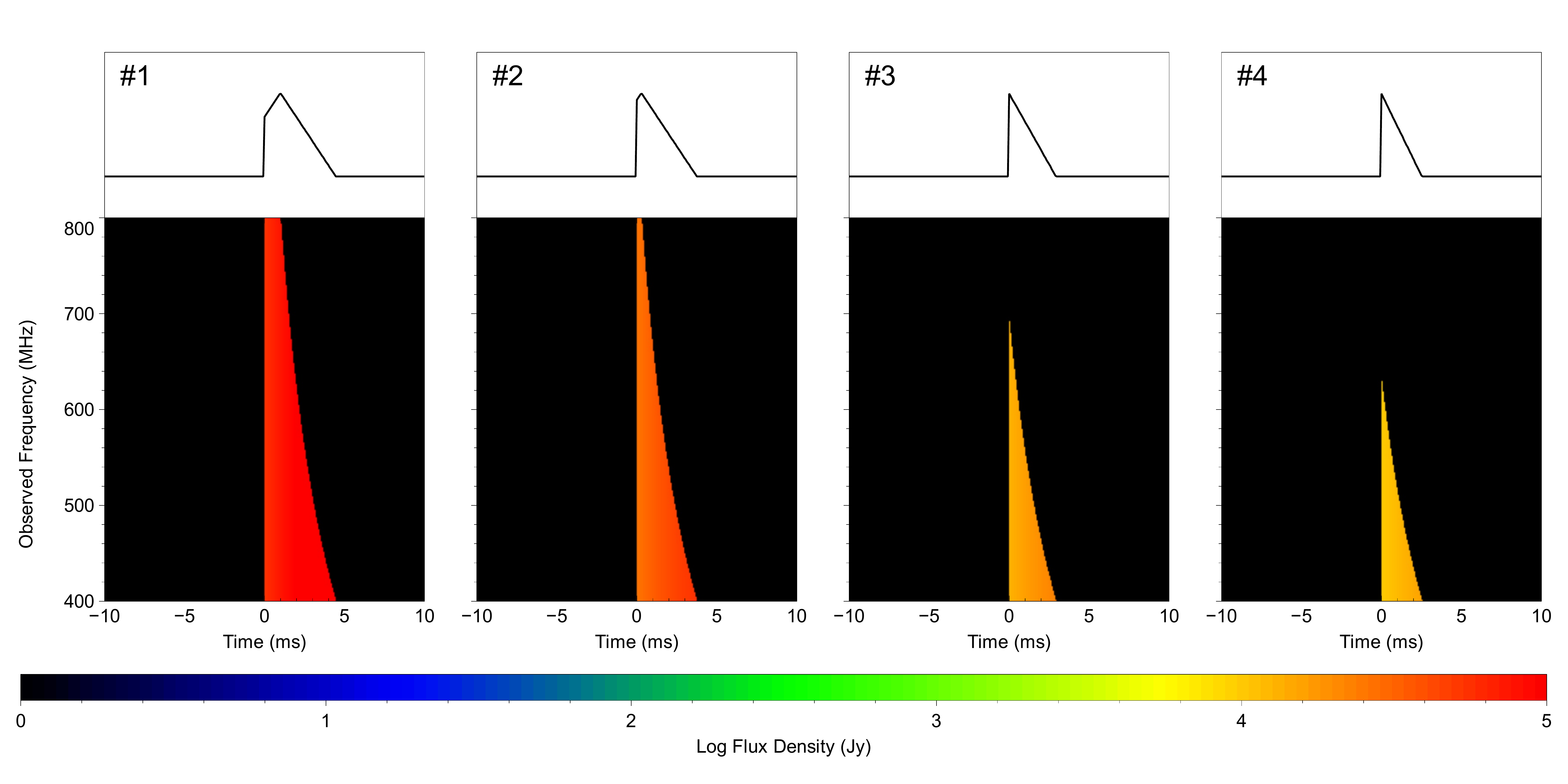}
  \caption{Same as in Figure \ref{figure:FRBs-waterfall-repeating} for the chunks listed 
   in Table \ref{table:FRBs-waterfall-scenarios-2}.}
  \label{figure:FRBs-waterfall-scenarios-2}
 \end{figure*}

\begin{figure*}
 \vskip 2.0in
 \centering
 \includegraphics[scale=0.25]{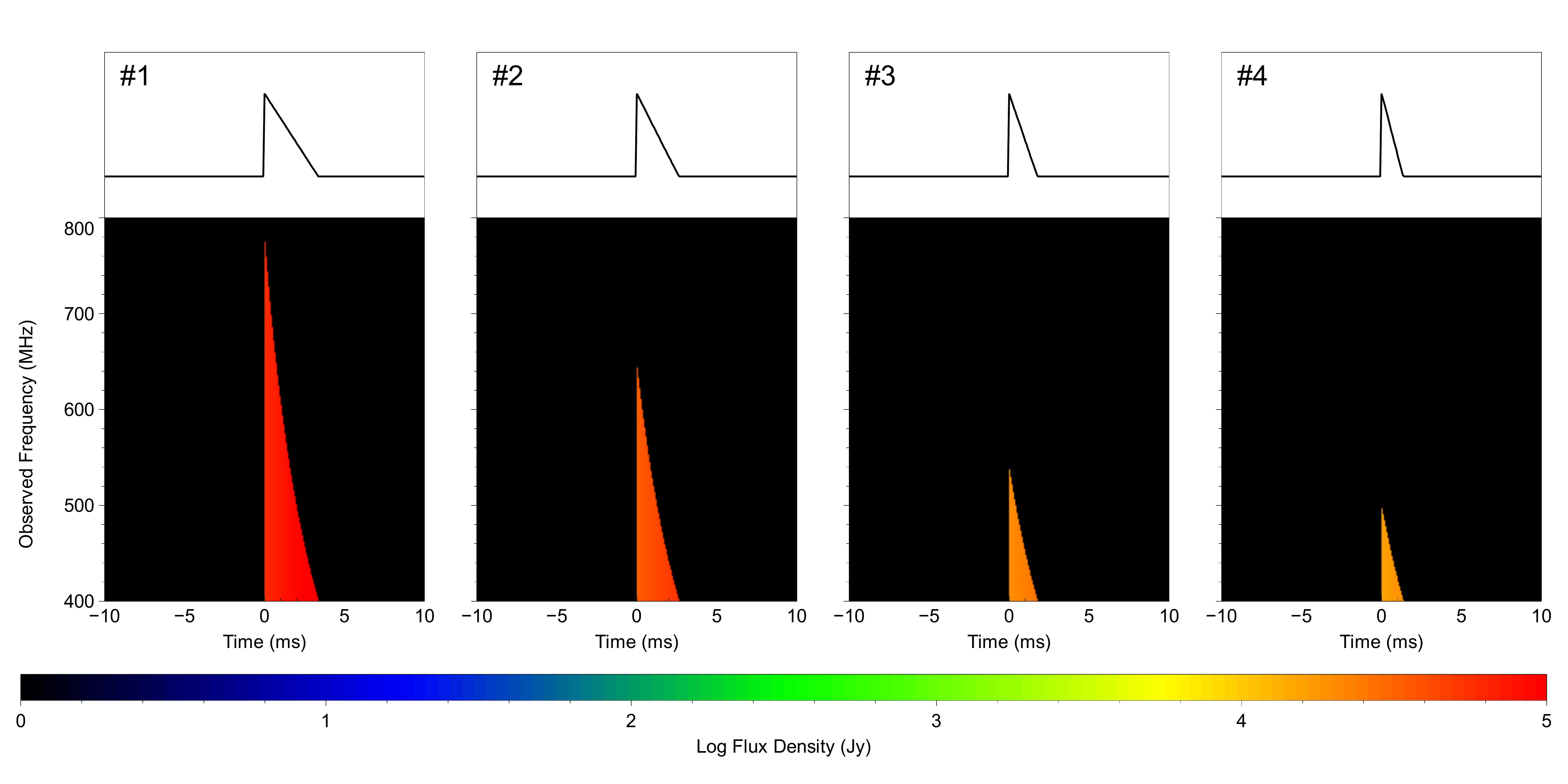}
  \caption{Same as in Figure \ref{figure:FRBs-waterfall-repeating} for the chunks listed 
   in Table \ref{table:FRBs-waterfall-scenarios-3}.}
    \label{figure:FRBs-waterfall-scenarios-3}
 \end{figure*}

Figures \ref{figure:FRBs-waterfall-scenarios} and \ref{figure:FRBs-waterfall-scenarios-2} show
waterfall plots for the repeating FRBs listed in Tables \ref{table:FRBs-waterfall-scenarios} and  \ref{table:FRBs-waterfall-scenarios-2}.
Each pixel in the waterfall plot is the flux density, i.e. $f_{\nu^{\rm obs.}}(\theta_{\rm c},t^{\rm obs.}$)
given in Eq. (\ref{appendix:eq:pureflux}) with $L_{\rm CSE}$ given by Eq.(\ref{eq:LCSE}). 
The resulting band(frequency)-summed flux density is shown in  the upper sub-panels and matches  the  analytically derived one (see Appendix \ref{appendix:flux-fluence} and related Figure \ref{figure:band-integrated-flux}). 
To obtain the integrated  flux density plot we add up the flux in each pixel 
 (i.e. over  the  detector's frequency band) along the vertical axis for each time with $f_{\nu^{\rm obs.}}(\theta_{\rm c},t^{\rm obs.})=0$
when $\nu_{\rm CSE}^{\rm obs.}(t)<\nu_{\rm pixel}^{\rm det.}$.
 Figure \ref{figure:FRBs-waterfall-scenarios-3} shows an example where for all chunks
  the maximum CSE frequency falls within the detector's band (here CHIME); see Table \ref{table:FRBs-waterfall-scenarios-3}
  for the corresponding simulations.

\subsection{Non-repeating vs repeating FRBs}
\label{appendix:repeats-non-repeats}  

In our model, FRBs are intrinsically all repeaters because each chunk gives an FRB beamed in a specific direction.
 Observed single (i.e. non-repeating) FRBs are an artifact of the detector's bandwidth and sensitivity. 
Consider a detector with maximum and minimum frequency $\nu_{\rm max.}^{\rm det.}$ and $\nu_{\rm min.}^{\rm det.}$,
respectively, and a fluence sensitivity threshold $F_{\rm min.}^{\rm det.}$.  The two conditions which 
must be simultaneously satisfied for repeats to occur are

\begin{equation}
\label{eq:repeat-condition}
\nu_{\rm CSE, max.}^{\rm obs.}(\bar{\theta}_{\rm S}) > \nu_{\rm min.}^{\rm det.} \quad {\rm and}\quad F(\bar{\theta}_{\rm S},\delta_{\rm m-WI},0)> F_{\rm min.}^{\rm det.}\ ,
\end{equation}
where $\bar{\theta}_{\rm S}$ is the average viewing angle for secondary chunks (see Eq. (\ref{eq:mean-angles}))\footnote{The secondary and tertiary chunks consist of a group of chunks with roughly a similar $\theta_{\rm c}$ and different azimuths (see Figure \ref{figure:honeycomb}).}.
Box ``A" in Table \ref{table:ICM-QNe-1} shows an example of FRBs where only
a few detectors can see the primary chunk (the shaded cells). In Box ``A" example,
while the $\nu_{\rm CSE, max.}^{\rm obs.}(\bar{\theta}_{\rm S}) > \nu_{\rm min.}^{\rm det.}$ is satisfied, the
fluence is below threshold for most detectors. Box ``B" shows the case where only CHIME sees repeats since the condition
 $\nu_{\rm CSE, max.}^{\rm obs.}(\bar{\theta}_{\rm S}) > \nu_{\rm min.}^{\rm det.}$  in Eq. (\ref{eq:repeat-condition}) 
 is violated by the secondary chunks for most detectors (the ``N/A" cells).  This is also the reason why  $\mathcal{G}(\theta_{\rm c},\delta_{\rm m-WI},0)=0$  in Table \ref{table:G-fluence} for $N_{\rm c}=10^{5}$  and $\Gamma_{\rm c}=10^{2.5}$.

In general ``non-repeats" occur for $f(\theta_{\rm c})>>1$ which is 
 the case for  high $\Gamma_{\rm c}$ ($\ge 10^{2.5}$) and/or low $N_{\rm c}$ ($<10^{5.5})$ as in  Boxes ``A" and ``B".
 In this regime, with $f(\theta_{\rm c})\sim (\Gamma_{\rm c}\theta_{\rm c})^2$, $n_{\rm cc}\propto \Gamma_{\rm c}^{12/5}, R_{\rm cc}\propto \Gamma_{\rm c}^{-4/5}$ and $\beta_{\rm cc}\propto \Gamma_{\rm c}^{2/5}$ we get

\begin{align}
\label{eq:scaling}
       \nu_{\rm CSE, max.}^{\rm obs.}(\theta_{\rm c})&\propto \Gamma_{\rm c}^{11/5} f(\theta_{\rm c})^{-1}\propto \Gamma_{\rm c}^{1/5}\theta_{\rm c}^{-2}\nonumber\\
       \mathcal{F}(\theta_{\rm c},0)&\propto \Gamma_{\rm c}^{-2/5} f(\theta_{\rm c})^{-2}\propto  \Gamma_{\rm c}^{-22/5}\theta_{\rm c}^{-4}\ .
\end{align}

 The maximum CSE frequency is weakly dependent on $\Gamma_{\rm c}$.
 Because $\mathcal{G}(\theta_{\rm c},\delta_{\rm m-WI},0)\propto \nu_{\rm CSE, max.}^{\rm obs.}(\theta_{\rm c})^2 \propto \Gamma_{\rm c}^{22/5}\theta_{\rm c}^{-4}$ when $\nu_{\rm CSE, max.}^{\rm obs.}(\theta_{\rm c})>\nu_{\rm max.}^{\rm det.}$ (see Eq. (\ref{appendix:eq:fluence-flat-2}); see Appendix \ref{appendix:flat-spectrum}),  the fluence $F(\theta_{\rm c},\delta_{\rm m-WI},0)=\mathcal{F}(\theta_{\rm c},0)\times \mathcal{G}(\theta_{\rm c},\delta_{\rm m-WI},0)$ is independent of the Lorentz factor and strongly dependent on the viewing angle  as $\theta_{\rm c}^{-8}$.  

The average viewing angle of the secondary and tertiary chunks as  
derived in Eq. (\ref{eq:mean-angles}) can be expressed in terms of the primary chunk as 
 $\bar{\theta}_{\rm S}\simeq (7/3)\bar{\theta}_{\rm P}$ and  $\bar{\theta}_{\rm T}\simeq 6\bar{\theta}_{\rm P}$  with the consequence that 
 $\nu_{\rm CSE, max.}^{\rm obs.}(\bar{\theta}_{\rm S})= (3/7)^2 \nu_{\rm CSE, max.}^{\rm obs.}(\bar{\theta}_{\rm P})$
and $\nu_{\rm CSE, max.}^{\rm obs.}(\bar{\theta}_{\rm T})=(1/36)\times \nu_{\rm CSE, max.}^{\rm obs.}(\bar{\theta}_{\rm P})$. Also, 
  $F(\bar{\theta}_{\rm S},\delta_{\rm m-WI},0)\simeq (3/7)^8 F(\bar{\theta}_{\rm P},\delta_{\rm m-WI},0)$
  and $F(\bar{\theta}_{\rm T},\delta_{\rm m-WI},0)\simeq (1/6)^8 F(\bar{\theta}_{\rm P},\delta_{\rm m-WI},0)$
  which demonstrates that only the primary chunk would fall within most FRB detector bands and above the sensitivity threshold.
Boxes ``A" and  ``B"  in Table \ref{table:ICM-QNe-1} show that the frequency and the fluence  
 for the secondary and tertiary chunks, in the non-repeating FRBs, do follow the 
 $\theta_{\rm c}^{-2}$ and  $\theta_{\rm c}^{-8}$ dependencies, respectively. In general,  the scaling follows the
 more general form of the dependency given as  $f(\theta_{\rm c})^{-1}$ and $f(\theta_{\rm c})^{-4}$, respectively.

    Repeating FRBs  are obtained  for relatively lower values of $f(\theta_{\rm c}$) for the secondary and tertiary chunks which is the case for 
    higher $N_{\rm c}$ values.  Boxes ``D" and ``E" in Table  \ref{table:ICM-QNe-1} 
show that most detectors would see the secondary chunks with a few detectors capable of detecting also the 
 tertiary chunks  (shaded cells). Boxes ``C" and ``F" correspond to the
 low $\Gamma_{\rm c}$ scenario (in this case $10^2$) with the maximum CSE frequency ($\nu_{\rm CSE, max.}^{\rm obs.}\propto \Gamma_{\rm c}^{11/5}$ for $f(\theta_{\rm c})\sim 1$) being in the
 sub-GHz regime thus eliminating ASKAP, Parkes and Arecibo detections. In this regime,
  CHIME can detect many repeats for a range in $N_{\rm c}$.

\section{Case study}
\label{appendix:case-studies}

   Overall, our model can reproduce   general properties of observed non-repeating and repeating  FRBs.  
   In this appendix, we focus particularly on  FRB 180916.J0158$+$65 and FRB 121102.

\subsection{FRB 180916.J0158$+$65}
\label{sec:frb-16day}

  A year long observation of FRB 180916.J0158$+$65 led to the detection  of tens of  bursts with a regular $\sim 16$ day cycle 
with  bursts arriving in a 4-day phases (\citet{chime_2020a}). In our model, repetition is set by the angular separation between
emitting chunks which yields a roughly constant time delay between bursts (see discusion
around Eq. (\ref{eq:Deltac})). Boxes A, B and C in Table \ref{table:ICM-QNe-1} (i.e. for $N_{\rm c}=10^5$ and $10^2\le \Gamma_{\rm c}\le 10^3$),  show that typical time delays between bursts within a repeating FRB is $12\ {\rm days}  < \Delta t_{\rm repeat}^{\rm obs.}< 20\ {\rm days}$.

The simulations use  randomly spaced chunks rather than  the
simple honeycomb geometry presented in Appendix \ref{appendix:honeycomb}. It is possible to view the QN such that we get FRBs
 from chunks arriving roughly periodically.  An example is given in
 Table \ref{table:FRBs-repeating-16days}  with  a  $\sim$16-day period repeating FRB.
  A 4-day window (a ``smearing" effect)  can  also obtained  by varying the 
chunk parameters  such as the mass and the Lorentz factor and/or the ambient
number density $n_{\rm amb.}^{\rm ns}$ for a given QN.

\subsection{FRB 121102}
\label{appendix:FRB121102}

FRB 121102 was discovered by PARKES at a redshift of $z\sim 0.1972$ (\citealt{spitler_2014}).
Its main properties include the quiescent and active periods on month-long scales   (\citealt{michilli_2018}),
 with hundreds of bursts so far detected (e.g. \citet{gajjar_2018,hessels_2019}). It has been associated with a star-forming
 region in an irregular, low-metallicity dwarf galaxy (\citet{bassa_2017}). 
  The high RM measured in FRB 121102 ($RM\sim 10^5$ rad m$^{-2}$; \citealt{michilli_2018}) sets it apart from other FRBs.

Table \ref{table:FRB121102} shows an example of an FRB from an ICM-QN in our model 
 lasting for $\sim 20$ years reminiscent of FRB 121102. This is obtained
 by setting a higher $\gamma_{\rm CSE}$ (here 40) and a low $\Gamma_{\rm c}$ (here 40)
compared to  fiducial values listed in Table \ref{table:parameters}. 
A variation in chunk mass is necessary to obtain the variability in width and
fluence seen in FRB 121102.

 We find that the unique properties of FRB 121102 mentioned above may be best explained in our model if 
  we assume that the QN responsible for it occurred inside a galaxy. 
  This would be the case for    NSs with small kick velocities. For example for a velocity of  
  $\sim 10$ km s$^{-1}$, the NS would have travelled only about a kilo-parsec in $\sim 10^8$ years
  by the time it experience a QN transition. Table \ref{table:FRB121102-galactic} shows an example of a galactic FRB,  
 lasting for $\sim 3$ years, obtained by considering an ambient density of $n_{\rm amb.}^{\rm ns}=10^{-2}$ cm$^{-3}$
  representative of a galactic/halo environment.
  
  If the QN occurs in the vicinity of a star forming region in the galaxy
  (i.e. probably rich in HII regions),  as seems to be the case for FRB 121102, 
  the CSE from the QN chunks would be susceptible to lensing  thus enhancing the number of bursts (\citealt{cordes_2019}). Lensing would 
        ``scramble" any regular cycle (i.e. the $\Delta t_{\rm repeat}^{\rm obs.}$ period) expected due to the spatial distribution of the QN chunk.
        An FRB from a galactic QN  at low redshift would mean a sensitivity to more chunks at higher $\theta_{\rm c}$;
       i.e. a bigger solid angle is accessible to detectors. 
       
       Finally, it may be possible that the high RM associated with FRB 121102 is 
       intrinsic to the QN chunks.  The rotation measure is  $RM =0.81\int_0^{d} n_{\rm e} B_{\parallel} dl_{\parallel}$ with
    the magnetic field along the line-of-sight  in units of $\mu$G and $l_{\parallel}$ in parsecs.
    With $n_{\rm e}=n_{\rm cc}$, $B_{\parallel}=B_{\rm p-WI,s}$ (see Eq.(\ref{eq:Bcc,s})) and $d\sim 2R_{\rm cc}$, 
    the RM  induced by a chunk during the CSE phase we estimate to be $RM_{\rm cc} \simeq 822.2\ {\rm rad\ m}^{-2}\times n_{\rm cc, 1}^{3/2}R_{\rm cc, 15}$. Or, 
  \begin{align}
  \label{eq:RMcc}
  RM_{\rm cc} &\simeq 2.7\times 10^5\ {\rm rad\ m}^{-2}\times\nonumber\\
   &\times  \frac{m_{\rm c, 22.3}^{9/20}\sigma_{\rm HH, -27}^{7/5}\Gamma_{\rm c, 2.5}^{14/5}{n^{\rm ns}_{\rm amb., -1}}^{7/5}}{\kappa_{\rm c, -1}^{21/20}}\ ,
  \end{align}
   for $n_{\rm amb.}=0.1$ cm$^{-3}$ representative of the hot ISM component within galaxies (\citealt{cox_2005}).

\section{Predictions}
\label{appendix:predictions}

 \subsection{FRBs in LOFAR}
 \label{sec:FRBs-in-LOFAR}

       Our simulations  show that on average  CHIME detects  5 times more FRBs than
   ASKAP and Parkes.  This is due to the fact that the CSE frequency in our model decreases with an increase in
   $f(\theta_{\rm c})$ (i.e. with higher viewing angle $\theta_{\rm c}$) making CHIME more sensitive to secondary chunks (i.e. sees a bigger solid angle)   for a given QN. The number of chunks $N_{\nu^{\rm obs.}}^{\rm obs.}$ (i.e. FRBs per QN) detectable at any frequency is given in 
     Appendix \ref{appendix:chunks-per-frequency} and expressed in  Eq. (\ref{appendix:eq:dNc-nu})  as
   \begin{equation}
     \label{eq:dNc-nu}
     \frac{dN_{\nu^{\rm obs.}}^{\rm obs.}}{d{\nu^{\rm obs.}}}\propto {\nu^{\rm obs.}}^{-2}\ .
     \end{equation}
       Applying the above to CHIME and ASKAP detectors, for example, we get

      \begin{equation}
      \label{eq:CHIME-vs-ASKAP}
      \frac{N_{\rm CHIME}^{\rm obs.}}{N_{\rm ASKAP}^{\rm obs.}}= \frac{\Delta \nu_{\rm CHIME}}{\Delta \nu_{\rm ASKAP}}\times \frac{\nu_{\rm ASKAP, p}^2}{\nu_{\rm CHIME, p}^2}
      \sim \frac{0.4}{0.4}\times \frac{1.3^2}{0.6^2}\sim 4.7\ ,
      \end{equation}
      independently of $\Gamma_{\rm c}$ (i.e. for a given QN)  in agreement with the simulation results; 
      the subscript ``p" refers to the band's peak frequency (see Table \ref{table:detectors}).

      Past CHIME's band the FRBs will  drift  into the LOFAR's band. In addition, 
      emission from chunks at high viewing angles will be visible to LOFAR.
      Using Eq. (\ref{eq:dNc-nu}) to compare LOFAR (high-band antenna) to CHIME we arrive at
       
        \begin{equation}
        \label{eq:LOFAR-CHIME-ratio}
      \frac{N_{\rm LOFAR}^{\rm obs.}}{N_{\rm CHIME}^{\rm obs.}}= \frac{\Delta \nu_{\rm LOFAR}}{\Delta \nu_{\rm CHIME}}\times \frac{\nu_{\rm CHIME, p}^2}{\nu_{\rm LOFAR, p}^2}
      \sim \frac{0.18}{0.4}\times \frac{0.6^2}{0.18^2}\sim 5\ .
      \end{equation}
      LOFAR should thus detect on average 5 times more bursts than CHIME
      from a given QN.   Our simulations do not yield LOFAR's detections too often 
        except in a few cases when the chunk is massive
        and very close to the observer's line-of-sight  such as in the
        simulations shown in Tables \ref{table:FRBs-waterfall-scenarios}-\ref{table:FRBs-waterfall-scenarios-3} 
         with LOFAR's fluence very close to the threshold of $10^3$ Jy ms (see also cases in Table \ref{table:ICM-QNe-1}).
      This is understandable because for a given QN, an $f(\theta_{\rm c})\sim 100$ is necessary for the CSE frequency
       to fall within LOFAR's band. However, these high $f(\theta_{\rm c})$ values
       yield a fluence ($\propto f(\theta_{\rm c})^{-4}$)  below the LOFAR's sensitivity limit.
             The ratio given in Eq. (\ref{eq:LOFAR-CHIME-ratio})
        is likely to be reduced by: (i) dispersion effects (which are
      more pronounced at MHz frequencies); (ii) the Earth's ionosphere which affects signals in the tens of MHz range.

 \subsection{FRBs from IGM-QNe?}
  \label{appendix:IGM-QNe}

  Table \ref{table:IGM-equations} 
  summarizes the equations relevant to FRBs from IGM-QNe. These were derived from Table \ref{table:ICM-equations}  
  using  $n_{\rm amb.}^{\rm ns}\simeq 2\times 10^{-7}\ {\rm cm}^{-3}\times (1+z)^3$
  for the IGM  (e.g. \citealt{mcquinn_2016}).  The maximum CSE frequency is
    
     \begin{equation}
     \label{eq:CSE-IGM}
     \nu_{\rm CSE, max.}^{\rm obs.} \simeq \frac{36.6\ {\rm MHz}}{(1+z)f(\theta_{\rm c})}\times \delta_{\rm CSE, -1}\gamma_{\rm CSE, 1}^{2}\Gamma_{\rm c, 2.5}n_{\rm cc, -4}^{1/2}\ ,
     \end{equation} 
     which falls below most radio detectors/receivers except may be for LOFAR's low-band antenna for which $\nu_{\rm min.}^{\rm det.}=30$ MHz (\citet{vanhaarlem_2013}).  Because $f(\theta_{\rm c})>>1$ for non-repeating FRBs 
 (see Appendix (\ref{appendix:repeats-non-repeats})), the maximum CSE frequency will fall below LOFAR minimum frequency. 
 Also, repeating FRBs (i.e. with low $\Gamma_{\rm c}$)  from IGM-QNe at high high-redshift would
 yield frequencies below the LOFAR's band. Thus FRBs from IGM-QNe may not be detectable with current detectors.

  Besides the CSE frequency which would likely fall below the LOFAR band, we also  argue that 
   IGM-QNe may not occur in nature. Isolated massive NS in field galaxies (with halos extending up to $\sim 100$ kpc or more) would need to
  travel long  distances before they enter the IGM. 
   For a NS with a typical  kick velocity of $300$ km s$^{-1}$, nucleation timescales of at least $\sim 10^{9}$ years
  would be required for the NS to enter the IGM prior to the QN event.  For typical quark nucleation timescales  of $\sim 10^8$ years
  (and a narrow nucleation timescale distribution), even NSs with a 
  kick velocity of $\sim 10^3$ km s$^{-1}$ would travel only about 100 kpc 
  reaching at most the edge of their galaxies.  While we cannot with full certainty rule out FRBs from IGM-QNe
   they seem unlikely. Instead, in field galaxies it is likely that FRBs would be associated with halo-QNe (see Appendix  \ref{appendix:FRB121102}), meaning that in field galaxies old NSs would experience the QN phase (yielding FRBs) while still embedded  in the halo. 
   
     {\bf Monster FRBs from IGM-QNe}:   FRBs
  from chunks seen very close to the line-of-sight (i.e. $f(\theta_{\rm c})\sim 1$) could reach a fluence
   in the millions of Jy ms (see Table \ref{table:IGM-equations}). 
   Several effects conspire to make FRBs from IGM-QNe much brighter than those
  from galactic- and ICM-QNe. The low IGM density means the chunks must travel 
  large distance, and thus reaching larger radii, and becoming colder (i.e. associated with lower
  $\beta_{\rm cc}$ values) when they become collisionless (see Table \ref{table:IGM-equations}).
  There is also the band effect with the lower frequency ones contributing  higher values of $\mathcal{G}(\theta_{\rm c},\delta_{\rm m-WI},0)$  to the total fluence, $F(\theta_{\rm c},\delta_{\rm m-WI},0)= \mathcal{F}(\theta_{\rm c},0)\times \mathcal{G}(\theta_{\rm c},\delta_{\rm m-WI},0)$ (see
  Appendix \ref{appendix:flat-spectrum}  and the corresponding Table \ref{table:G-fluence}).   However, FRBs from IGM-QNe if they occur would be rare events and even so their 
    frequencies may fall outside the LOFAR's band (i.e. $\nu_{\rm CSE, max.}^{\rm obs.}< 30$ MHz); see discussion in  \S \ref{appendix:IGM-QNe}.
    
    \subsection{The pre-CSE phase}
    \label{appendix-pre-CSE}
   
   There are  plausible emission mechanisms prior to the CSE phase:
 
 (i) {\it Thermal Bremsstrahlung (TB)} emission from the
 chunks before they enter the collisionless phase (see Appendix \ref{appendix:ionized-chunk}).
 The corresponding spectrum is flat and has  a maximum frequency $\nu_{\rm TB}^{\rm obs.}=D(\Gamma_{\rm c},\theta_{\rm c})T_{\rm c, ic}/(1+z)$ with $T_{\rm ic}\simeq 13.6$ eV the chunk's temperature when it becomes ionized by hadronic collisions with the
ambient medium. This gives

\begin{equation}
\nu_{\rm TB}^{\rm obs.}\simeq
  2.1\times 10^{18}\ {\rm Hz}\times \frac{\Gamma_{\rm c, 2.5}}{(1+z)f(\theta_{\rm c})}\ ,
  \end{equation}
  which is in the keV range. The corresponding maximum X-ray luminosity, given by Eq. (\ref{appendix:eq:Lic}), is:

  \begin{align}
  L_{\rm TB, max.}^{\rm obs.}\simeq 4.4\times 10^{37}\ {\rm erg\ s}^{-1}\times \frac{1}{(1+z)^2f(\theta_{\rm c})^4}\times \sigma_{\rm HH, -27}^{3}\Gamma_{\rm c, 2.5}^{10}m_{\rm c, 22.3}^2{n_{\rm amb., -3}^{\rm ns}}^{3} \ .
  \end{align}
 
  The TB phase would persist for $\Delta t_{\rm TB}^{\rm obs.}\sim t_{\rm cc}^{\rm obs.}$ which is of the order
  of  days  (see Eq. (\ref{appendix:eq:tccobs})).

(ii) {\it Incoherent synchrotron emission (ISE)} in the very early stages of filament merging phase, preceding the CSE phase.
The corresponding ISE frequency in the observer's
frame ($D(\Gamma_{\rm c},\theta_{\rm c})\nu_{\rm ISE}/(1+z)$) would be 
  
  \begin{equation}
  \label{eq:nu-ISE}
  \nu_{\rm ISE}^{\rm obs.}\simeq \frac{115.7\ {\rm GHz}}{(1+z)f(\theta_{\rm c})}\times \Gamma_{\rm c, 2.5}\gamma_{\rm CSE, 1}^2n_{\rm cc, 1}^{1/2}\ .
  \end{equation}

  The maximum luminosity (which assumes contribution form all chunk's electrons) is $L_{\rm ISE, max.}=(m_{\rm c}/m_{\rm H})\times P_{\rm e}$
  with the ISE power per electron $P_{\rm e}=1.6\times 10^{-15} \gamma_{\rm CSE}^2B_{\rm p-WI,s}^2$ (e.g. \citet{lang_1999}).
  The observed maximum ISE luminosity, $L_{\rm ISE, max.}^{\rm obs.}=D(\Gamma_{\rm c},\theta_{\rm c})^4L_{\rm ISE, max.}/(1+z)^2$, is thus
  
  \begin{align}
   \label{eq:L-ISE}
  L_{\rm ISE, max.}^{\rm obs.}\simeq \frac{5.8\times 10^{34}\ {\rm erg\ s}^{-1}}{(1+z)^2f(\theta_{\rm c})^4}\times  \Gamma_{\rm c, 2.5}^4m_{\rm c, 22.3}\gamma_{\rm CSE, 1}^2n_{\rm cc, 1}\ ,
  \end{align}
   which is much dimmer than the  subsequent CSE phase. The ISE phase is 
    short lived ($<< t_{\rm m-WI}$) compared to the CSE phase  and may be hard to detect.

    \subsection{FRBs and Ultra-High Energy Cosmic Rays (UHECRs)}
    \label{appendix:FRBs-UHECRs}
    
  Once the Weibel shock forms following proton trapping, the chunk's Lorentz factor $\Gamma_{\rm c}$ decreases rapidly  with the sweeping of ambient protons. 
 Half of the chunk's kinetic energy is 
  converted into heat after sweeping $m_{\rm c}/\Gamma_{\rm c}$ of material (e.g. \citealt{piran_1999}).
  In the chunk's frame we have $m_{\rm c}/\Gamma_{\rm c}=A_{\rm cc} \Gamma_{\rm c} n_{\rm amb.}m_{\rm H} c \Delta t_{\rm c, sw.}$
  with $\Delta t_{\rm c, sw.}$ the characteristic deceleration timescale. A  slowdown of a QN chunk would occur after it travels a
  distance of a few parsecs  ($c\Gamma_{\rm c} \Delta t_{\rm c, sw.}$) from the FRB site.  
  In the observer's frame it occurs on a timescale of 
  
  \begin{align}
  \label{eq:tsw-obs}
 \Delta t_{\rm c, sw.}^{\rm obs.}&\sim 2\ {\rm s}\times (1+z)f(\theta_{\rm c})\times \frac{m_{\rm c, 22.3}}{R_{\rm cc, 15}^2\Gamma_{\rm c, 2.5}^{3}n_{\rm amb.,-3}}\ .
  \end{align}
    
  The Weibel shock (which ends the BI-WI process), may be inductive to Fermi acceleration  (\citealt{fermi_1949}).
The particles in the ambient medium and/or in the chunk
can be boosted  by  $\sim 2\Gamma_{\rm c}^2$ (e.g. \citealt{gallant_1999}) reaching energies of the order of
 \begin{equation}
 E_{\rm UHECR}\sim 2\times 10^{15}\ {\rm eV}\times A\times  \Gamma_{\rm c, 2.5}^2\ ,
 \end{equation}
  where $A$ is the atomic weight of the accelerated particles (i.e. the
 chemical imprint of both  the ambient medium and of the chunk material).
 A distribution in $\Gamma_{\rm c}$ (with $10^{1.5} < \Gamma_{\rm c} < 10^{3.5}$ as suggested by our fits to FRB data)
 would allow a range in UHECR of $2\times 10^{13}\ {\rm eV} < E_{\rm UHECR}/A < 2\times 10^{17}\ {\rm eV}$.
 
 A rate of one QN per thousand years per galaxy means an available power of  $\sim 10^{48}$ erg yr$^{-1}$ (i.e.
 $E_{\rm QN}\sim 10^{51}$ erg per thousand year) per
galaxy which  should be enough power to account for UHECRs (e.g. \citet{berezinsky_2008,murase_2009}
and references therein).   Thus collisionless QN chunks could potentially act as efficient UHECR accelerators. These 
    are tiny regions (of size $R_{\rm cc}\sim 10^{15}$ cm)   spread over a very large volume 
which would make it hard for detectors to resolve.

\subsection{Other predictions}
\label{sec:additional-predictions}

\begin{itemize}

\item {\bf FRBs from galactic/halo-QNe}: These FRBs could be   associated with field galaxies as well as galaxy clusters. 
  While in galaxy clusters they would be induced by QNe  from NSs with a low kick velocity, in field galaxies with extended haloes, 
 isolated old  NSs would likely experience the QN event before reaching the IGM (see Appendix \ref{appendix:IGM-QNe}).
    A possible differentiator between FRBs from ICM-QNe and those from galactic/halo-QNe  may be the high RM in the latter ones (Eq.(\ref{eq:RMcc}));

     \item {\bf Super FRBs from halo- and ICM-QNe}:     FRBs  from the primary chunk 
  would be extremely bright with a fluence in the tens of thousands of Jy ms for CHIME's band 
   and hundreds of Jy ms for LOFAR's high-band antenna (see
  examples in boxes ``D" and ``E"  in Table \ref{table:ICM-QNe-1}). However these events 
   may be rare if a typical ICM-QN yields  $N_{\rm c}<10^{5.5}$ based on our model's fits  to FRB data;

 \item  {\bf QN compact remnant in X-rays}:  The QS is born with a  surface magnetic field of the order of 
  $\sim 10^{14}$ G owing to strong fields generated during the hadronic-to-quark-matter
phase transition (\citealt{iwazaki_2005,dvornikov_2016a,dvornikov_2016b}). Despite such high magnetic field, 
 QSs according to the QN model do not pulse in radio since they are
born as aligned rotators (\citealt{ouyed_2004,ouyed_2006}). Instead, during the quark star spin-down, vortices (and the magnetic field they confine) are expelled (\citet{ouyed_2004,niebergal_2010b}). 
 The subsequent magnetic field reconnection leads to the production of
X-rays at a rate of $L_{\rm X}\sim 2\times 10^{34}\ {\rm erg\ s}^{-1}\times \eta_{\rm X, -1}\dot{P}_{-11}^2$
where $\eta_{\rm X}$ is an efficiency parameter related to the rate of conversion of  
 magnetic energy to radiation and $\dot{P}$ the period derivative  (see \S 5 in \citealt{ouyed_2007a});

\item {\bf FRBs in Low-Mass Xray Binaries}: For a QN in a binary (see \citealt{ouyed_2014}), chunks
that manage to escape the binary through low-density regions should yield FRBs. 
Thus our model predicts the plausible  connection of some FRBs with Type-Ia SNe
 though statistically such an association should be very weak due to FRB beaming effects.

  \end{itemize}

    \section{Model's limitations}
  \label{appendix:limitations}
  
  \begin{itemize}
  
  \item  {\bf The frequency-time diagram}: Patchiness (i.e. gaps) in  the frequency-time diagram during drifting (in the milli-second timescales) has been
 observed. It may be a consequence of scintillation effects  induced by the ambient medium  as suggested 
  in the literature (\citealt{macquart_2018}) from the  comparison of the bright nearby ASKAP FRBs to the
 dimmer farther away Parkes FRBs (i.e. based on the DM-brightness relation; \citealt{shannon_2018}). 
 However,  there remains the possibility that the patchiness may be intrinsic to the chunk and may be a result of 
 different parts of the chunk acting at different times.  This is beyond the scope of this paper and will be explored elsewhere;

 \item {\bf  Polarization}: In its current form, our model cannot explain the
 degree of polarization associated with some FRBs. We can only argue that 
 the WI saturated magnetic field may play a role.  The filament's magnetic field  strength at saturation is
 $B_{\rm p-WI, s}\sim 0.12\ {\rm G}\times n_{\rm cc, 1}^{1/2}$ (see Eq. (\ref{eq:Bcc,s}))
 and may induce polarization at some level. At the
 beginning of filament merging,   the many independent (i.e. non-communicating) bunches should yield a relatively less polarized CSE
  despite the high $B_{\rm p-WI, s}$.  CSE may show more  polarization 
 towards the end of filament merging when emission from the reduced number of
  (and thus larger size)  bunches  is expected to be more synchronized. 
   Alternatively, if one bunch triggers another they may emit in the same polarization. This will be explored elsewhere;
   
     \item {\bf FRB 121102 high RM}:  FRB 121102 high rotation measure of $RM\sim 10^5$ rad m$^{-2}$ (\citealt{michilli_2018}) sets it apart from other FRBs.   The RM induced by the chunk on the CSE
  is given by  Eq. (\ref{eq:RMcc}) which shows that in our model  high RM values 
  can be obtained for FRBs from galactic-QNe with a high ambient medium density $n_{\rm amb.}^{\rm ns}> 10^{-3}$ cm$^{-3}$. However, 
   in the high ambient medium density case, and for fiducial  parameter values,  our simulations yield repeating FRBs lasting at most only a few years   (Tables \ref{table:FRB121102} and \ref{table:FRB121102-galactic}). A parameter survey is needed
    which may yield longer timescales.  It may also be the case that the
   high RM associated with FRB 121102 is due to plasma within
    the associated galaxy. This issue will be investigated elsewhere;

 \item  {\bf FRB 121102 persistent radio source}: FRB 121102 has also been associated with a persistent radio source
 with luminosity $L \sim 10^{39}$ erg s$^{-1}$ (\citealt{tendulkar_2017,bassa_2017,chatterjee_2017,marcote_2017})
 hinting at a pulsar. This would seem to support our suggestion that  this FRB may be from a galactic-QN in a star-forming dwarf galaxy 
 (see Appendix \ref{appendix:FRB121102}). In this case, we would argue that the radio source (may  be a pulsar) 
  is independent of the FRB proper;

 \item {\bf The minimum CSE frequency}: It is set by the chunk's plasma
 frequency $\nu_{\rm CSE, min.}^{\rm obs.}(\theta_{\rm c})=\nu_{\rm p, e}^{\rm obs.}(\theta_{\rm c})$  in our model (see Eq. (\ref{eq:nu-plasma-obs}))   and  is below the minimum frequency of most FRB detectors (see Table \ref{table:detectors}).  A parameter survey will be performed
  in the future to determine which parameters can yield scenarios with $\nu_{\rm CSE, min.}^{\rm obs.}(\theta_{\rm c})> \nu_{\rm min.}^{\rm det.}$.
  There is the possibility that the CSE may be suppressed before the CSE frequency drops below the plasma frequency; e.g. if Weibel filaments do not grow  beyond a size of $\sim c/\nu_{\rm p, e}$ during the merging process;

    \item  {\bf  Chunk's composition}:  The extremely neutron-rich, relativistically expanding, QN ejecta is converted to unstable r-process
material in a fraction of a second following the explosion (\citealt{jaikumar_2007,kostka_2014}; for details, see Appendix B.2 in \citealt{ouyed_2020}).  Here, we assumed that the chunk  is  dissociated into its hadronic constituents yielding the background ${\it (e^{-},p^{+})}$ plasma. A future avenue would consist of taking into account the ionic composition of the chunk.
  
  \end{itemize}

\section{Implications}   
\label{appendix:implications}
      
        \subsection{FRBs as probes of collisionless plasma instabilities}
        \label{appendix:implications-plasmas}

FRBs can become a laboratory for studying collisionless plasma instabilities if indeed,
 as suggested by our model, the Buneman and the thermal Weibel instabilities are at the heart of
  this phenomenon. FRBs from QNe may provide some guidance to models and PIC simulations  
  of inter-penetrating plasma instabilities. In particular:

\begin{itemize}

  \item  {\bf   Buneman saturation}:  Our fits to FRB data suggests a BI saturation parameter $\zeta_{\rm B}\sim 10^{-1}$
  which translates to about 10\% of the beam electron kinetic energy (in the chunk's frame) being converted to heating chunk's
  electrons. These numbers are comparable to those derived from PIC simulations (e.g. \citealt{dieckmann_2012,moreno_2018});
  
  \item {\bf Filament merging}: FRBs in our model can shed light on the filament merging  process. 
  For example,  our simulations of FRB data suggests $\delta_{\rm m-WI}\ge 1.0$ and $\gamma_{\rm CSE}\ge  10$, 
  in line with recent PIC simulations (e.g. \citet{takamoto_2019}) and may further be used to
  inform future models and  PIC simulations of the filament merging process; 

\item {\bf The Weibel shock}:  The plausible association of FRBs with UHECRs  (see Appendix \ref{appendix:FRBs-UHECRs}), would
confirm that  the Weibel shock took place. Comparing the energy in UHECRs
to the kinetic energy of a typical QN  ejecta $\sim 10^{51}$-$10^{52}$ erg could 
 in principle provide an estimate of the efficiency of particle acceleration in Weibel shocks;

   \item {\bf Micro-bunching instability}:  Perturbations to the bunch density can be amplified by the
  interaction with the CSE proper which may result in a ``sawtooth" instability (\citealt{heifets_2002,venturini_2002}).
  One possible manifestation of the instability is by inducing spikeness in  FRB lightcurves which
  if confirmed by observations would support our model and would offer a unique in-sight into the micro-bunching mechanism 
   in inter-penetrating plasmas.

  \end{itemize}

\subsection{FRBs as probes of the QCD phase diagram}
\label{appendix:implications-QCD}

Of relevance to Quantum-Chromo-Dynamics (QCD) and its phase
diagram, in particular to the still poorly known phases of quark matter (e.g. \citealt{rajagopal_1999} and references therein), we note:

\begin{itemize}

\item {\bf Quark nucleation timescales}: Our model's fits to FRB data hint at a quark nucleation timescale of $\sim 10^8$ years.
This may constrain models of nucleation in dense matter and in neutron stars (e.g. \citealt{bombaci_2004,harko_2004})
 and may be used to constraint  quark deconfinement density;

\item {\bf Quark nucleation in cold and hot NSs}: The energy release during the conversion of
a NS to a QS is of the order of $\sim 3.8\times 10^{53}\ {\rm erg}\times (M_{\rm NS}/2M_{\odot})\times \Delta E_{\rm con.,-4}$
for a $2M_{\odot}$ NS and a conversion energy release, $\Delta E_{\rm con.}$, of about $100$ MeV ($\sim 10^{-4}$ erg) per neutron converted (e.g. \citealt{weber_2005}). Our model for FRBs (involving slowly rotating, old and cold NSs) 
and for GRBs (involving rapidly rotating, young and hot NSs; see \citet{ouyed_2020}) 
 suggests two nucleation regimes.  The hot NS case  (with trapped neutrinos)  releases an important fraction
   (up-to $\sim 30$\%)  of the conversion energy as kinetic energy of the QN ejecta (on average $E_{\rm QN}\sim 5\times 10^{52}$ ergs) while for the cold NS case (with free-streaming neutrinos)  a substantial fraction of the conversion
   energy is lost to neutrinos before the QN event; the kinetic energy of the QN ejecta
   in this case is about a percent of the conversion energy with  $E_{\rm QN}\sim 5\times 10^{51}$ erg;

\item {\bf Color super-conductivity}: A future detection of the radio-quiet ICM-QN  compact remnant via its X-ray emission
 (see Appendix \ref{sec:additional-predictions}), would mean that the QS 
 is likely born  in a  superconducting state (i.e. the Color-Flavor-Locked phase; \citealt{alford_1999}).

\end{itemize}

\subsection{Implications to astrophysics}
\label{appendix:implications-astro}

Implications of QN to astrophysics have been reviewed in \citet{ouyed_2018a,ouyed_2018b}. 
If the model is a correct representation of FRBs then it would particularly strengthen the idea  that:

\begin{itemize}

\item {\bf Quark stars} exist in nature and form mainly from  old NSs exploding as QNe at a rate
of  about 10\% of the core-collapse SN rate;

\item {\bf Missing pulsars}: The formed quark star is {\bf radio quiet} owing to the quark-matter Meissner effect 
which forces the magnetic dipole field to be aligned with the spin axis
(\citealt{ouyed_2004,ouyed_2006,niebergal_2010b}).  Because an important fraction of these old NSs
are potential galactic/halo-QN and ICM-QN  candidates (i.e. becoming radio-quiet after
 the FRB phase),  it would thus appear as if  these went missing from the outskirsts of galaxies;

\item {\bf QNe within a few years of a core-collapse SN} of massive stars may  be at the origin of LGRBs as demonstrated in \citet{ouyed_2020}; 
see \S 7.4 in that paper for short duration GRBs.  Thus the same engine, the exploding NS, is responsible
for GRBs and FRBs in our model. For the FRBs case, the QN occurs hundreds of million of years after the SN;

\item {\bf QNe in binaries} may be of relevance  to cosmology. When the companion
of the exploding NS is a CO white dwarf, a Type-Ia QN results. A
QN-Ia is effectively a Type-Ia SN triggered by the QN ejecta impacting the WD. The QN
 is triggered by accretion onto the NS from the companion which drives
the NS core density above the deconfinement value. The properties of Type-Ia QNe, and 
  the lightcurve, are redshift dependent (see Figure 3 in \citet{ouyed_2014})\footnote{The Phillips 
relationship (\citet{phillips_1993}) is a natural outcome of Type-Ia QNe: In addition to the energy from the $^{56}$Ni decay
powering the QN-exploded CO white dwarf, a QN-Ia is
 powered by spin-down from the Quark star (the QN compact remnant which ends up buried within the
expanding CO ejecta). This results in QN-Ia obeying a Phillips-like
relation where the variation in luminosity is due to the QS spin-down 
power (\citealt{ouyed_2014}); see in particular \S 4.1 and Figure 1 in that paper  where
it is shown that the correlation between peak absolute magnitude and light curve shape is redshift-dependent}.
  If Type-Ia QNe  contaminate Type-Ia SNe samples, the latter may not be standardizable (\citealt{ouyed_2014}).
\citet{kang_2020} provide a recent analysis of the impact of the luminosity evolution on 
the light-curve fitters used by the SNe Ia community.

\end{itemize}



\newpage 
\setcounter{section}{0}

 \begin{table*}

\begin{center}
\caption{{\bf Fiducial parameters}}
\begin{tabular}{|c|c|c|c||c|c||c|c|c|c||c|c|c|}\hline
\multicolumn{4}{|c||}{Chunk} & \multicolumn{2}{|c||}{Ambient medium} & \multicolumn{4}{|c||}{BI-WI$^{1}$} & \multicolumn{3}{|c|}{CSE$^{2}$} \\\hline 
 $N_{\rm c}$ & $m_{\rm c}$ (gm) & $\Gamma_{\rm c}$ &  $\kappa_{\rm c}$  (cm$^{2}$ gm$^{-1}$) & $n_{\rm amb.}^{\rm ns}$ (cm$^{-3}$) & $\sigma_{\rm HH}$ (cm$^{2}$) & $\zeta_{\rm BI}$ &  $\beta_{\rm WI}$  & $\zeta_{\rm m-WI}$ & $\delta_{\rm m-WI}$  & $\delta_{\rm CSE}$ & $\gamma_{\rm CSE}$ & $\alpha_{\rm CSE}$\\\hline
$10^6$ & $10^{22.3}$  & $10^{2.5}$ & 0.1  &  $10^{-3}$& $10^{-27}$ & $10^{-1}$ & $10^{-1}$ & $10^2$ & 1.0 & $10^{-1}$ & $10^1$ & $0.0$\\\hline
\end{tabular}\\
\label{table:parameters}
\end{center}
$^{1}$ The Buneman-Weibel Instabilities phase.\\
$^{2}$ Coherent Synchrotron Emission phase.\\
$\mathbf{N_{\rm c}}$ is the total number of chunks per QN.\\
$\mathbf{m_{\rm c}}=M_{\rm QN}/N_{\rm c}$  is the chunk's mass with $M_{\rm QN}=N_{\rm c}m_{\rm c}$ the QN ejecta mass
(the NS outermost crust ejected during the QN). \\ 
$\mathbf{\Gamma_{\rm c}}$ is the Lorentz factor of the QN ejecta (the chunk's Lorentz factor). The ejecta's kinetic energy $\Gamma_{\rm c}\times(N_{\rm c}m_{\rm c})c^2$ erg is  
a few percents of the NS to QS conversion energy (see \S \ref{sec:QN}).\\
\boldmath{$\kappa_{\rm c}$} is the chunk's opacity.\\
  $\mathbf{n_{\rm amb.}^{\rm ns}}$ is the baryon number density of the ambient medium (representative of the ICM) in the NS frame.\\
  $\mathbf{\sigma_{\rm HH}}$ is the hadronic collision cross-section.\\ 
   $\mathbf{\zeta_{\rm BI}}$  is the percentage of the beam's electron energy (in the chunk's frame) converted to heating the chunk electrons by the BI.\\
     $\mathbf{\beta_{\rm WI}}=\beta_{\perp}/\beta_{\parallel}$ the ratio of transverse to longitudinal 
 thermal speed of electron chunks at the onset of the WI (Eq. (\ref{eq:lambdaWI})).\\
      $\mathbf{\zeta_{\rm m-WI}}$ sets the filament merging characteristic timescale (Eq. (\ref{eq:tmWI})).\\
       $\mathbf{\delta_{\rm m-WI}}$  controls the filament merging rate (Eq. (\ref{eq:filament-merging})).\\
        $\mathbf{\delta_{\rm CSE}}$ sets the CSE frequency  (Eq. (\ref{eq:nu-CSE})) which also sets the bunch's scaling parameter 
        $\mathbf{\delta_{\rm b}}$ (Eq. (\ref{eq:deltab})).\\
         $\mathbf{\gamma_{\rm CSE}}$ is the electron's Lorentz factor at CSE trigger during filament merging (Eq. (\ref{eq:nu-CSE})).\\
         $\mathbf{\alpha_{\rm CSE}}$ the positive power-law spectral index ($\alpha_{\rm CSE}=0.0$ corresponds to a flat spectrum).\\
\end{table*}

\clearpage
   
  \begin{table*}
\begin{center}
\caption{{\bf FRBs from ICM-QNe}: Key equations describing the properties (baryon number density, radius and sound
speed) of the collisionless QN chunks in the
ICM and the resulting CSE features (frequency, duration and fluence).
Also shown is the time since the QN, $t_{\rm cc}^{\rm obs.}$, and the time separation between emitting 
 chunks $\Delta t_{\rm repeat}^{\rm obs.}$ (see Appendix \ref{appendix:honeycomb}). The fiducial parameter values 
  are given in Table \ref{table:parameters}.}
\begin{tabular}{|c|c|c|}\hline
  \multicolumn{3}{|c|}{FRBs from ICM-QNe}\\\hline\hline
  \multicolumn{3}{|c|}{Collisionless chunk (``cc") properties}\\ \hline\hline
 Number density &  $n_{\rm cc}$ (cm$^{-3}$)&$\simeq  14.6\times \frac{m_{\rm c, 22.3}^{1/10}}{\kappa_{\rm c, -1}^{9/10}}\times (\sigma_{\rm HH, -27}\Gamma_{\rm c, 2.5}^{2}{n_{\rm amb., -3}^{\rm ns}})^{6/5}$  $\quad$ [Eq. (\ref{appendix:eq:cc2})]\\\hline
  Radius & $R_{\rm cc}$ (cm)& $\simeq  5.9\times 10^{14}\times\frac{(m_{\rm c, 22.3}\kappa_{\rm c, -1})^{3/10}}{ (\sigma_{\rm HH, -27}\Gamma_{\rm c, 2.5}^2{n_{\rm amb., -3}^{\rm ns}})^{2/5}}$ $\quad$ [Eq. (\ref{appendix:eq:cc3})]\\\hline
  Thermal speed & $\beta_{\rm cc}=v_{\rm cc}/c$  & $\simeq 1.6\times 10^{-2} \times (m_{\rm c, 22.3}\kappa_{\rm c, -1})^{1/10} (\sigma_{\rm HH, -27}\Gamma_{\rm c, 2.5}^{2}{n_{\rm amb.,-3}^{\rm ns}})^{1/5}$ $\quad$ [Eq. (\ref{appendix:eq:betaecc})] \\\hline
 Time since QN &   $t_{\rm cc}^{\rm obs.}$ (days) &$\simeq  2.6\times \frac{(1+z)f(\theta_{\rm c})}{\Gamma_{\rm c, 2.5}}\times  \frac{(m_{\rm c, 22.3}\kappa_{\rm c, -1})^{1/5}}{(\sigma_{\rm HH, -27}\Gamma_{\rm c, 2.5}^2{n_{\rm amb., -3}^{\rm ns}})^{3/5}}$  $\quad$ [Eq. (\ref{appendix:eq:tcc})]\\\hline
  \multicolumn{3}{|c|}{Coherent synchrotron emission (CSE) properties}\\ \hline\hline
  Frequency$^{1}$ & $\nu_{\rm CSE, max.}^{\rm obs.}(\theta_{\rm c})$ (GHz)& $\simeq 11.6\times \frac{1}{(1+z)f(\theta_{\rm c})}\times \delta_{\rm CSE, -1}\Gamma_{\rm c, 2.5}\gamma_{\rm CSE, 1}^2n_{\rm cc, 1}^{1/2}$   $\quad$ [Eq. (\ref{eq:nu-CSEobs})]\\\hline
 Width &  $\Delta t_{\rm FRB}^{\rm obs.}$ (ms) & $\simeq 2.4\times (1+z)f(\theta_{\rm c})\times \frac{\zeta_{\rm m-WI,3}}{\Gamma_{\rm c, 2.5}n_{\rm cc, 1}^{1/2}}    \times {\rm Min}(..,..)^{2}$  $\quad$ [Eq. (\ref{eq:dtFRB})]\\\hline
 Fluence$^{3}$ &  $\mathcal{F}(\theta_{\rm c},0)$ (Jy ms) & $\simeq 810\ {\rm Jy\ ms}\   \frac{1}{f(\theta_{\rm c})^2d_{\rm L, 27.5}^2}\times  \frac{\zeta_{\rm BI, -1}\beta_{\rm WI, -1}}{\delta_{\rm CSE, -1}\gamma_{\rm CSE,1}^2}\times \frac{\Gamma_{\rm c, 2.5}^4 R_{\rm cc, 15}^2 {n_{\rm amb., -3}^{\rm ns}}}{n_{\rm cc, 1}\beta_{\rm cc, -2}}$  $\quad$ [Eq. (\ref{appendix:eq:fluence-flat-4})]\\\hline 
 Repeat  time$^{4}$ &  $\Delta t_{\rm repeat}^{\rm obs.}$  (days) &$\simeq  1.3\times (1+z)\times \frac{1}{N_{\rm c, 6}}\times  \left(\frac{m_{\rm c, 22.3}\kappa_{\rm c, -1}}{\sigma_{\rm HH, -27}^3\Gamma_{\rm c, 2.5}{n_{\rm amb., -3}^{\rm ns}}^{3}}\right)^{1/5}$  $\quad$ (Eq. (\ref{eq:dtrepeat}))\\\hline
\end{tabular}\\
\label{table:ICM-equations}
\end{center}
$^{1}$ The frequency drifts in time to a minimum value  set by the chunk' plasma frequency $\nu_{\rm p, e}^{\rm obs.}(\theta_{\rm c})\simeq 18\ {\rm MHz}\times \frac{1}{(1+z)f(\theta_{\rm c})}\times \Gamma_{\rm c, 2.5}n_{\rm cc, 1}^{1/2}$.\\
$^{2}$ ${\rm Min}\left[\left(  \left(642.7\delta_{\rm CSE, -1}\gamma_{\rm CSE,1}^2\right)^{\frac{1}{\delta_{\rm m-WI}}} -1\right) ,\left(\left( \frac{\nu_{\rm CSE, max.}^{\rm obs.}(\theta_{\rm c})}{\nu_{\rm min.}^{\rm det.}} \right)^{1/\delta_{\rm m-WI}} - \left(\frac{\nu_{\rm CSE, max.}^{\rm obs.}(\theta_{\rm c})}{\nu_{\rm max.}^{\rm det.}} \right)^{1/\delta_{\rm m-WI}}\right)\right]$; see \S \ref{sec:FRB-ICM-QNe}.\\
$^{3}$  $F(\theta_{\rm c},\delta_{\rm m-WI},0)=\mathcal{F}(\theta_{\rm c},0)\times \mathcal{G}(\theta_{\rm c},\delta_{\rm m-WI},0)$ for the flat spectrum case ($\alpha_{\rm CSE}=0$) with $\mathcal{G}(\theta_{\rm c},\delta_{\rm m-WI},0)$ given  in Eq. (\ref{appendix:eq:fluence-flat-2}) and Table \ref{table:G-fluence}.\\
$^{4}$ Independent of the viewing angle $\theta_{\rm c}$ (i.e. $f(\theta_{\rm c})$) due to geometry and the spatial distribution
of chunks (see Appendix \ref{appendix:honeycomb} and Eq. (\ref{eq:Deltac})).\\
\end{table*}

%

  \clearpage

\begin{table*}
\begin{center}
\caption{{\bf $\mathcal{G}(\theta_{\rm c},\delta_{\rm m-WI},0)$} (see Eq. (\ref{appendix:eq:fluence-flat-2})) values for  fiducial parameters.}
\resizebox{0.5\textwidth}{!}{
\begin{tabular}{|c||ccc|}\hline
\multicolumn{4}{|c|}{$N_{\rm c}=10^6$ ($N_{\rm c}=10^5$)} \\\hline
 & Primary & Secondary$^1$ & Tertiary$^1$  \\\hline
Arecibo &  5.5 (1.0)& 2.0 (N/A)$^3$& 0.12 (N/A) \\\hline
Parkes &  6.6 (1.2) & 2.4 (N/A) & 0.12 (N/A) \\\hline
ASKAP &  7.1 (1.3) & 2.6 (N/A) & 0.17 (N/A)  \\\hline
CHIME &  76.2 (13.7) & 27.3 (0.9) & 1.8 (N/A) \\\hline
LOFAR$^2$ & $10^3$ (179.8) & 358.3 (12.2) & 23.6 (0.13) \\\hline
\end{tabular}
}
\label{table:G-fluence}
\end{center}
$^1$ These are chunks with a  similar $\theta_{\rm c}$ but different azimuths.\\
$^2$ In all tables, the LOFAR's fluence listed is for the high-band antenna bandwidth (see Table \ref{table:detectors}).\\
$^3$ ``N/A" (not applicable) means the maximum CSE frequency, $\nu_{\rm CE, max.}^{\rm obs.}(\theta_{\rm c})$, is below the detector's
minimum frequency $\nu_{\rm min.}^{\rm det.}$  (see Table \ref{table:detectors}).\\
\end{table*}

\clearpage

\begin{table*}
\begin{center}
\caption{{\bf FRBs from ICM-QNe}: FRB properties (frequency, duration and fluence; see  Table \ref{table:ICM-equations}) for the detectors listed in Table \ref{table:detectors}. The redshift is $z=0.2$
which corresponds to a luminosity distance of $d_{\rm L}\simeq 1$ Gpc. The time delay between repeats is $\Delta t_{\rm repeat}^{\rm obs.}$.
The fluences per detector  are 
 given with the shaded cells showing the fluence values within detector's sensitivity  (listed in Table \ref{table:detectors}).}
\resizebox{\textwidth}{!}{
\begin{tabular}{|c||c|c|c||c|c|c|}\hline
  Varied parameter$^{1}$ & \multicolumn{3}{|c||}{{\bf Box A :} ($N_{\rm c}=10^5,\Gamma_{\rm c}=10^{3})$} & \multicolumn{3}{|c|}{{\bf Box D :} ($N_{\rm c}=10^6,\Gamma_{\rm c}=10^{3})$}\\ \hline\hline
 Chunk type & 1 (primary) & 6 (secondaries)$^{2}$ & 12 (tertiaries)$^{2}$ & 1 (primary) & 6 (secondaries) & 12 (tertiaries)\\\hline
$f(\bar{\theta}_{\rm c})$  & 18.78   & 97.8  & 667.95
 & 2.78   & 10.68  & 67.69 \\\hline
  $\nu_{\rm CSE, max.}^{\rm obs.}(\theta_{\rm c})\ {\rm (GHz)}\rightarrow {\nu_{\rm p, e}^{\rm obs.}}^{3}$ (MHz) & 
   $\simeq 7.8\rightarrow 12.1$ & $\simeq 1.5\rightarrow 2.3$ & $\simeq 0.22\rightarrow 0.34$  &
   $\simeq 52.7\rightarrow 82.0$ & $\simeq 13.7\rightarrow 21.3$ & $\simeq 2.2\rightarrow 3.4$\\\hline
  $t_{\rm m-WI}^{\rm obs.}$ (ms) &  
   $\simeq 3.5$  &  $\simeq 18.4$ & $\simeq 125.7$ &
    $\simeq 0.52$ & $\simeq 2.0$& $\simeq 12.7$ \\\hline
   Fluence (Jy ms) [Arecibo] &  
    \cellcolor{blue!25}$\simeq  0.7$ &   $\simeq 8.1\times 10^{-4}$ & N/A$^4$ &
     \cellcolor{blue!25}\cellcolor{blue!25}$\simeq 1.5\times 10^3$ & \cellcolor{blue!25}$\simeq 7.0$ & $\simeq 4.3\times 10^{-3}$\\\hline 
  Fluence (Jy ms) [Parkes] & 
   $\simeq 0.9$  &  $\simeq 8.4\times 10^{-4}$& N/A &
    \cellcolor{blue!25}$\simeq 1.8\times 10^3$ & \cellcolor{blue!25}$\simeq 8.5$ & $\simeq 5.2\times 10^{-3}$\\\hline 
     Fluence (Jy ms) [ASKAP] & 
   $\simeq 1.0$  &  $\simeq 1.3\times 10^{-3}$& N/A &
    \cellcolor{blue!25}$\simeq 2\times 10^3$ & $\simeq 9.1$ & $\simeq 5.6\times 10^{-3}$\\\hline 
  Fluence (Jy ms) [CHIME] &  
   \cellcolor{blue!25}$\simeq 10.2$  &  $\simeq 0.014$& N/A &
    \cellcolor{blue!25}$\simeq 2.1\times 10^4$ & \cellcolor{blue!25}$\simeq 97.3$ & $\simeq 0.06$\\\hline 
  Fluence (Jy ms) [LOFAR] &  
   $\simeq 133.7$  &  $\simeq 0.18$&  $\simeq 7.1\times 10^{-5}$ &
    \cellcolor{blue!25}$\simeq 2.8\times 10^{5}$ & \cellcolor{blue!25}$\simeq 1.3\times 10^3$ & $\simeq 0.8$\\\hline 
$\Delta t_{\rm repeat}^{\rm obs.}$ (days)  &
  $\simeq 12.5$ & $\simeq 12.5$ &  $\simeq 12.5$  &
   $\simeq 1.3$ & $\simeq 1.3$& $\simeq 1.3$\\\hline\hline
  Varied parameter$^{1}$ & \multicolumn{3}{|c||}{{\bf Box B :} ($N_{\rm c}=10^5,\Gamma_{\rm c}=10^{2.5})$} & \multicolumn{3}{|c|}{{\bf Box E :} ($N_{\rm c}=10^6,\Gamma_{\rm c}=10^{2.5})$}\\ \hline\hline
 Chunk type & 1 (primary) & 6 (secondaries) & 12 (tertiaries) & 1 (primary) & 6 (secondaries) & 12 (tertiaries)\\\hline
$f(\bar{\theta}_{\rm c})$  & 2.78   & 10.68  & 67.7 &
 1.18   & 1.97  & 7.67 \\\hline
  $\nu_{\rm CSE, max.}^{\rm obs.}(\theta_{\rm c})\ {\rm (GHz)}\rightarrow {\nu_{\rm p, e}^{\rm obs.}}$ (MHz) & 
  $\simeq 4.2\rightarrow 6.5$ & $\simeq 1.1\rightarrow 1.7$ & $\simeq 0.18\rightarrow 0.27$  & 
   $\simeq 9.9\rightarrow 15.4$  & $\simeq 5.9\rightarrow 9.2$ & $\sim 1.5\rightarrow 2.4$  \\\hline 
  $t_{\rm m-WI}^{\rm obs.}$ (ms) &  
  $\simeq 6.6$ & $\simeq 25.3$& $\simeq 160.4$ &
   $\simeq 2.8$  &  $\simeq 4.7$ & $\simeq 18.2$ \\\hline
   Fluence (Jy ms) [Arecibo] &  
    \cellcolor{blue!25}$\simeq 15.3$ & N/A & N/A &
    \cellcolor{blue!25}$\simeq 473.2$  &  \cellcolor{blue!25}$\simeq 60.7$& \cellcolor{blue!25}$\simeq 0.24$  \\\hline 
  Fluence (Jy ms) [Parkes] & 
   \cellcolor{blue!25}$\simeq 18.5$ & N/A & N/A  &
   \cellcolor{blue!25}$\simeq 571.5$  &  \cellcolor{blue!25}$\simeq 73.3$& $\simeq 0.24$ \\\hline 
     Fluence (Jy ms) [ASKAP] & 
      \cellcolor{blue!25}$\simeq 19.9$ & N/A & N/A &
   \cellcolor{blue!25}$\simeq 617.1$  &  \cellcolor{blue!25}$\simeq 79.2$& $\simeq 0.34$     \\\hline 
  Fluence (Jy ms) [CHIME] &  
  \cellcolor{blue!25}$\simeq 212.5$ & \cellcolor{blue!25}$\simeq 1.0$ & N/A &
   \cellcolor{blue!25}$\simeq 6.6\times 10^3$  &  \cellcolor{blue!25}$\simeq 843.8$& \cellcolor{blue!25}$\simeq 3.7$  \\\hline 
  Fluence (Jy ms) [LOFAR] &  
  \cellcolor{blue!25}$\simeq 2.8\times 10^3$ & $\simeq 12.8$ & $\simeq 3.5\times 10^{-3}$  &
   \cellcolor{blue!25}$\simeq 8.6\times 10^{4}$  &  \cellcolor{blue!25}$\simeq 1.1\times 10^4$ & $\simeq 48.0$  \\\hline 
$\Delta t_{\rm repeat}^{\rm obs.}$ (days)  &
$\simeq 15.8$ & $\simeq 15.8$& $\simeq 15.8$ &
  $\simeq 1.6$ & $\simeq 1.6$ &  $\simeq 1.6$  \\\hline
  Varied parameter$^{1}$ & \multicolumn{3}{|c||}{{\bf Box C :} ($N_{\rm c}=10^5,\Gamma_{\rm c}=10^{2})$} & \multicolumn{3}{|c|}{{\bf Box F :} ($N_{\rm c}=10^6,\Gamma_{\rm c}=10^{2})$}\\ \hline\hline
 Chunk type & 1 (primary) & 6 (secondaries) & 12 (tertiaries) & 1 (primary) & 6 (secondaries) & 12 (tertiaries)\\\hline
$f(\bar{\theta}_{\rm c})$  & 1.18   & 1.97  & 7.67 &
 1.018   & 1.097  & 1.667 \\\hline
  $\nu_{\rm CSE, max.}^{\rm obs.}(\theta_{\rm c})\ {\rm (GHz)}\rightarrow {\nu_{\rm p, e}^{\rm obs.}}$ (MHz)& 
  $\simeq 0.78\rightarrow 1.21$ & $\simeq 0.47\rightarrow 0.73$ & $\simeq 0.12\rightarrow 0.19$  & 
   $\simeq 0.9\rightarrow 1.4$  & $\simeq 0.8\rightarrow 1.3$ & $\sim 0.6\rightarrow 0.9$  \\\hline 
  $t_{\rm m-WI}^{\rm obs.}$ (ms) &  
  $\simeq 35.13$ & $\simeq 58.69$& $\simeq 228.74$ &
   $\simeq 30.4$  &  $\simeq 32.7$ & $\simeq 49.7$ \\\hline
   Fluence (Jy ms) [Arecibo] &  
    N/A & N/A & N/A &
    N/A  &  N/A& N/A  \\\hline 
  Fluence (Jy ms) [Parkes] & 
   N/A & N/A & N/A  &
   N/A  &  N/A& N/A \\\hline 
     Fluence (Jy ms) [ASKAP] & 
      N/A & N/A & N/A &
   N/A  &  N/A& N/A     \\\hline 
  Fluence (Jy ms) [CHIME] &  
  \cellcolor{blue!25}$\simeq 63.2$ & \cellcolor{blue!25}$\simeq 0.74$ & N/A &
   \cellcolor{blue!25}$\simeq 117.9$  &  \cellcolor{blue!25}$\simeq 87.4$& \cellcolor{blue!25}$\simeq 5.1$  \\\hline 
  Fluence (Jy ms) [LOFAR] &  
  $\simeq 863.6$ & $\simeq 110.8$ & $\simeq 0.012$  &
   \cellcolor{blue!25}$\simeq 1.5\times 10^{3}$  &  \cellcolor{blue!25}$\simeq 1.1\times 10^3$&  $\simeq 215.2$  \\\hline 
$\Delta t_{\rm repeat}^{\rm obs.}$ (days)  &
$\simeq 19.8$ & $\simeq 19.8$& $\simeq 19.8$ &
  $\simeq 2.0$ & $\simeq 2.0$ &  $\simeq 2.0$  \\\hline
\end{tabular}
}\\
\label{table:ICM-QNe-1}
\end{center}
$^{1}$ Other parameters are kept to their fiducial values listed in Table \ref{table:parameters}.\\
$^{2}$ Similar $\theta_{\rm c}$ but different azimuths.\\
$^{3}$ The arrow indicates frequency drifts in time to a minimum value  
given by the chunk's plasma frequency $\nu_{\rm p, e}^{\rm obs.}(\theta_{\rm c})\simeq \frac{18\ {\rm MHz}}{(1+z)f(\theta_{\rm c})}\times \Gamma_{\rm c, 2.5}n_{\rm cc, 1}^{1/2}$ (see Eq. (\ref{eq:nu-plasma-obs})).\\
$^{4}$ ``N/A" (not applicable) means the maximum CSE frequency, $\nu_{\rm CE, max.}^{\rm obs.}(\theta_{\rm c})$, is below the detector's
minimum frequency $\nu_{\rm min.}^{\rm det.}$.\\
\end{table*}

\clearpage

\begin{table*}
\begin{center}
\caption{{\bf Simulations}: example of a repeating FRB with the time delays between bursts  of minutes 
 and a few hours shown as shaded cells. The chunk mass distribution has a mean of $\bar{m}_{\rm c}=10^{22.32}$ gm and standard deviation $\sigma_{\rm \log{m_{\rm c}}}=1.0$.}
 
  \begin{minipage}{0.8\textwidth}
  \begin{center}
  {\bf Parameters}
  \end{center}
\end{minipage}%
  \vskip 0.05in
\begin{tabular}{cccccccccccc}\hline
     z        &  $d_{\rm L}$ (Gpc)       &  $N_{\rm c}$       &  $\log \Gamma_{\rm c}$       &  $\log \bar{m}_{\rm c}$ (gm)      &   $n_{\rm amb.}^{\rm ns}$ (cm$^{-3}$) & $\delta_{\rm m-WI}$   & $\delta_{\rm CSE}$     &  $\log \gamma_{\rm CSE}$    & $\zeta_{\rm BI}$   &  $\beta_{\rm WI}$ & $\log \sigma_{\rm HH}$ (cm$^{2}$) \\\hline 
     0.20   &    0.99   &    5.0E5    & 2.02    &   22.32    &     6.00E-3  &  1.00   &    0.10    &   10.00    &  0.10    &   0.10   &    -27.00 \\\hline 
 \end{tabular}
 ~\\
  
   \vskip 0.05in
\begin{minipage}{0.8\textwidth}
  \begin{center}
  {\bf Detections} ($\theta_{\rm c}(\# 0)=2.19$E-3)$^{1}$
  \end{center}
\end{minipage}%
  \vskip 0.05in
   \hskip -0.2in
  \scalebox{0.95}{
\begin{tabular}{cccccccccc}\hline
    \#        &  $\log{m_{\rm c}}$  (gm)     &      $\Delta \theta_{\rm c}$$^{2}$       &  $f(\theta_{\rm c})$      & 
    $t_{\rm OA}^{\rm obs.}$ (days)      & $\Delta t_{\rm OA}^{\rm obs.}$ (days)$^{3}$    & Frequency (MHz)$^{4}$     &  Width (ms) &  Fluence (Jy ms)$^{5}$\\\hline  
     0&21.01&0.00 &1.05 &0.00 &0.00&2.52E3&1.10&CHIME (28.32)
\\\hline
1&21.74&8.76E-4&1.11&3.17&3.17&2.61E3&1.06&CHIME (53.92)
\\\hline
2&21.25&4.30E-3&1.61 &4.80 &1.63 &1.70E3&1.64&CHIME (6.88)
\\\hline
3&21.70&-1.62E-3&1.37&5.34&0.54&2.10E3&1.33&CHIME (21.82)
\\\hline
4&21.99&7.65E-4&1.48&8.13&2.79&2.01E3&1.38&CHIME (22.62)
\\\hline
5&21.28&3.48E-3&2.12&8.69&0.56&1.29E3&2.15&CHIME (2.36)
\\\hline
6&20.85&2.10E-3&2.64&8.99&0.30&9.87E2&2.82&CHIME (0.59)
\\\hline
7&22.67&-7.87E-3&1.20&9.77&0.78&2.68E3&1.04&CHIME (1.13E2)
\\\hline
8&21.14&7.17E-3&2.45&10.03&0.26&1.10E3&2.53&CHIME (1.12)
\\\hline
9&21.99&-2.58E-3&1.87&12.17&2.13&1.59E3&1.75&CHIME (8.81)
\\\hline
10&22.92&-2.85E-3&1.40&14.95&\cellcolor{blue!25}2.79&2.36E3&1.18&CHIME (82.03)
\\\hline
11&22.51&1.95E-3&1.70&15.07&\cellcolor{blue!25}0.12&1.85E3&1.50&CHIME (23.20)
\\\hline
12&22.19&2.17E-3&2.14&17.00&\cellcolor{blue!25}1.94&1.42E3&1.96&CHIME (6.46)
\\\hline
13&22.30&-4.35E-4&2.05&17.11&\cellcolor{blue!25}0.11&1.51E3&1.85&CHIME (8.82)
\\\hline
14&21.20&4.97E-3&3.40&17.12&\cellcolor{blue!25}4.96E-3&7.99E2&3.48&CHIME (0.33)
\\\hline
15&21.03&1.90E-3&4.06&19.65&2.53&6.55E2&4.25&CHIME (0.12)
\\\hline
16&21.76&-2.87E-3&3.09&21.37&1.72&9.37E2&2.97&CHIME (0.91)
\\\hline
17&21.64&1.12E-3&3.45&22.96&1.59&8.28E2&3.36&CHIME (0.51)
\\\hline
18&22.03&-1.60E-3&2.95&23.62&0.66&1.01E3&2.74&CHIME (1.50)
\\\hline
19&22.21&1.46E-4&2.99&26.64&3.01&1.02E3&2.73&CHIME (1.72)
\\\hline
20&23.31&-4.76E-3&1.82&27.10&\cellcolor{blue!25}0.46&1.90E3&1.46&CHIME (44.73)
\\\hline
21&22.74&2.50E-3&2.37&27.10&\cellcolor{blue!25}3.63E-3&1.37E3&2.04&CHIME (8.04)
\\\hline
22&22.17&3.07E-3&3.24&28.90&1.80&9.37E2&2.97&CHIME (1.21)
\\\hline
23&22.76&-2.38E-3&2.55&30.01&1.11&1.27E3&2.18&CHIME (6.20)
\\\hline
24&21.70&5.19E-3&4.22&30.62&0.60&6.82E2&4.08&CHIME (0.24)
\\\hline
25&22.64&-3.88E-3&2.91&32.96&2.35&1.10E3&2.53&CHIME (3.16)
\\\hline
26&22.59&4.72E-4&3.05&33.97&1.00&1.04E3&2.67&CHIME (2.47)
\\\hline
27&22.20&2.30E-3&3.81&35.61&1.64&7.99E2&3.48&CHIME (0.65)
\\\hline
28&22.83&-2.89E-3&2.88&36.07&\cellcolor{blue!25}0.46&1.14E3&2.44&CHIME (4.12)
\\\hline
29&23.09&-1.17E-3&2.55&36.18&\cellcolor{blue!25}0.11&1.32E3&2.10&CHIME (9.04)
\\\hline
30&23.91&-3.42E-3&1.78&36.77&0.59&2.08E3&1.34&CHIME (96.97)
\\\hline
31&21.97&0.01&5.09&44.29&\cellcolor{blue!25}7.52&5.83E2&4.77&CHIME (0.14)
\\\hline
32&22.12&-8.00E-4&4.75&44.33&\cellcolor{blue!25}0.04&6.35E2&4.38&CHIME (0.23)
\\\hline
33&22.42&-1.15E-3&4.29&46.18&1.85&7.27E2&3.82&CHIME (0.51)
\\\hline
34&22.04&2.36E-3&5.27&47.83&1.65&5.68E2&4.90&CHIME (0.12)
\\\hline
35&22.72&-2.67E-3&4.18&52.36&4.52&7.74E2&3.59&CHIME (0.82)
\\\hline
36&22.58&1.39E-3&4.73&56.11&3.75&6.73E2&4.13&CHIME (0.41)
\\\hline
37&22.81&-2.68E-4&4.62&61.52&5.41&7.08E2&3.93&CHIME (0.60)
\\\hline
38&22.70&8.87E-4&4.98&63.17&1.66&6.47E2&4.30&CHIME (0.37)
\\\hline
39&22.63&7.70E-4&5.32&65.44&\cellcolor{blue!25}2.27&6.02E2&4.62&CHIME (0.25)
\\\hline
40&23.51&-4.55E-3&3.55&65.59&\cellcolor{blue!25}0.15&9.99E2&2.78&CHIME (3.93)
\\\hline
41&23.30&1.17E-3&3.95&66.48&0.89&8.74E2&3.18&CHIME (1.99)
\\\hline
42&23.55&-9.81E-4&3.61&68.21&1.73&9.86E2&2.82&CHIME (3.83)
\\\hline
43&23.48&7.73E-4&3.88&71.30&3.09&9.10E2&3.06&CHIME (2.65)
\\\hline
44&23.31&1.82E-3&4.57&78.05&6.75&7.57E2&3.67&CHIME (1.12)
\\\hline
45&22.86&3.88E-3&6.29&88.22&10.17&5.22E2&5.32&CHIME (0.13)
\\\hline
46&23.55&-3.74E-3&4.63&89.05&0.83&7.68E2&3.62&CHIME (1.41)
\\\hline
47&23.12&2.65E-3&5.77&91.50&2.45&5.86E2&4.74&CHIME (0.31)
\\\hline
48&23.26&-4.42E-4&5.57&94.60&3.10&6.18E2&4.50&CHIME (0.44)
\\\hline
49&23.32&2.22E-4&5.67&99.15&4.55&6.11E2&4.55&CHIME (0.44)
\\\hline
50&23.84&-1.97E-3&4.81&1.07E2&8.13&7.64E2&3.64&CHIME (1.68)
\\\hline
51&23.76&3.11E-3&6.21&1.35E2&27.92&5.87E2&4.74&CHIME (0.48)
\\\hline
52&24.03&3.72E-3&8.16&2.04E2&69.13&4.61E2&6.04&CHIME (0.12)
\\\hline
53&24.92&-1.16E-3&7.51&2.87E2&82.75&5.54E2&5.02&CHIME (0.79)
\\\hline
     \\
\end{tabular}
\label{table:FRBs-repeating-minutes}
}\\
\end{center}
$^{1}$ $\theta_{\rm c}(\# 0)$ is the viewing angle in radians of the first detected chunk.\\
$^{2}$ $\Delta \theta_{\rm c}$ is the difference between the current chunk's $\theta_{\rm c}$ and the previous one that arrived.\\
$^{3}$ $\Delta t_{\rm OA}^{\rm obs.}$ is the time-delay (difference in time-of-arrival, $t_{\rm OA}^{\rm obs.}$)
between successive bursts.\\
$^{4}$ Shown  is the maximum CSE frequency $\nu_{\rm CSE}^{\rm obs.}(\theta_{\rm c})$ (Eq. (\ref{eq:nu-CSEobs})). \\
$^{5}$ Only detectors with fluence above sensitivity threshold (see Table \ref{table:detectors}) are shown. 
\end{table*}

\clearpage

\begin{table*}
\vskip 3.0in
\begin{center}
\caption{{\bf Simulations}: example of a repeating FRB yielding the waterfall plot
 in Figure \ref{figure:FRBs-waterfall-scenarios}.}
 
  \begin{minipage}{0.5\textwidth}
  \begin{center}
  {\bf Parameters}
  \end{center}
\end{minipage}%
  \vskip 0.05in
\begin{tabular}{cccccccccccccc}\hline
     z        &  $d_{\rm L}$ (Gpc)       &  $N_{\rm c}$       &  $\log \Gamma_{\rm c}$       &  $m_{\rm c}$ (gm)      &   $n_{\rm amb.}^{\rm ns}$ (cm$^{-3}$) & $\delta_{\rm m-WI}$   & $\delta_{\rm CSE}$     &  $\gamma_{\rm CSE}$    & $\zeta_{\rm BI}$   &  $\beta_{\rm WI}$ & $\log \sigma_{\rm HH}$ (cm$^{2}$) \\\hline 
     0.20   &    0.99   &    1.0E5    & 2.30    &   22.75    &    1.00E-3  &  1.00   &    0.10    &   10.00    &  0.10    &   0.10   &    -27.00 \\\hline 
 \end{tabular}
 ~\\
  
   \vskip 0.05in
\begin{minipage}{0.5\textwidth}
  \begin{center}
  {\bf Detections} ($\theta_{\rm c}(\# 0)=5.47$E-3)
  \end{center}
\end{minipage}%
  \vskip 0.05in
\begin{tabular}{ccccccccccc}\hline
    \#        &       $\Delta \theta_{\rm c}$       &  $f(\theta_{\rm c})$      & 
    $t_{\rm OA}^{\rm obs.}$ (days)      & $\Delta t_{\rm OA}^{\rm obs.}$ (days)    & Frequency (MHz)     &  Width (ms) &  Fluence (Jy ms)\\\hline         
     0           &      0.00        &  2.19        &  0.00        &  0.00        &  2.03E3      &  1.37        &  ASKAP (13.19)\\\hline  
                 &                                         &              &              &              &              &              &  CHIME (1.41E2)\\\hline  
                 &                                       &              &              &              &              &              &  LOFAR (1.85E3)\\\hline  
     1           &      2.05E-3     &  3.25        &  11.10       &  11.10       &  1.37E3      &  2.03        &  CHIME (28.96)\\\hline  
     2           &      1.84E-3     &  4.49        &  24.04       &  12.94       &  9.90E2      &  2.81        &  CHIME (7.98)\\\hline  
     3           &      7.71E-4     &  5.09        &  30.30       &  6.26        &  8.74E2      &  3.18        &  CHIME (4.84)\\\hline  
     4           &      4.54E-3     &  9.57        &  77.21       &  46.91       &  4.65E2      &  5.99        &  CHIME (0.18)\\\hline  
     \\
\end{tabular}
\label{table:FRBs-waterfall-scenarios}
\end{center}
\end{table*}

\clearpage

\begin{table*}
\vskip 3.0in
\begin{center}
\caption{{\bf Simulations}: example of a repeating FRB yielding  the waterfall plot
in Figure \ref{figure:FRBs-waterfall-scenarios-2}.}
 
  \begin{minipage}{0.5\textwidth}
  \begin{center}
  {\bf Parameters}
  \end{center}
\end{minipage}%
  \vskip 0.05in
\begin{tabular}{cccccccccccccc}\hline
     z        &  $d_{\rm L}$ (Gpc)       &  $N_{\rm c}$       &  $\log \Gamma_{\rm c}$       &  $\log m_{\rm c}$ (gm)      &  $n_{\rm amb.}^{\rm ns}$ (cm$^{-3}$) & $\delta_{\rm m-WI}$   & $\delta_{\rm CSE}$     &  $\gamma_{\rm CSE}$    & $\zeta_{\rm BI}$   &  $\beta_{\rm WI}$ & $\log \sigma_{\rm HH}$ (cm$^{2}$) \\\hline 
     0.20   &    0.99   &    1.0E5    & 2.10    &   22.95    &    1.00E-3  &  1.00   &    0.10    &   10.00    &  0.10    &   0.10   &    -27.00 \\\hline 
 \end{tabular}
 ~\\
  
   \vskip 0.05in
\begin{minipage}{0.5\textwidth}
  \begin{center}
  {\bf Detections} ($\theta_{\rm c}(\# 0)=5.47$E-3)
  \end{center}
\end{minipage}%
  \vskip 0.05in
\begin{tabular}{cccccccccc}\hline
    \#        &     $\Delta \theta_{\rm c}$       &  $f(\theta_{\rm c})$      & 
    $t_{\rm OA}^{\rm obs.}$ (days)      & $\Delta t_{\rm OA}^{\rm obs.}$ (days)    & Frequency (MHz)     &  Width (ms) &  Fluence (Jy ms)\\\hline         
     0          &      0.00        &  1.47        &  0.00        &  0.00        &  1.12E3      &  2.48        &  CHIME (1.37E2)\\\hline
                 &                                         &              &              &              &              &              &  LOFAR (1.80E3)\\\hline
     1           &      2.05E-3     &  1.90        &  13.35       &  13.35       &  8.71E2      &  3.19        &  CHIME (49.94)\\\hline
     2           &      1.84E-3     &  2.39        &  28.90       &  15.55       &  6.91E2      &  4.02        &  CHIME (19.35)\\\hline
     3           &      7.71E-4     &  2.63        &  36.42       &  7.52        &  6.29E2      &  4.42        &  CHIME (12.56)\\\hline  
     \\
\end{tabular}
\label{table:FRBs-waterfall-scenarios-2}
\end{center}
\end{table*}

\clearpage

\begin{table*}
\vskip 3.0in
\begin{center}
\caption{{\bf Simulations}: example of a repeating FRB yielding  the waterfall plot
in Figure \ref{figure:FRBs-waterfall-scenarios-3}.}
 
  \begin{minipage}{0.5\textwidth}
  \begin{center}
  {\bf Parameters}
  \end{center}
\end{minipage}%
  \vskip 0.05in
\begin{tabular}{cccccccccccccc}\hline
     z        &  $d_{\rm L}$ (Gpc)       &  $N_{\rm c}$       &  $\log \Gamma_{\rm c}$       &  $\log m_{\rm c}$ (gm)      &  $n_{\rm amb.}^{\rm ns}$ (cm$^{-3}$) & $\delta_{\rm m-WI}$   & $\delta_{\rm CSE}$     &  $\gamma_{\rm CSE}$    & $\zeta_{\rm BI}$   &  $\beta_{\rm WI}$ & $\log \sigma_{\rm HH}$ (cm$^{2}$) \\\hline 
     0.20   &    0.99   &    1.0E5    & 2.00    &   23.05    &    1.00E-3  &  1.00   &    0.10    &   10.00    &  0.10    &   0.10   &    -27.00 \\\hline 
 \end{tabular}
 ~\\
  
   \vskip 0.05in
\begin{minipage}{0.5\textwidth}
  \begin{center}
  {\bf Detections} ($\theta_{\rm c}(\# 0)=5.47$E-3)
  \end{center}
\end{minipage}%
  \vskip 0.05in
\begin{tabular}{cccccccccc}\hline
    \#        &     $\Delta \theta_{\rm c}$       &  $f(\theta_{\rm c})$      & 
    $t_{\rm OA}^{\rm obs.}$ (days)      & $\Delta t_{\rm OA}^{\rm obs.}$ (days)    & Frequency (MHz)     &  Width (ms) &  Fluence (Jy ms)\\\hline         
     0          &      0.00        &  1.30        &  0.00        &  0.00        &  7.75E2      &  3.59        &  CHIME (1.01E2)\\\hline 
                 &                                        &              &              &              &              &              &  LOFAR (1.33E3)\\\hline 
     1           &      2.05E-3     &  1.57        &  14.63       &  14.63       &  6.43E2      &  4.32        &  CHIME (45.16)\\\hline 
     2           &      1.84E-3     &  1.88        &  31.69       &  17.05       &  5.37E2      &  5.18        &  CHIME (17.67)\\\hline 
     3           &      7.71E-4     &  2.03        &  39.94       &  8.25        &  4.97E2      &  5.60        &  CHIME (10.73)\\\hline 
\\
\end{tabular}
\label{table:FRBs-waterfall-scenarios-3}
\end{center}
\end{table*}


\clearpage

\begin{table*}
\vskip 3.0in
\begin{center}
\caption{{\bf Simulations}: example of a $\sim$ 16-day period  FRB.
The chunk mass distribution has a mean of $\bar{m}_{\rm c}=10^{22.64}$ gm and variance $\sigma_{\rm m}=1.0$.}
 
  \begin{minipage}{0.5\textwidth}
  \begin{center}
  {\bf Parameters}
  \end{center}
\end{minipage}%
  \vskip 0.05in
\begin{tabular}{cccccccccccccc}\hline
     z        &  $d_{\rm L}$ (Gpc)       &  $N_{\rm c}$       &  $\log \Gamma_{\rm c}$       &  $\log \bar{m}_{\rm c}$ (gm)      &   $n_{\rm amb.}^{\rm ns}$ (cm$^{-3}$) & $\delta_{\rm m-WI}$   & $\delta_{\rm CSE}$     &  $\gamma_{\rm CSE}$    & $\zeta_{\rm BI}$   &  $\beta_{\rm WI}$ & $\log \sigma_{\rm HH}$ (cm$^{2}$) \\\hline 
     0.20   &    0.99   &    1.01E5    & 2.41    &   22.64    &     1.00E-3  &  1.00   &    0.10    &   10.00    &  0.10    &   0.10   &    -27.00 \\\hline 
 \end{tabular}
 ~\\
  
   \vskip 0.05in
\begin{minipage}{0.5\textwidth}
  \begin{center}
  {\bf Detections} ($\theta_{\rm c}(\# 0)=5.05$E-3)
  \end{center}
\end{minipage}%
  \vskip 0.05in
\begin{tabular}{cccccccccc}\hline
    \#        &  $\log{m_{\rm c}}$ (gm)      &      $\Delta \theta_{\rm c}$       &  $f(\theta_{\rm c})$      & 
    $t_{\rm OA}^{\rm obs.}$ (days)      & $\Delta t_{\rm OA}^{\rm obs.}$ (days)    & Frequency (MHz)     &  Width (ms) &  Fluence (Jy ms)\\\hline         
     0           &  23.42              &  0.00        &  2.66        &  0.00        &  0.00        &  3.09E3      &  0.90        &  ASKAP (35.25)\\\hline
                 &                                  &              &              &              &              &              &              &  CHIME (3.76E2)\\\hline
                 &                                     &              &              &              &              &              &              &  LOFAR (4.93E3)\\\hline
     1           &  24.11       &      9.77E-4     &  3.36        &  16.34       &  16.34       &  2.65E3      &  1.05        &  ASKAP (30.51)\\\hline
                 &                                      &              &              &              &              &              &              &  CHIME (3.25E2)\\\hline
                 &                                     &              &              &              &              &              &              &  LOFAR (4.27E3)\\\hline
     2           &  22.41       &      6.20E-3     &  10.70       &  33.75       &  17.42       &  6.84E2      &  4.07        &  CHIME (0.43)\\\hline
     3           &  23.23       &      -7.78E-4    &  9.50        &  50.31       &  16.56       &  8.46E2      &  3.29        &  CHIME (1.83)\\\hline
     4           &  23.36       &      1.01E-3     &  11.06       &  67.46       &  17.15       &  7.37E2      &  3.77        &  CHIME (1.15)\\\hline
     5           &  23.09       &      2.11E-3     &  14.75       &  83.77       &  16.31       &  5.36E2      &  5.18        &  CHIME (0.21) \\\hline    
     \\
\end{tabular}
\label{table:FRBs-repeating-16days}
\end{center}
\end{table*}

\clearpage
\begin{table*}
\caption{{\bf Simulations}: Example of an FRB from an ICM-QN (here $n_{\rm amb.}^{\rm ns}=3.87\times 10^{-4}$ cm$^{-3}$) active for $\sim 20$ years with properties reminiscent of FRB121102. Similar FRBs
can be generated with a high number of chunks (here $N_{\rm c}=3\times 10^5$), a low Lorentz factor (here $\Gamma_{\rm c}=40.27$)
and chunk electrons accelerated to high Lorentz factor (here $\gamma_{\rm CSE}= 40$) during filament merging; other
parameters are kept to their fiducial values  given in Table \ref{table:parameters}.   The mass is randomly selected from a Gaussian distribution with mean mass $\bar{m}_{\rm c}=10^{22.96}$ gm
 and standard deviation $\sigma_{m}=1.0$. The first detected chunk viewing angle is  $\theta_{\rm c}(\# 0)=1.11$E-2.}
\begin{tabular}{cccccccccccc}\hline
    \#        &  $\log{m_{\rm c}}$ (gm)      &     $\Delta \theta_{\rm c}$       &  $f(\theta_{\rm c})$      & 
    $t_{\rm OA}^{\rm obs.}$ (days)      & $\Delta t_{\rm OA}^{\rm obs.}$ (days)    & Frequency (MHz)     &  Width (ms) &  Fluence (Jy ms)\\\hline   
     0           &  20.74       &        0.00        &  1.20        &  0.00        &  0.00        &  7.86E2      &  56.57       &  CHIME (4.10)\\\hline
     1           &  21.54       &        -6.29E-3    &  1.04        &  75.05       &  75.05       &  9.98E2      &  44.57       &  CHIME (18.55)\\\hline
     2           &  21.06       &        0.01        &  1.46        &  1.23E2      &  47.79       &  6.71E2      &  66.29       &  CHIME (2.62)\\\hline
     3           &  21.92       &        -0.01       &  1.06        &  1.54E2      &  31.39       &  1.02E3      &  43.51       &  CHIME (26.50)\\\hline
     4           &  21.56       &        6.75E-3     &  1.26        &  1.59E2      &  4.76        &  8.22E2      &  54.14       &  CHIME (8.60)\\\hline
     5           &  21.73       &        -2.42E-3    &  1.17        &  1.60E2      &  1.33        &  9.02E2      &  49.30       &  CHIME (14.06)\\\hline
     6           &  21.03       &        0.01        &  1.83        &  2.22E2      &  61.49       &  5.33E2      &  83.53       &  CHIME (0.79)\\\hline
     7           &  21.08       &        -8.40E-5    &  1.83        &  2.33E2      &  11.04       &  5.38E2      &  82.74       &  CHIME (0.87)\\\hline
     8           &  22.19       &        -0.01       &  1.17        &  2.67E2      &  34.43       &  9.57E2      &  46.48       &  CHIME (24.56)\\\hline
     9           &  21.50       &        0.01        &  1.72        &  3.09E2      &  41.72       &  5.99E2      &  74.30       &  CHIME (2.07)\\\hline
     10          &  22.14       &        -7.36E-3    &  1.31        &  3.18E2      &  9.47        &  8.49E2      &  52.40       &  CHIME (14.61)\\\hline
     11          &  21.31       &       0.01        &  2.05        &  3.64E2      &  45.30       &  4.91E2      &  90.67       &  CHIME (0.55)\\\hline
     12          &  22.36       &        -0.01       &  1.28        &  3.73E2      &  8.79        &  8.88E2      &  50.10       &  CHIME (20.40)\\\hline
     13          &  21.65       &        9.74E-3     &  1.85        &  4.02E2      &  28.97       &  5.66E2      &  78.54       &  CHIME (1.72)\\\hline
     14          &  22.38       &        -7.90E-3    &  1.37        &  4.22E2      &  20.69       &  8.35E2      &  53.29       &  CHIME (16.13)\\\hline
     15          &  22.27       &        1.45E-3     &  1.44        &  4.24E2      &  2.06        &  7.82E2      &  56.88       &  CHIME (11.53)\\\hline
     16          &  22.31       &        -1.41E-4    &  1.43        &  4.36E2      &  11.62       &  7.90E2      &  56.29       &  CHIME (12.41)\\\hline
     17          &  22.47       &        -4.27E-4    &  1.41        &  4.80E2      &  43.71       &  8.18E2      &  54.40       &  CHIME (15.89)\\\hline
     18          &  23.30       &        -0.01       &  1.03        &  5.31E2      &  51.24       &  1.23E3      &  36.11       &  CHIME (1.44E2)\\\hline
     19          &  21.94       &        0.02        &  1.96        &  5.47E2      &  16.63       &  5.52E2      &  80.59       &  CHIME (1.82)\\\hline
     20          &  23.43       &        -0.02       &  1.02        &  5.77E2      &  29.63       &  1.26E3      &  35.30       &  CHIME (1.74E2)\\\hline
     21          &  22.24       &        0.02        &  1.80        &  5.92E2      &  15.23       &  6.24E2      &  71.24       &  CHIME (4.23)\\\hline
     22          &  22.34       &        -1.32E-4    &  1.79        &  6.29E2      &  36.33       &  6.35E2      &  70.07       &  CHIME (4.90)\\\hline
     23          &  21.80       &        6.19E-3     &  2.29        &  6.31E2      &  2.02        &  4.66E2      &  95.56       &  CHIME (0.51)\\\hline
     24          &  23.30       &        -0.02       &  1.16        &  6.35E2      &  4.09        &  1.10E3      &  40.63       &  CHIME (90.17)\\\hline
     25          &  23.38       &        -6.53E-4    &  1.14        &  6.56E2      &  21.71       &  1.13E3      &  39.52       &  CHIME (1.07E2)\\\hline
     26          &  22.72       &        9.62E-3     &  1.57        &  6.73E2      &  16.37       &  7.53E2      &  59.10       &  CHIME (13.43)\\\hline
     27          &  22.97       &        -2.51E-3    &  1.43        &  6.96E2      &  22.98       &  8.53E2      &  52.15       &  CHIME (26.56)\\\hline
     28          &  22.27       &        8.40E-3     &  1.99        &  7.02E2      &  6.43        &  5.66E2      &  78.63       &  CHIME (2.62)\\\hline
     29          &  22.83       &        -6.50E-3    &  1.54        &  7.02E2      &  0.03        &  7.81E2      &  56.96       &  CHIME (16.91)\\\hline
     30          &  23.52       &        -8.95E-3    &  1.14        &  7.21E2      &  18.56       &  1.14E3      &  38.96       &  CHIME (1.25E2)\\\hline
     31          &  23.01       &        7.27E-3     &  1.44        &  7.23E2      &  1.80        &  8.50E2      &  52.35       &  CHIME (26.87)\\\hline
     32          &  22.00       &        0.01        &  2.31        &  7.25E2      &  2.14        &  4.73E2      &  94.08       &  CHIME (0.67)\\\hline
     33          &  22.77       &        -8.63E-3    &  1.63        &  7.37E2      &  12.50       &  7.30E2      &  60.90       &  CHIME (12.34)\\\hline
     34          &  23.39       &        -7.47E-3    &  1.25        &  7.50E2      &  12.60       &  1.03E3      &  43.25       &  CHIME (74.91)\\\hline
     35          &  22.22       &        0.01        &  2.14        &  7.53E2      &  3.30        &  5.24E2      &  84.89       &  CHIME (1.63)\\\hline
     36          &  22.30       &        -4.24E-5    &  2.13        &  7.91E2      &  37.58       &  5.30E2      &  83.97       &  CHIME (1.84)\\\hline
     37          &  23.14       &        -8.43E-3    &  1.53        &  8.48E2      &  56.84       &  8.16E2      &  54.52       &  CHIME (25.00)\\\hline
     38          &  22.91       &        2.78E-3     &  1.70        &  8.49E2      &  1.23        &  7.13E2      &  62.42       &  CHIME (12.20)\\\hline
     39          &  22.26       &        7.59E-3     &  2.31        &  8.57E2      &  8.30        &  4.88E2      &  91.15       &  CHIME (1.03)\\\hline
     40          &  22.50       &        -2.36E-3    &  2.10        &  8.71E2      &  13.68       &  5.51E2      &  80.73       &  CHIME (2.65)\\\hline
     41          &  22.43       &        1.07E-3     &  2.19        &  8.84E2      &  13.40       &  5.24E2      &  84.94       &  CHIME (1.87)\\\hline
     42          &  23.29       &        -9.68E-3    &  1.49        &  9.02E2      &  18.05       &  8.50E2      &  52.35       &  CHIME (32.67)\\\hline
     43          &  22.30       &        0.01        &  2.37        &  9.11E2      &  9.18        &  4.77E2      &  93.30       &  CHIME (0.88)\\\hline
     44          &  22.36       &        -1.93E-5    &  2.37        &  9.40E2      &  28.51       &  4.80E2      &  92.67       &  CHIME (0.97)\\\hline
     45          &  23.39       &        -0.01       &  1.49        &  9.52E2      &  12.11       &  8.63E2      &  51.58       &  CHIME (37.04)\\\hline
     46          &  22.38       &        0.01        &  2.43        &  9.90E2      &  37.85       &  4.69E2      &  94.89       &  CHIME (0.81)\\\hline
     47          &  24.09       &        -0.02       &  1.17        &  1.06E3      &  67.28       &  1.19E3      &  37.35       &  CHIME (2.19E2)\\\hline
     48          &  23.79       &        5.60E-3     &  1.40        &  1.12E3      &  58.11       &  9.57E2      &  46.47       &  CHIME (73.87)\\\hline
     49          &  23.02       &        0.01        &  2.08        &  1.18E3      &  61.52       &  5.92E2      &  75.18       &  CHIME (5.58)\\\hline
     50          &  22.86       &        2.79E-3     &  2.32        &  1.23E3      &  54.82       &  5.19E2      &  85.69       &  CHIME (2.38)\\\hline

         \\
     \end{tabular}
\label{table:FRB121102}
\end{table*}

\clearpage
\begin{table*}
... Table \ref{table:FRB121102} continued.
\begin{tabular}{ccccccccccc}\\\hline\hline
    \#        &  $\log{m_{\rm c}}$  (gm)     &      $\Delta \theta_{\rm c}$       &  $f(\theta_{\rm c})$      & 
    $t_{\rm OA}^{\rm obs.}$ (days)      & $\Delta t_{\rm OA}^{\rm obs.}$ (days)    & Frequency (MHz)     &  Width (ms) &  Fluence (Jy ms)\\\hline
     51          &  22.56       &        3.68E-3     &  2.68        &  1.24E3      &  8.02        &  4.34E2      &  1.03E2      &  CHIME (0.38)\\\hline
     52          &  22.59       &        4.94E-4     &  2.74        &  1.29E3      &  53.82       &  4.27E2      &  1.04E2      &  CHIME (0.30)\\\hline
     53          &  23.29       &        -7.53E-3    &  2.03        &  1.33E3      &  37.48       &  6.24E2      &  71.28       &  CHIME (8.71)\\\hline
     54          &  23.51       &        -1.18E-3    &  1.93        &  1.42E3      &  93.24       &  6.72E2      &  66.24       &  CHIME (14.27)\\\hline
     55          &  23.60       &        -9.10E-5    &  1.93        &  1.49E3      &  64.10       &  6.81E2      &  65.34       &  CHIME (16.11)\\\hline
     56          &  23.16       &        5.33E-3     &  2.39        &  1.51E3      &  17.65       &  5.22E2      &  85.16       &  CHIME (3.04)\\\hline
     57          &  23.72       &        -6.37E-3    &  1.85        &  1.51E3      &  6.71        &  7.20E2      &  61.83       &  CHIME (22.24)\\\hline
     58          &  23.93       &        -2.08E-3    &  1.70        &  1.54E3      &  22.59       &  8.01E2      &  55.55       &  CHIME (39.90)\\\hline
     59          &  23.28       &        7.95E-3     &  2.34        &  1.58E3      &  40.93       &  5.41E2      &  82.28       &  CHIME (4.09)\\\hline
     60          &  23.29       &        7.56E-4     &  2.41        &  1.64E3      &  63.89       &  5.25E2      &  84.71       &  CHIME (3.45)\\\hline
     61          &  23.09       &        2.40E-3     &  2.65        &  1.65E3      &  8.24        &  4.67E2      &  95.25       &  CHIME (1.28)\\\hline
     62          &  23.64       &        -4.62E-3    &  2.21        &  1.79E3      &  1.41E2      &  5.98E2      &  74.44       &  CHIME (9.00)\\\hline
     63          &  23.03       &        7.72E-3     &  2.99        &  1.83E3      &  39.87       &  4.11E2      &  1.08E2      &  CHIME (0.16)\\\hline
     64          &  23.04       &        2.08E-4     &  3.01        &  1.86E3      &  26.67       &  4.09E2      &  1.09E2      &  CHIME (0.12)\\\hline
     65          &  23.85       &        -9.34E-3    &  2.09        &  1.87E3      &  15.18       &  6.48E2      &  68.69       &  CHIME (15.28)\\\hline
     66          &  23.31       &        6.73E-3     &  2.72        &  1.91E3      &  42.03       &  4.66E2      &  95.43       &  CHIME (1.46)\\\hline
     67          &  23.92       &        -7.01E-3    &  2.06        &  1.92E3      &  4.64        &  6.61E2      &  67.36       &  CHIME (17.56)\\\hline
     68          &  24.23       &        -3.44E-3    &  1.80        &  1.93E3      &  11.06       &  7.86E2      &  56.60       &  CHIME (45.66)\\\hline
     69          &  24.30       &        -5.88E-4    &  1.75        &  1.95E3      &  20.80       &  8.11E2      &  54.84       &  CHIME (54.45)\\\hline
     70          &  23.70       &        7.15E-3     &  2.34        &  1.97E3      &  16.06       &  5.68E2      &  78.36       &  CHIME (7.15)\\\hline
     71          &  23.33       &        4.27E-3     &  2.76        &  1.97E3      &  3.92        &  4.60E2      &  96.61       &  CHIME (1.32)\\\hline
     72          &  24.00       &        -6.76E-3    &  2.12        &  2.06E3      &  86.10       &  6.49E2      &  68.52       &  CHIME (17.11)\\\hline
     73          &  23.58       &        5.67E-3     &  2.65        &  2.13E3      &  75.28       &  4.94E2      &  90.10       &  CHIME (2.78)\\\hline
     74          &  23.59       &        1.08E-4     &  2.66        &  2.15E3      &  21.00       &  4.92E2      &  90.38       &  CHIME (2.74)\\\hline
     75          &  24.53       &        -0.01       &  1.76        &  2.20E3      &  48.17       &  8.29E2      &  53.64       &  CHIME (69.31)\\\hline
     76          &  23.76       &        8.82E-3     &  2.51        &  2.20E3      &  2.95        &  5.33E2      &  83.40       &  CHIME (5.25)\\\hline
     77          &  24.70       &        -0.01       &  1.63        &  2.21E3      &  10.73       &  9.11E2      &  48.82       &  CHIME (1.14E2)\\\hline
     78          &  23.62       &        0.01        &  2.91        &  2.41E3      &  2.01E2      &  4.52E2      &  98.37       &  CHIME (1.34)\\\hline
     79          &  23.92       &        -3.03E-3    &  2.59        &  2.48E3      &  65.37       &  5.27E2      &  84.45       &  CHIME (5.44)\\\hline
     80          &  24.08       &        1.53E-4     &  2.60        &  2.71E3      &  2.30E2      &  5.33E2      &  83.41       &  CHIME (6.55)\\\hline
     81          &  24.06       &        1.29E-3     &  2.74        &  2.84E3      &  1.27E2      &  5.06E2      &  87.91       &  CHIME (4.63)\\\hline
     82          &  23.75       &        4.42E-3     &  3.24        &  2.91E3      &  76.51       &  4.13E2      &  1.08E2      &  CHIME (0.30)\\\hline
     83          &  24.29       &        -5.22E-3    &  2.65        &  3.08E3      &  1.63E2      &  5.36E2      &  83.02       &  CHIME (7.77)\\\hline
     84          &  23.83       &        5.77E-3     &  3.30        &  3.11E3      &  28.36       &  4.08E2      &  1.09E2      &  CHIME (0.19)\\\hline
     85          &  24.36       &        -5.59E-3    &  2.67        &  3.21E3      &  1.09E2      &  5.37E2      &  82.92       &  CHIME (8.22)\\\hline
     86          &  24.00       &        4.88E-3     &  3.22        &  3.28E3      &  66.28       &  4.27E2      &  1.04E2      &  CHIME (0.81)\\\hline
     87          &  24.10       &        -1.24E-3    &  3.07        &  3.29E3      &  10.75       &  4.53E2      &  98.18       &  CHIME (1.91)\\\hline
     88          &  24.18       &        1.09E-3     &  3.20        &  3.57E3      &  2.78E2      &  4.39E2      &  1.01E2      &  CHIME (1.37)\\\hline
     89          &  24.11       &        1.08E-3     &  3.33        &  3.60E3      &  28.32       &  4.18E2      &  1.06E2      &  CHIME (0.55)\\\hline
     90          &  24.53       &        -3.91E-3    &  2.87        &  3.78E3      &  1.83E2      &  5.08E2      &  87.54       &  CHIME (6.56)\\\hline
     91          &  25.12       &        -3.07E-3    &  2.55        &  4.47E3      &  6.90E2      &  6.13E2      &  72.52       &  CHIME (28.49)\\\hline
     92          &  24.66       &       7.99E-3     &  3.45        &  4.92E3      &  4.50E2      &  4.29E2      &  1.04E2      &  CHIME (1.39)\\\hline
     93          &  25.79       &        -4.48E-3    &  2.92        &  7.14E3      &  2.22E3      &  5.78E2      &  76.94       &  CHIME (33.59)\\\hline
  \\
     \end{tabular}
\label{table:FRB121102-continued}
\end{table*}

\clearpage
\begin{table*}
\caption{{\bf Simulations}: Example of an FRB from a galactic QN  active for $\sim 3$ years with properties reminiscent of FRB121102. 
The main difference from the FRB in Table \ref{table:FRB121102} is the higher ambient density (here $n_{\rm amb.}^{\rm ns}=10^{-2}$ cm$^{-3}$) representative of the hot ISM component within galaxies. The first detected chunk viewing angle is  $\theta_{\rm c}(\# 0)=1.11$E-2.}
\begin{tabular}{ccccccccccc}\\\hline\hline
    \#        &  $\log{m_{\rm c}}$  (gm)     &      $\Delta \theta_{\rm c}$       &  $f(\theta_{\rm c})$      & 
    $t_{\rm OA}^{\rm obs.}$ (days)      & $\Delta t_{\rm OA}^{\rm obs.}$ (days)    & Frequency (MHz)     &  Width (ms) &  Fluence (Jy ms)\\\hline

     0           &  20.75       &        0.00        &  1.20        &  0.00        &  0.00        &  5.42E2      &  8.21        &  CHIME (0.31)\\\hline
     1           &  21.55       &        -6.29E-3    &  1.04        &  11.08       &  11.08       &  6.86E2      &  6.49        &  CHIME (1.74)\\\hline
     2           &  21.07       &        0.01        &  1.45        &  17.67       &  6.58        &  4.64E2      &  9.58        &  CHIME (0.13)\\\hline
     3           &  21.93       &        -0.01       &  1.06        &  22.61       &  4.94        &  7.03E2      &  6.33        &  CHIME (2.51)\\\hline
     4           &  21.56       &        6.75E-3     &  1.26        &  23.07       &  0.46        &  5.67E2      &  7.85        &  CHIME (0.70)\\\hline
     5           &  21.73       &        -2.42E-3    &  1.17        &  23.36       &  0.29        &  6.22E2      &  7.16        &  CHIME (1.26)\\\hline
     6           &  22.20       &        -1.81E-4    &  1.16        &  38.93       &  15.58       &  6.59E2      &  6.75        &  CHIME (2.28)\\\hline
     7           &  22.14       &        3.61E-3     &  1.30        &  46.20       &  7.26        &  5.86E2      &  7.59        &  CHIME (1.24)\\\hline
     8           &  22.36       &        -5.66E-4    &  1.28        &  54.09       &  7.89        &  6.13E2      &  7.26        &  CHIME (1.81)\\\hline
     9           &  22.38       &        1.84E-3     &  1.36        &  61.19       &  7.10        &  5.77E2      &  7.71        &  CHIME (1.35)\\\hline
     10          &  22.27       &        1.45E-3     &  1.43        &  61.40       &  0.21        &  5.41E2      &  8.23        &  CHIME (0.88)\\\hline
     11          &  22.32       &        -1.41E-4    &  1.42        &  63.09       &  1.69        &  5.47E2      &  8.14        &  CHIME (0.96)\\\hline
     12          &  22.48       &        -4.27E-4    &  1.40        &  69.46       &  6.37        &  5.65E2      &  7.87        &  CHIME (1.30)\\\hline
     13          &  23.30       &        -0.01       &  1.03        &  77.62       &  8.16        &  8.47E2      &  5.26        &  CHIME (13.91)\\\hline
     14          &  23.44       &        -5.93E-4    &  1.02        &  84.39       &  6.77        &  8.66E2      &  5.14        &  CHIME (16.70)\\\hline
     15          &  22.25       &        0.02        &  1.78        &  85.36       &  0.97        &  4.33E2      &  10.27       &  CHIME (0.13)\\\hline
     16          &  22.34       &        -1.32E-4    &  1.77        &  90.62       &  5.26        &  4.40E2      &  10.10       &  CHIME (0.18)\\\hline
     17          &  23.30       &        -0.01       &  1.15        &  92.45       &  1.83        &  7.54E2      &  5.90        &  CHIME (8.74)\\\hline
     18          &  23.39       &        -6.53E-4    &  1.13        &  95.66       &  3.21        &  7.75E2      &  5.74        &  CHIME (10.39)\\\hline
     19          &  22.72       &        9.62E-3     &  1.56        &  97.26       &  1.60        &  5.21E2      &  8.54        &  CHIME (0.96)\\\hline
     20          &  22.98       &        -2.51E-3    &  1.42        &  1.01E2      &  3.54        &  5.90E2      &  7.54        &  CHIME (2.29)\\\hline
     21          &  22.84       &        1.91E-3     &  1.53        &  1.02E2      &  0.78        &  5.41E2      &  8.23        &  CHIME (1.29)\\\hline
     22          &  23.02       &        -1.68E-3    &  1.43        &  1.05E2      &  3.09        &  5.88E2      &  7.57        &  CHIME (2.31)\\\hline
     23          &  23.53       &        -7.27E-3    &  1.14        &  1.05E2      &  0.37        &  7.86E2      &  5.66        &  CHIME (12.11)\\\hline
     24          &  22.78       &        0.01        &  1.62        &  1.07E2      &  1.48        &  5.06E2      &  8.79        &  CHIME (0.82)\\\hline
     25          &  23.40       &        -7.47E-3    &  1.24        &  1.09E2      &  2.49        &  7.10E2      &  6.27        &  CHIME (7.21)\\\hline
     26          &  22.91       &        8.48E-3     &  1.68        &  1.23E2      &  13.57       &  4.94E2      &  9.00        &  CHIME (0.76)\\\hline
     27          &  23.15       &        -2.78E-3    &  1.51        &  1.23E2      &  0.08        &  5.65E2      &  7.88        &  CHIME (2.05)\\\hline
     28          &  23.30       &        -6.11E-4    &  1.48        &  1.31E2      &  7.98        &  5.88E2      &  7.57        &  CHIME (2.81)\\\hline
     29          &  23.40       &        -9.76E-5    &  1.47        &  1.38E2      &  7.23        &  5.97E2      &  7.46        &  CHIME (3.24)\\\hline
     30          &  24.10       &        -7.18E-3    &  1.16        &  1.54E2      &  16.07       &  8.21E2      &  5.42        &  CHIME (21.29)\\\hline
     31          &  23.79       &        5.60E-3     &  1.39        &  1.62E2      &  7.79        &  6.62E2      &  6.72        &  CHIME (6.97)\\\hline
     32          &  23.29       &        9.48E-3     &  2.01        &  1.92E2      &  30.09       &  4.34E2      &  10.26       &  CHIME (0.27)\\\hline
     33          &  23.52       &        -1.18E-3    &  1.91        &  2.05E2      &  13.60       &  4.66E2      &  9.54        &  CHIME (0.73)\\\hline
     34          &  23.60       &        -9.10E-5    &  1.91        &  2.15E2      &  9.27        &  4.73E2      &  9.41        &  CHIME (0.87)\\\hline
     35          &  23.72       &        -1.03E-3    &  1.83        &  2.18E2      &  3.65        &  4.99E2      &  8.91        &  CHIME (1.44)\\\hline
     36          &  23.93       &        -2.08E-3    &  1.69        &  2.22E2      &  3.55        &  5.55E2      &  8.01        &  CHIME (3.21)\\\hline
     37          &  23.65       &        6.49E-3     &  2.18        &  2.58E2      &  35.83       &  4.16E2      &  10.71       &  CHIME (0.15)\\\hline
     38          &  23.86       &        -1.41E-3    &  2.06        &  2.70E2      &  11.98       &  4.50E2      &  9.88        &  CHIME (0.64)\\\hline
     39          &  23.93       &        -2.84E-4    &  2.04        &  2.76E2      &  6.78        &  4.59E2      &  9.69        &  CHIME (0.83)\\\hline
     40          &  24.24       &        -3.44E-3    &  1.78        &  2.79E2      &  2.13        &  5.45E2      &  8.16        &  CHIME (3.57)\\\hline
     41          &  24.31       &        -5.88E-4    &  1.74        &  2.82E2      &  3.11        &  5.63E2      &  7.91        &  CHIME (4.47)\\\hline
     42          &  24.00       &        4.66E-3     &  2.09        &  2.96E2      &  14.58       &  4.51E2      &  9.86        &  CHIME (0.73)\\\hline
     43          &  24.53       &        -4.57E-3    &  1.74        &  3.18E2      &  21.64       &  5.75E2      &  7.74        &  CHIME (5.87)\\\hline
     44          &  24.70       &        -1.88E-3    &  1.62        &  3.20E2      &  2.34        &  6.31E2      &  7.05        &  CHIME (10.53)\\\hline
     45          &  25.13       &        0.01        &  2.51        &  6.43E2      &  3.23E2      &  4.27E2      &  10.42       &  CHIME (0.76)\\\hline
     46          &  25.80       &        3.51E-3     &  2.88        &  1.03E3      &  3.84E2      &  4.03E2      &  11.04       &  CHIME (0.12)\\\hline
\end{tabular}         \\
\label{table:FRB121102-galactic}
\end{table*}

\clearpage

\begin{table*}
\vskip 3.0in
\begin{center}
\caption{{\bf FRBs from IGM-QNe}: Key equations describing the properties (baryon number density, radius and sound
speed) of the collisionless QN chunks in the
IGM and the resulting CSE features (frequency, duration and fluence).
Also shown is the time since the QN, $t_{\rm cc}^{\rm obs.}$, and the time separation between emitting 
 chunks $\Delta t_{\rm repeat}^{\rm obs.}$ (see Appendix \ref{appendix:honeycomb}).}
\begin{tabular}{|c|c|c|}\hline
     \multicolumn{3}{|c|}{FRBs from IGM-QNe}\\\hline\hline
  \multicolumn{3}{|c|}{Collisionless chunk (``cc") properties}\\ \hline\hline
 Number density & $n_{\rm cc}$ (cm$^{-3}$) & $\simeq 2.7\times 10^{-4}\times (1+z)^{18/5}\times \frac{m_{\rm c, 22.3}^{1/10}}{\kappa_{\rm c, -1}^{9/10}}\sigma_{\rm HH, -27}^{6/5}\Gamma_{\rm c, 2.5}^{12/5}$\\\hline
 Radius & $R_{\rm cc}$ (cm)& $\simeq  2.4\times10^{16}\times \frac{1}{(1+z)^{6/5}}\times \frac{(m_{\rm c, 22.3}\kappa_{\rm c, -1})^{3/10}}{\sigma_{\rm HH, -27}^{2/5}\Gamma_{\rm c, 2.5}^{4/5}}$ \\\hline
Thermal speed &   $\beta_{\rm cc}$  & $\simeq 2.7\times 10^{-3} \times (1+z)^{3/5}\times (m_{\rm c, 22.3}\kappa_{\rm c, -1})^{1/10} (\sigma_{\rm HH, -27}\Gamma_{\rm c, 2.5}^{2})^{1/5}$\\\hline
  Time since QN &  $t_{\rm cc}^{\rm obs.}$ (days)  & $\simeq 439.2\times \frac{f(\theta_{\rm c})}{(1+z)^{4/5}}\times \frac{(m_{\rm c, 22.3}\kappa_{\rm c, -1})^{1/5}}{\sigma_{\rm HH, -27}^{3/5}\Gamma_{\rm c, 2.5}^{11/5}}$\\\hline
  \multicolumn{3}{|c|}{Coherent synchrotron emission (CSE) properties}\\ \hline\hline
Frequency$^{1}$ &  $\nu_{\rm CSE, max.}^{\rm obs.}(\theta_{\rm c})$ (MHz)  & $\simeq 36.6\times \frac{1}{(1+z)f(\theta_{\rm c})}\times \delta_{\rm CSE, -1}\gamma_{\rm CSE, 1}^{2}\Gamma_{\rm c, 2.5}n_{\rm cc, -4}^{1/2}$\\\hline
Width &  $\Delta t_{\rm CSE}^{\rm obs.}$ (ms)  &  $\simeq 76.0\times (1+z)f(\theta_{\rm c})\times \frac{\zeta_{\rm m-WI,2}}{\Gamma_{\rm c, 2.5}n_{\rm cc, -4}^{1/2}}    \times {\rm Min}(..,..)^{2}$\\\hline
  $\mathcal{F}$ (Jy ms)
Fluence$^{3}$ & $\mathcal{F}(\theta_{\rm c},0)$ (Jy ms)  &  $\simeq 1.6\times 10^6\times   \frac{(1+z)^3}{f(\theta_{\rm c})^2d_{\rm L, 27.5}^2}\times \frac{\zeta_{\rm BI, -1}\beta_{\rm WI, -1}}{\delta_{\rm CSE, -1}\gamma_{\rm CSE,1}^2} \times \frac{\Gamma_{\rm c, 2.5}^4 R_{\rm cc, 16}^2}{n_{\rm cc, -4}\beta_{\rm cc, -3}}$\\\hline 
 Repeat  time$^{4}$ &   $\Delta t_{\rm repeat}^{\rm obs.}$  (days) & $\simeq 223.6\times \frac{1}{(1+z)^{4/5}}\times \frac{1}{N_{\rm c, 6}}\times  \left( \frac{m_{\rm c, 22.3}\kappa_{\rm c, -1}}{\sigma_{\rm HH, -27}^{3}\Gamma_{\rm c, 2.5}}\right)^{1/5}$\\\hline
\end{tabular}\\
\label{table:IGM-equations}
\end{center}
$^{1}$ The  frequency drifts to a minimum value  
given by the chunk's plasma frequency $\nu_{\rm p, e}^{\rm obs.}(\theta_{\rm c})\simeq \frac{56.9\ {\rm kHz}}{(1+z)f(\theta_{\rm c})}\times \Gamma_{\rm c, 2.5}n_{\rm cc, -4}^{1/2}$.\\
$^{2}$ ${\rm Min}\left[\left(  \left(642.7\delta_{\rm CSE, -1}\gamma_{\rm CSE,1}^2\right)^{\frac{1}{\delta_{\rm m-WI}}} -1\right) ,\left(\left( \frac{\nu_{\rm CSE, max.}^{\rm obs.}(\theta_{\rm c})}{\nu_{\rm min.}^{\rm det.}} \right)^{1/\delta_{\rm m-WI}} - \left(\frac{\nu_{\rm CSE, max.}^{\rm obs.}(\theta_{\rm c})}{\nu_{\rm max.}^{\rm det.}} \right)^{1/\delta_{\rm m-WI}}\right)\right]$; see \S \ref{sec:FRB-ICM-QNe}.\\
$^{3}$  $F(\theta_{\rm c},\delta_{\rm m-WI},0)=\mathcal{F}(\theta_{\rm c},0)\times \mathcal{G}(\theta_{\rm c},\delta_{\rm m-WI},0)$ for the flat spectrum case ($\alpha_{\rm CSE}=0$) with $\mathcal{G}(\theta_{\rm c},\delta_{\rm m-WI},0)$ given   in Eq. (\ref{appendix:eq:fluence-flat-2}) and Table \ref{table:G-fluence}.\\
$^{4}$ Independent of the viewing angle $\theta_{\rm c}$ (i.e. $f(\theta_{\rm c})$) due to geometry and the spatial distribution
of chunks (see Appendix \ref{appendix:honeycomb} and Eq. (\ref{eq:Deltac})).\\
\end{table*}


\clearpage


 \clearpage

\clearpage

\clearpage


 \clearpage

 \clearpage
 
\clearpage

\clearpage

 \clearpage  


\begin{thebibliography}{}


\bibitem[Bassa et al.(2017)]{bassa_2017} Bassa, C. G. et al. 2017, \apj, 843, L8

\bibitem[Barthelmy et al.(2020)]{barthelmy_2020} Barthelmy, S. D. et al., 2020, GRB Circular Network 27657, 1 

\bibitem[Bochenek et al.(2020)]{bochenek_2020} Bochenek, C. D., Ravi, V., Below, K. V. et al. 2020, arXiv:2005.10828

\bibitem[Bombaci et al.(2004)]{bombaci_2004}  Bombaci, I. Parenti, I. \& Vidana, I. 2004, \apj, 614, 314 

\bibitem[Caleb et al.(2019)]{caleb_2019} Caleb, M., Stappers, B. W., Rajwade, K., \& Flynn, C. 2019, MNRAS, 484, 5500

\bibitem[Cavaliere(2000)]{cavaliere_2016}  Cavaliere, A., Fusco-Femiano, R., \& Lapi, A. 2016, \apj, 824, 145

\bibitem[CHIME/FRB Collaboration(2018)]{chime_2018} CHIME/FRB Collaboration, Amiri, M., Bandura, K., et al.\ 2018, \apj, 863, 48

\bibitem[CHIME/FRB Collaboration(2019a)]{chime_2019a} CHIME/FRB Collaboration, Amiri, M., Bandura, K., et al.\ 2019a, \nat, 566, 235

\bibitem[CHIME/FRB Collaboration(2019b)]{chime_2019b} CHIME/FRB Collaboration, Andersen, B.~C., Bandura, K. et al.\ 2019b, \apj, 885, L24

\bibitem[CHIME/FRB Collaboration(2020a)]{chime_2020a}  CHIME/FRB Collaboration, Amiri, M. et al. 2020a, [arXiv:2001.10275]

\bibitem[CHIME/FRB Collaboration(2020b)]{chime_2020b} CHIME/FRB Collaboration, Andersen, B. C., Bandura, K., et al. 2020b [arXiv:2005.10324v2]

\bibitem[Champion et al.(2016)]{champion_2016} Champion, D. J., Petroff, E., Karmer, M., et al. 2016, \mnras, 460, L30

\bibitem[Cordes \& Chatterjee(2019)]{cordes_2019} Cordes, J. M. \& Chatterjee, S., 2019, \araa, 57, 579 

\bibitem[Deserno(2004)]{deserno_2004} Deserno, M. 2004 [\url{http://www.cmu.edu/biolphys/deserno/pdf/sphere_equi.pdf}]
  
\bibitem[Fabian(1994)]{fabian_1994} Fabian, A. C.\ 1994, Ann. Rev. Astron. Astrophys., 32, 277

\bibitem[Faucher-Gigu{\`e}re \& Kaspi(2006)]{faucher_2006} Faucher-Gigu{\`e}re, C.-A., \& Kaspi, V.~M.\ 2006, \apj, 643, 332

\bibitem[Gajjar et al.(2018)]{gajjar_2018} Gajjar, V., Siemion, A.~V., Price, D.~C., et al.\ 2018, \apj, 863, 2

\bibitem[Hessels et al.(2019)]{hessels_2019} Hessels, J.~W.~T. et al. 2019, \apj, 876, L23

\bibitem[Harko et al.(2004)]{harko_2004} Harko, T. \& Cheng, K.~S. and Tang, P.~S., 2004, \apj, 608, 945

\bibitem[Iwazaki(2005)]{iwazaki_2005}  Iwazaki, A. 2005, Phys. Rev. D, 72, 114003

\bibitem[Katz(2014)]{katz_2014} Katz, J.~I.\ 2014, \prd, 89, 103009

\bibitem[Kellermann \& Pauliny-Toth(1969)]{kellermann_1969} Kellermann, K. I. \& Pauliny-Toth, I. I. K. 1969, \apj, 55, L71

\bibitem[Ker\"anen et al.(2005)]{keranen_2005} Ker\"anen, P., Ouyed, R., \& Jaikumar, P.\ 2005, \apj, 618, 485

\bibitem[Kulkarni et al.(2014)]{kulkarni_2014} Kulkarni, S.~R., Ofek, E.~O., Neill, J.~D. et al.\ 2014, \apj, 797, 70

\bibitem[Lang(1999)]{lang_1999} Lang, K. R. 1999, Astrophysical formulae, Third edition  (New York: Springer)

\bibitem[Letaw et al.(1983)]{letaw_1983} Letaw, J. R., Silberberg, R. \& Tsao, C. H. 1983, \apj,  51, 271
 
\bibitem[Lorimer et al.(2007)]{lorimer_2007} Lorimer, D.~R., Bailes, M., McLaughlin, M.~A., Narkevic, D.~J., \& Crawford, F.\ 2007, Science, 318, 777
 
\bibitem[Lorimer(2018)]{lorimer_2018}  Lorimer, D.~R.\ 2018, Nature Astronomy, 2, 860
         
 \bibitem[McQuinn(2016)]{mcquinn_2016} McQuinn, M. 2016, \araa,  2016, 54, 313

\bibitem[Metzger et al.(2017)]{metzger_2017} Metzger B. D., Berger E., Margalit B., 2017, \apj, 841, 14
 
 \bibitem[Michilli et al.(2018)]{michilli_2018} Michilli, D., Seymour, A., Hessels, J.~W.~T., et al.\ 2018, \nat, 553, 182
 
\bibitem[Niebergal et al.(2010a)]{niebergal_2010a} Niebergal, B., Ouyed, R., \& Jaikumar, P. 2010a, \prc, 82, 062801

\bibitem[Niebergal et al.(2010b)]{niebergal_2010b} Niebergal, B.,  Ouyed, R.,  Negreiros, R. \&   Weber, F., 2010b, \prd,  81, 043005 

\bibitem[Niebergal(2011)]{niebergal_phd_2011} Niebergal, B., 2011, ``Hadronic-to-Quark-Matter Phase Transition: Astrophysical Implications", Thesis (Ph.D.), University of Calgary (Canada), 2011.; Publication Number: AAT NR81856; ISBN: 9780494818565


\bibitem[Ouyed et al.(2004)]{ouyed_2004} Ouyed, R., Elgar\o y, \O., Dahle, H., \& Ker\"anen, P. 2004, A\&A, 420, 1025

\bibitem[Ouyed et al.(2006)]{ouyed_2006} Ouyed, R., Niebergal, B., Dobler, W., \& Leahy, D.\ 2006, \apj, 653, 558

\bibitem[Ouyed et al.(2007a)]{ouyed_2007a} Ouyed, R.,  Leahy, D. \& Niebergal, B.,\ 2007a, \aap, 473, 357

\bibitem[Ouyed et al.(2007b)]{ouyed_2007b} Ouyed, R.,  Leahy, D. \& Niebergal, B.,\ 2007b, \aap, 475, 63

\bibitem[Ouyed \& Leahy(2009)]{ouyed_leahy_2009} Ouyed, R., \& Leahy, D. 2009, \apj, 696, 562

\bibitem[Ouyed al.(2020)]{ouyed_2020}  Ouyed, R., Leahy, D., \& Koning, N. 2020, \raa, 20, 27

\bibitem[Ouyed, A. et al.(2018a)]{ouyed_amir_2018a} Ouyed, A., Ouyed, R., \& Jaikumar, P. 2018a, Physics Letters B, 777, 184

\bibitem[Ouyed, A. et al.(2018b)]{ouyed_amir_2018b} Ouyed, A., Ouyed, R., \& Jaikumar, P. 2018b, Universe, 4, 51


\bibitem[Ouyed, A.(2018)]{ouyed_amir_phd_2018} Ouyed, A. 2018, ``The Neutrino Sector in Hadron-Quark Combustion: Physical and Astrophysical Implications", Thesis (Ph.D.), University of Calgary (Canada), 2018 [\url{http://dx.doi.org/10.11575/PRISM/27841}]

 

\bibitem[Petroff et al.(2016)]{petroff_2016} Petroff, E., Barr, E.~D., Jameson, A., et al.  2016, \pasa, 33, e045

\bibitem[Petroff et al.(2019)]{petroff_2019} Petroff, E., Hessels, J.~W.~T. \& Lorimer, D.~R., 2019,  Astronomy and Astrophysics Reviews, 27, 4

\bibitem[Platts et al.(2019)]{platts_2019} Platts, E., Weltman, A., Walters, A., et al.\ 2019, \physrep, 821, 1 

\bibitem[Popov et al.(2018)]{popov_2018} Popov, S.~B., Postnov, K.~A., \& Pshirkov, M.~S. 2018, Physics Uspekhi, 61, 965

\bibitem[Ravi et al.(2016)]{ravi_2016} Ravi, V., Shannon, R.~M., Bailes, M., et al.\ 2016, Science, 354, 1249 

\bibitem[Ravi(2019)]{ravi_2019} Ravi, V. 2019, Nature Astronomy, 3, 928 

\bibitem[Richardson(2019)]{richardson_2019} Richardson, A. S. 2019, NRL Plasma Formulary 

\bibitem[Salpeter(1955)]{salpeter_1955} Salpeter, E. E. 1955, \apj, 121, 161

\bibitem[Scholz et al.(2016)]{sholz_2016}  Scholz, P., Spitler, L.~G., Hessels, J.~W.~T., et al.\ 2016, \apj, 833, 177

\bibitem[Scholz et al.(2020)]{scholz_2020} Scholz, P. et al., 2020, The Astronomer's Telegram 13681, 1

\bibitem[Spitler et al.(2014)]{spitler_2014} Spitler, L.~G., Cordes, J.~M., Hessels, J.~W.~T., et al.\ 2014, \apj, 790, 101

\bibitem[Spitler et al.(2016)]{spitler_2016} Spitler, L. G., Scholz, P., Hessels, J. W. T., et al.  2016, Nature 2016, 531, 202

\bibitem[Staff et al.(2006)]{staff_2006}Staff, J., Ouyed, R., \& Jaikumar, P., 2006, \apj, 645, L145
 
\bibitem[Thornton et al.(2013)]{thornton_2013} Thornton, D., Stappers, B., Bailes, M., et al.\ 2013, Science, 341, 53

  \bibitem[Tully(1987)]{tully_1987}  Tully R. B., 1987, ApJ, 321, 280

\bibitem[van Haarlem et al.(2013)]{vanhaarlem_2013} van Haarlem, M. P., Wise, M. W., Gunst, A. W., et al. 2013, A\&A, 556, A2
  
  \bibitem[Vogt et al.(2004)]{vogt_2004} Vogt C., Rapp R., \& Ouyed R., 2004, Nuclear Physics A, 735, 543
  
\bibitem[Weber(2005)]{weber_2005}   Weber, F. 2005, Progress in Particle and Nuclear Physics, 54, 193

\bibitem[Witze(2020)]{witze_2020} Witze, A., \nat, 2020, 583, 322

\end{thebibliography}

\begin{thebibliography}{}


\bibitem[Achterberg et al.(2007)]{achterberg_2007} Achterberg, A. Wiersma, J. \& C. A. Norman, C. A  2007, A\&A 475, 19

\bibitem[Alford et al.(1999)]{alford_1999}  Alford, M.,  Rajagopal, K., \&  Wilczek, Frank, 1999, Nuclear Physics B, 537, 443
  
\bibitem[Bailes et al.(2017)]{bailes_2017} Bailes, M. et al. 2017, Publ. Astron. Soc. Australia, 34, e045     

\bibitem[Bassa et al.(2017)]{bassa_2017} Bassa, C. G. et al. 2017, \apj, 843, L8

\bibitem[Barthelmy et al.(2020)]{barthelmy_2020} Barthelmy, S. D. et al., 2020, GRB Circular Network 27657, 1 

\bibitem[Berezinsky(2008)]{berezinsky_2008} Berezinsky, V. 2008, Propagation and origin of ultra high-energy cosmic rays. Adv. Space Res.41, 2071

\bibitem[Bochenek et al.(2020)]{bochenek_2020} Bochenek, C. D., Ravi, V., Below, K. V. et al. 2020, arXiv:2005.10828

\bibitem[Bombaci et al.(2004)]{bombaci_2004}  Bombaci, I. Parenti, I. \& Vidana, I. 2004, \apj, 614, 314 

\bibitem[Bret(2009)]{bret_2009} Bret, A.\ 2009, \apj, 699, 990

\bibitem[Bret et al.(2010)]{bret_2010}  Bret, A., Gremillet, L., \& Dieckmann, M. E. 2010, Phys. Plasmas 17, 120501
  
\bibitem[Bret et al.(2016)]{bret_2016} Bret, A., Stockem Novo, A., Narayan, R., et al.\ 2016, Laser and Particle Beams, 34, 362 

\bibitem[Buneman(1958)]{buneman_1958} Buneman, O.\ 1958, Phys. Rev., 115, 503

\bibitem[Buneman(1959)]{buneman_1959} Buneman O 1959 Phys. Rev. 115 503

\bibitem[Caleb et al.(2019)]{caleb_2019} Caleb, M., Stappers, B. W., Rajwade, K., \& Flynn, C. 2019, MNRAS, 484, 5500

\bibitem[Cavaliere(2000)]{cavaliere_2016}  Cavaliere, A., Fusco-Femiano, R., \& Lapi, A. 2016, \apj, l, 824, 145

\bibitem[CHIME/FRB Collaboration(2018)]{chime_2018} CHIME/FRB Collaboration, Amiri, M., Bandura, K., et al.\ 2018, \apj, 863, 48

\bibitem[CHIME/FRB Collaboration(2019a)]{chime_2019a} CHIME/FRB Collaboration, Amiri, M., Bandura, K., et al.\ 2019, \nat, 566, 235

\bibitem[CHIME/FRB Collaboration(2019b)]{chime_2019b} CHIME/FRB Collaboration, Andersen, B.~C., Bandura, K. et al.\ 2019, \apj, 885, L24

\bibitem[CHIME/FRB Collaboration(2020)]{chime_2020a}  CHIME/FRB Collaboration, Amiri, M. et al. 2020a, [arXiv:2001.10275]

\bibitem[CHIME/FRB Collaboration(2020)]{chime_2020b} CHIME/FRB Collaboration, Andersen, B. C., Bandura, K., et al. 2020b
[arXiv:2005.10324v2]

\bibitem[Champion et al.(2016)]{champion_2016} Champion, D. J., Petroff, E., Kramer, M., et al. 2016, MNRAS, 460, L30

\bibitem[Chatterjee et al.(2017)]{chatterjee_2017} Chatterjee, S., Law, C. J., Wharton, R. S., et al. 2017, Nature, 541, 58

\bibitem[Cordes et al.(2006)]{cordes_2006} Cordes, J. M., Freire, P. C. C., Lorimer, D. R., et al. 2006, \apj, 637, 446

\bibitem[Cordes \& Chatterjee(2019)]{cordes_2019} Cordes, J. M. \& Chatterjee, S., 2019, \araa, 57, 579 

 \bibitem[Cox(2005)]{cox_2005}   Cox, D. P. 2005, \araa, 43, 337

\bibitem[Davidson(1970)]{davidson_1970}  Davidson, R. C, Krall, N. A., Papadopoulos, K., \& Shanny, R.\ 1970, Phys.
Rev. Letters, 24,579

\bibitem[Davidson(1974)]{davidson_1974} Davidson, R. C. 1974, Frontiers in Physics, 43 (Reading: W. A. Benjamin)

\bibitem[Deserno(2004)]{deserno_2004} Deserno, M. 2004 [\url{http://www.cmu.edu/biolphys/deserno/pdf/sphere_equi.pdf}]

\bibitem[Dieckmann et al.(2012)]{dieckmann_2012}  Dieckmann, M. E. et al. 2012, Plasma Phys. Control. Fusion 54, 085015
  
\bibitem[Dvornikov(2016a)]{dvornikov_2016a}   Dvornikov, M. 2016a, Physics Letters B, 760, 406
  
\bibitem[Dvornikov(2016b)]{dvornikov_2016b}  Dvornikov, M. 2016b, Nuclear Physics B, 913, 79
  
\bibitem[Fabian(1994)]{fabian_1994} Fabian, A. C.\ 1994, Ann. Rev. Astron. Astrophys., 32, 277

\bibitem[Faucher-Gigu{\`e}re \& Kaspi(2006)]{faucher_2006} Faucher-Gigu{\`e}re, C.-A., \& Kaspi, V.~M.\ 2006, \apj, 643, 332

\bibitem[Fermi,(1949)]{fermi_1949} Fermi, E. 1949, Phys. Rev., 75, 1169
 
\bibitem[Frederiksen et al.(2004)]{frederiksen_2004}    {{Frederiksen}, J. Trier and {Hededal}, C.~B. and {Haugb{\o}lle}, T. \& 
         {Nordlund}, {\r{A}}.} 2004,   \apj, 608, L13
  
\bibitem[Fried(1959)]{fried_1959} Fried, B. D. 1959, Physics of Fluids, 2, 337

\bibitem[Gajjar et al.(2018)]{gajjar_2018} Gajjar, V., Siemion, A.~P.~V., Price, D.~C., et al.\ 2018, \apj, 863, 2

\bibitem[Gallant \& Achterberg(1999)]{gallant_1999} Gallant, Y. A. \& Achterberg, A. 1999, MNRAS, 305, L06

\bibitem[Ginzburg \& Syrovatskii(1965)]{ginzburg_1965} Ginzburg, V. L. \& Syrovatskii, S. Y. 1965, Ann. Rev. Astron. Astrophys., 3, 297 

\bibitem[Gruzinov(2001)]{gruzinov_2001} Gruzinov, A. 2001, \apj, 563, L15

\bibitem[Gunn \& Gott(1972)]{gunn_1972} Gunn J. E., \& Gott J. R., III, 1972, ApJ, 176, 1

\bibitem[Heifets \& Stupakov(2002)]{heifets_2002} Heifets, S. \& Stupakov, G. 2002, Physical Review Special Topics - Accelerators and Beams, 5, 05

\bibitem[Hessels et al.(2019)]{hessels_2019} Hessels, J.~W.~T. et al. 2019, \apj, 876, L23

\bibitem[Harko et al.(2004)]{harko_2004} {{Harko}, T. \& {Cheng}, K.~S. and {Tang}, P.~S.}, 2004, \apj, 608, 945

\bibitem[Hirose(1978)]{hirose_1978} Hirose, A.\ 1978, Plasma Physics 20, 481 

\bibitem[Iwazaki(2005)]{iwazaki_2005}  Iwazaki, A. 2005, Phys. Rev. D, 72, 114003

\bibitem[Jaikumar et al.(2007)]{jaikumar_2007} Jaikumar, P., Meyer, B. S., Otsuki, K, \& Ouyed, R. 2007, A\&A, 471, 227

\bibitem[Johnston et al.(2008)]{johnston_2008} Johnston, S. et al. 2008, Exp. Astron., 22, 151

\bibitem[Kang et al.(2020)]{kang_2020}    {{Kang}, Yijung and {Lee}, Young-Wook and {Kim}, Young-Lo and
         {Chung}, Chul and {Ree}, Chang Hee}, 2020,  \apj, 889, 8 [arXiv:1912.04903]

 \bibitem[Kato(2005)]{kato_2005} Kato, T. N. 2005, Phys. Plasmas, 12, 080705

 \bibitem[Kato(2007)]{kato_2007} Kato, T. N., 2007, \apj, 668, 974

\bibitem[Katz(2014)]{katz_2014} Katz, J.~I.\ 2014, \prd, 89, 103009

\bibitem[Kellermann \& Pauliny-Toth(1969)]{kellermann_1969} Kellermann, K. I. \& Pauliny-Toth, I. I. K. 1969, \apj, 55, L71

\bibitem[Ker\"anen et al.(2005)]{keranen_2005} Ker\"anen, P., Ouyed, R., \& Jaikumar, P. 2005, \apj, 618, 485

\bibitem[Kostka et al.(2014)]{kostka_2014} Kostka, M., Koning, K., Shand, Z. Ouyed, R., \& Jaikumar, P. 2014, A\&A 568, A97

\bibitem[Kulkarni et al.(2014)]{kulkarni_2014} Kulkarni, S.~R., Ofek, E.~O., Neill, J.~D., Zheng, Z., \& Juric, M.\ 2014, \apj, 797, 70

\bibitem[Lang(1999)]{lang_1999} Lang, K. R. 1999, Astrophysical formulae, Third edition  (New York: Springer)

\bibitem[Larson et al.(1980)]{larson_1980}  Larson R. B., Tinsley B. M., \& Caldwell C. N., 1980, ApJ, 237, 692

\bibitem[Lee \& Lampe(1973)]{lee_2001} Lee, R., \& Lampe, M. 1973, \prl, 31, 1390

\bibitem[Letaw et al.(1983)]{letaw_1983} Letaw, J. R., Silberberg, R. \& Tsao, C. H. 1983, \apj,  51, 271
 
\bibitem[Lorimer et al.(2007)]{lorimer_2007} Lorimer, D.~R., Bailes, M., McLaughlin, M.~A., Narkevic, D.~J., \& Crawford, F.\ 2007, Science, 318, 777
 
\bibitem[Lorimer(2018)]{lorimer_2018}  Lorimer, D.~R.\ 2018, Nature Astronomy, 2, 860

\bibitem[Macquart et al.(2018)]{macquart_2018} {Macquart}, J. -P., {Shannon}, R.~M., {Bannister}, K.~W., 
         {James}, C.~W. \& {Ekers}, R.~D. \& {Bunton}, J.~D., 2018, apj, 872, L19

\bibitem[Marcote et al.(2017)]{marcote_2017} Marcote, B., Paragi, Z., Hessels, J. W. T., et al. 2017, \apj, 834, L8

\bibitem[Marquez \& Menezes(2017)]{marquez_2017} Marquez, K. D. \& Menezes, D. 2017, \jcap, 12, 028 
      
 \bibitem[Masui et al.(2015)]{masui_2015} Masui, K., Lin, H.-H., Sievers, J., et al. 2015, \nat, 528, 523  
 
 \bibitem[McCarthy et al.(2008)]{mccarthy_2008}   McCarthy, I. G., Frenk, C. S., Font A. S., et al. 2008, MNRAS, 383, 593
         
 \bibitem[McQuinn(2016)]{mcquinn_2016} McQuinn, M. 2016, Annu. Rev. Astron. Astrophys. 2016, 54, 313
  
\bibitem[Medvedev \& Loeb(1999)]{medvedev_1999} Medvedev, M. V., \& Loeb, A 1999, \apj, 526, 697

\bibitem[Medvedev et al.(2005)]{medvedev_2005} Medvedev, M. V., \& Fiore, M., Fonseca, R. A., Silva L. O., \& Mori, W. B. 2005, \apj, 618, L75

\bibitem[Mellinger et al.(2017)]{mellinger_2017} Mellinger, R.,  Weber, F., Spinella, W., Contrera, G., \& Orsaria, Milva 2017, Universe, 3, 5 

\bibitem[Metzger et al.(2017)]{metzger_2017} Metzger B. D., Berger E., Margalit B., 2017, \apj, 841, 14
 
 \bibitem[Michilli et al.(2018)]{michilli_2018} Michilli, D., Seymour, A., Hessels, J.~W.~T., et al.\ 2018, \nat, 553, 182
 
 \bibitem[Milosavljevic \& Nakar(2006)]{milosavljevic_2006}Milosavljevi\'c, M. \& \& Nakar, E. 2006, \apj, 641, 978
 
 \bibitem[Moreno et al.(2018)]{moreno_2018}    Moreno, Q. et al. 2018, Phys. Plasmas, 25, 062125
 
 \bibitem[Motz(1951)]{motz_1951} Motz, H. 1951, J. Appl. Phys. 22, 527
 
\bibitem[Murase \& Takami(2009)]{murase_2009}  Murase, K., \& Takami, H. 2009, ApJ, 690, L14
 
 \bibitem[Murphy et a.(1997)]{murphy_1997} Murphy, J. B., Krinsky, S. \& Gluckstern, R. L., Part. Accel. 1997, 57, 9 
 
\bibitem[Niebergal et al.(2010a)]{niebergal_2010a} Niebergal, B., Ouyed, R., \& Jaikumar, P. 2010a, \prc, 82, 062801

\bibitem[Niebergal et al.(2010b)]{niebergal_2010b} Niebergal, B.,  Ouyed, R.,  Negreiros, R. \&   Weber, F., 2010b, \prd,  81, 043005 

\bibitem[Niebergal(2011)]{niebergal_phd_2011} Niebergal, B., 2011, ``Hadronic-to-Quark-Matter Phase Transition: Astrophysical Implications", Thesis
(Ph.D.), University of Calgary (Canada), 2011.; Publication Number: AAT NR81856; ISBN:
9780494818565

\bibitem[Nishikawa et al.(2009)]{nishikawa_2009} Nishikawa K.-I., Niemiec J., Hardee P. E., et al. 2009, \apjl, 698, L10 

\bibitem[Nodvick \& Saxon(1954)]{nodvick_1954} Nodvick, J. S. \& Saxon, D. S. 1954, Phys. Rev. 96, 180 

\bibitem[Ohira \& Takahara(2008)]{ohira_2008} Ohira, Y. \& Takahara, F.\ 2008, \apj, 688, 320



\bibitem[Ouyed et al.(2004)]{ouyed_2004} Ouyed, R., Elgar\o y, \O., Dahle, H.,
\& Ker\"anen, P. 2004, A\&A, 420, 1025

\bibitem[Ouyed et al.(2006)]{ouyed_2006} Ouyed, R., Niebergal, B., Dobler, W., \& Leahy, D.\ 2006, \apj, 653, 558

\bibitem[Ouyed et al.(2007a)]{ouyed_2007a} Ouyed, R.,  Leahy, D. \& Niebergal, B.,\ 2007a, \aap, 473, 357

\bibitem[Ouyed et al.(2007b)]{ouyed_2007b} Ouyed, R.,  Leahy, D. \& Niebergal, B.,\ 2007b, \aap, 475, 63

\bibitem[Ouyed \& Leahy(2009)]{ouyed_leahy_2009} Ouyed, R., \& Leahy, D. 2009, \apj, 696, 562

\bibitem[Ouyed et al.(2014)]{ouyed_2014} Ouyed, R., Koning, N., Leahy, D., Staff, J.~E., \& Cassidy, D.~T.\ 2014, \raa, 14, 497-519

\bibitem[Ouyed et al.(2018a)]{ouyed_2018a}  Ouyed, R., Leahy, D., \& Koning, N. 2018a, \apj, 818, 77, ``Quark-novae in binaries: Observational signatures and implications to astrophysics", Proceedings of the Fourteenth Marcel Grossmann Meeting, 12-18 July 2015, Rome, Italy. Eds.  M. Bianchi, R. T. Jansen and R. Ruffini (World Scientific Publishing Co. Pte. Ltd., ISBN \#9789813226609), 1877

\bibitem[Ouyed et al.(2018b)]{ouyed_2018b}  Ouyed, R., Leahy, D., \& Koning, N. 2018b, \apj, 818, 77, ``Quark-nova compact remnants: Observational signatures in astronomical data and implications to compact stars", Proceedings of the Fourteenth Marcel Grossmann Meeting, 12-18 July 2015, Rome, Italy. Eds.  M. Bianchi, R. T. Jansen and R. Ruffini (World Scientific Publishing Co. Pte. Ltd., ISBN \#9789813226609), 3387

\bibitem[Ouyed al.(2020)]{ouyed_2020}  Ouyed, R., Leahy, D., \& Koning, N. 2020, \raa, 20, 27

\bibitem[Ouyed, A et al.(2018a)]{ouyed_amir_2018a} Ouyed, A., Ouyed, R., \& Jaikumar, P. 2018a, Physics Letters B, 777, 184

\bibitem[Ouyed, A et al.(2018b)]{ouyed_amir_2018b} Ouyed, A., Ouyed, R., \& Jaikumar, P. 2018b, Universe, 4, 51


\bibitem[Ouyed, A(2018)]{ouyed_amir_phd_2018} Ouyed, A. 2018, ``The Neutrino Sector in Hadron-Quark Combustion: Physical and Astrophysical
Implications", Thesis (Ph.D.), University of Calgary (Canada), 2018 [\url{http://dx.doi.org/10.11575/PRISM/27841}]


\bibitem[Ouyed, A(2019)]{ouyed_amir_2019} Ouyed, A., Ouyed, R., \& Jaikumar, P. 2019, ?The Structure of the Hadron-Quark Reaction Zone?, in
Proceedings of the Compact Stars in the QCD Phase Diagram VII (CSQCD VII), June 11 - 15,
2018, NY, NY. Universe, 5(6), 136, eds. Vivian de la Incera, Efrain Ferrer, James Lattimer and David
Blaschke [arXiv:1906.08404]

 

\bibitem[Petroff et al.(2016)]{petroff_2016} Petroff, E., Barr, E.~D., Jameson, A., et al.\ 2016, \pasa, 33, e045

\bibitem[Petroff et al.(2019)]{petroff_2019} Petroff, E., Hessels, J.~W.~T. \& Lorimer, D.~R., 2019, Astronomy and Astrophysics Reviews, 27, 4

\bibitem[Phillips(1993)]{phillips_1993} Phillips, M. M. 1993, \apj, 413, L105

\bibitem[Piran(1999)]{piran_1999} Piran, T. 1999, Phys. Rep., 314, 575

\bibitem[Platts et al.(2018)]{platts_2019} Platts, E., Weltman, A., Walters, A., et al.\ 2019, \physrep, 821, 1 

\bibitem[Popov et al.(2018)]{popov_2018} Popov, S.~B., Postnov, K.~A., \& Pshirkov, M.~S. 2018, Physics Uspekhi, 61, 965

\bibitem[Quilis et al.(2000)]{quilis_2000}  Quilis V., Moore B., \& Bower R., 2000, Science, 288, 1617

\bibitem[Rajagopal(1999)]{rajagopal_1999} Rajagopal, K. 1999, \nphysa, 661, 150

\bibitem[Rajawat \& Sengupta(2016)]{rajawat_2016} Rajawat, R. S. \& Sengupta, S. 2016, Physics of Plasmas 23, 102110

\bibitem[Ravi et al.(2016)]{ravi_2016} Ravi, V., Shannon, R.~M., Bailes, M., et al.\ 2016, Science, 354, 1249 

\bibitem[Ravi(2019)]{ravi_2019} Ravi, V. 2019, Nature Astronomy, 3, 928 

\bibitem[Richardson(2019)]{richardson_2019} Richardson, A. S. 2019, NRL Plasma Formulary (Naval Research Lab Washington, DC, Pulsed Power Physics Branch)

\bibitem[Ryden(2016)]{ryden_2016} Ryden, B. S. 2016, "Introduction to Cosmology", Cambridge University Press

\bibitem[Salpeter(1955)]{salpeter_1955} Salpeter, E. E. 1955, \apj, 121, 161

\bibitem[Schwinger(1949)]{schwinger_1949} Schwinger, J. 1949, Phys. Rev. 75, 12, 1912 

\bibitem[Schiff(1946)]{schiff_1946}  Schiff, L. I., 1946, Rev. of Sci. Instr. Vol. 7, Num. 1, p. 6 

\bibitem[Scholz et al.(2016)]{sholz_2016}  Scholz, P., Spitler, L.~G., Hessels, J.~W.~T., et al.\ 2016, \apj, 833, 177

\bibitem[Scholz et al.(2020)]{scholz_2020} Scholz, P. et al., 2020, The Astronomer?s Telegram 13681, 1

\bibitem[Shannon et al.(2018)]{shannon_2018} Shannon, R. et al. 2018, \nat, 562, 386

  \bibitem[Spitkovsky(2008)]{spitkovsky_2008} Spitkovsky A.\ 2008, \apj,  673, L39

\bibitem[Spitler et al.(2014)]{spitler_2014} Spitler, L.~G., Cordes, J.~M., Hessels, J.~W.~T., et al.\ 2014, \apj, 790, 101

\bibitem[Spitler et al.(2016)]{spitler_2016} Spitler, L. G., Scholz, P., Hessels, J. W. T., et al.  2016, Nature 2016, 531, 202

\bibitem[Staff et al.(2006)]{staff_2006}Staff, J., Ouyed, R., \& Jaikumar, P., 2006, \apj, 645, L145

\bibitem[Staff et al.(2012)]{staff_2012} Staff, J.~E., Jaikumar, P., Chan, V., \& Ouyed, R.\ 2012, \apj, 751, 24

\bibitem[Tanabashi et al.(2018)]{tanabashi_2018}  Tanabashi, M. et al. (Particle Data Group), 2018, \prd 98, 010001 

\bibitem[Takamoto et al.(2018)]{takamoto_2018}  Takamoto, M., Matsumoto, Y., \& Kato, T. 2018, \apj, 860, L1 

\bibitem[Takamoto et al.(2019)]{takamoto_2019} Takamoto, M., \& Matsumoto, Y. \& {Kato}, T. N., 2019, \apj, 877, 137

\bibitem[Tendulkar et al.(2017)]{tendulkar_2017} Tendulkar, S. P. et al. 2017, \apj, 834, L7
 
\bibitem[Thornton et al.(2013)]{thornton_2013} Thornton, D., Stappers, B., Bailes, M., et al.\ 2013, Science, 341, 53

  \bibitem[Tully(1987)]{tully_1987}  Tully R. B., 1987, ApJ, 321, 280

\bibitem[van Haarlem et al.(2013)]{vanhaarlem_2013} van Haarlem, M. P., Wise, M. W., Gunst, A. W., et al. 2013, A\&A, 556, A2

  \bibitem[Venturini \& Warnock(2002)]{venturini_2002} Venturini, M. \& Warnock, R. 2002, \prl, 89, 224802
  
  \bibitem[Vogt et al.(2004)]{vogt_2004} Vogt C., Rapp R., \& Ouyed R., 2004, Nuclear Physics A, 735, 543
  
\bibitem[Weber(2005)]{weber_2005}   Weber, F. 2005, Progress in Particle and Nuclear Physics, 54, 193

\bibitem[Weibel(1959)]{weibel_1959} Weibel, E. 1959, \prl, 2, 83 

\bibitem[Witze(2020)]{witze_2020} Witze, A., \nat, 2020, 583, 322

  \bibitem[Yoon \& Davidson(1987)]{yoon_1987} Yoon,  P.  H.,  \&  Davidson,  R.  C, 1987, Phys. Rev. A, 35, 2718



\end{thebibliography}
\end{document}